\documentstyle[preprint,prd,aps,amsfonts,epsf]{revtex} % 12 pt
\tightenlines
\newcommand{\sigmaBr}{
	0.132 ^{+0.041}_{-0.037} \, ({\rm stat.}) \, 
		\pm 0.031 \, ({\rm syst.}) \,
	  ^{+0.032}_{-0.020} \, ({\rm lifetime})
}
\newcommand{\ctauBc}{
	137^{+53}_{-49}\,({\rm stat.}) 	\pm 9\,({\rm syst}) \: \mu m
}

\newcommand{\tauBc}{
	0.46^{+0.18}_{-0.16}\, ({\rm stat.}) \, 
	\pm 0.03 \, ({\rm syst.}) \: {\rm ps }
}

\newcommand{\dpr}{\prime \prime}
\newcommand{\Bc}{$B_c$}
\newcommand{\Bcp}{$B_c^+$}

\newcommand{\Jpsi}{$J/\psi$}

\newcommand{\Jpsimu}{$J/\psi \, \mu $}
\newcommand{\Jpsie}{$J/\psi  \, e   $}
\newcommand{\Jpsil}{$J/\psi  \, \ell$}
\newcommand{\JpsiK}{$J/\psi  \, K$}
\newcommand{\JpsiKpm}{$J/\psi  \, K^{\pm}$}
\newcommand{\JpsiX}{$J/\psi  \, X$}

\newcommand{\BJpsilnu}{$B_c  \rightarrow J/\psi \, \ell \, \nu$}
\newcommand{\BJpsimuX}{$B_c \rightarrow J/\psi \, \mu  \, X$}
\newcommand{\BJpsieX}{$B_c  \rightarrow J/\psi \, e    \, X$}
\newcommand{\BJpsilX}{$B_c  \rightarrow J/\psi \, \ell \, X$}
\newcommand{\BJpsimu}{$B_c \rightarrow J/\psi \, \mu$}
\newcommand{\BJpsie}{$B_c  \rightarrow J/\psi \, e$}
\newcommand{\BJpsil}{$B_c  \rightarrow J/\psi \, \ell \, \nu$}
\newcommand{\BJpsiK}{$B  \rightarrow J/\psi \, K$}
\newcommand{\BJpsiX}{$B  \rightarrow J/\psi \, X$}

\newcommand{\cc}{$c$}

\newcommand{\bbar}{$\overline{b}$}
\newcommand{\ccbar}{$c\overline{c}$}
\newcommand{\bbbar}{$b\overline{b}$}
\newcommand{\BBbar}{$B\overline{B}$}
\newcommand{\PbarP}{$p\overline{p}$}
\newcommand{\pbarp}{$p\overline{p}$}

\newcommand{\Pt}{$p_T$}

\newcommand{\etaphi}{$\eta$--$\phi$}

\newcommand{\ipb}{pb$^{-1}$}

\renewcommand{\deg}[1]{#1$^{\circ}$}
\newcommand{\um}{$\mu$m}

\newcommand{\chisq}{$\chi^2$}

\newcommand{\ctau}{$c\tau$}
\newcommand{\ct}{$ct$}
\newcommand{\ctstar}{$ct^{\ast}$}

\newcommand{\sigBRBcl}{$\sigma\cdot BR(B_c^+\rightarrow J/\psi\,\ell^+ \nu)$}

\begin{document}
\draft

\preprint{\begin{minipage}[t]{3in} 
		\flushright
		FERMILAB-PUB-98/121-E
	\\	\today 
		\end{minipage} }

\title{
\vspace*{0.2in} 
Observation of \mbox{\boldmath $B_c$} Mesons 
in \PbarP\ Collisions at $\sqrt{s} = 1.8$\ TeV}

\def\r#1{\ignorespaces $^{#1}$}
\author{
%The CDF Collaboration}
% ---------------------- \input{authors} -----------
\font\eightit=cmti8
\hfilneg
\begin{sloppypar}
\noindent
F.~Abe,\r {17} H.~Akimoto,\r {39}
A.~Akopian,\r {31} M.~G.~Albrow,\r 7 A.~Amadon,\r 5 S.~R.~Amendolia,\r {27} 
D.~Amidei,\r {20} J.~Antos,\r {33} S.~Aota,\r {37}
G.~Apollinari,\r {31}
% ...
\end{sloppypar}
\vskip .026in
\begin{center}
(CDF Collaboration)
\end{center}
\vskip .026in
\begin{center}
\r 1  {\eightit Argonne National Laboratory, Argonne, Illinois 60439} \\
\r 2  {\eightit Istituto Nazionale di Fisica Nucleare, University of Bologna,
I-40127 Bologna, Italy} \\
\r 3  {\eightit Brandeis University, Waltham, Massachusetts 02254} \\
\r 4  {\eightit University of California at Los Angeles, Los 
Angeles, California  90024} \\  
\r 5  {\eightit University of Chicago, Chicago, Illinois 60637} \\
\r 6  {\eightit Duke University, Durham, North Carolina  27708} \\
\r 7  {\eightit Fermi National Accelerator Laboratory, Batavia, Illinois 
60510} \\
\r 8  {\eightit University of Florida, Gainesville, FL  32611} \\
\r 9  {\eightit Laboratori Nazionali di Frascati, Istituto Nazionale di Fisica
               Nucleare, I-00044 Frascati, Italy} \\
\r {10} {\eightit University of Geneva, CH-1211 Geneva 4, Switzerland} \\
\r {11} {\eightit Harvard University, Cambridge, Massachusetts 02138} \\
\r {12} {\eightit Hiroshima University, Higashi-Hiroshima 724, Japan} \\
\r {13} {\eightit University of Illinois, Urbana, Illinois 61801} \\
\r {14} {\eightit Institute of Particle Physics, McGill University, Montreal 
H3A 2T8, and University of Toronto,\\ Toronto M5S 1A7, Canada} \\
\r {15} {\eightit The Johns Hopkins University, Baltimore, Maryland 21218} \\
\r {16} {\eightit Institut f\"{u}r Experimentelle Kernphysik, 
Universit\"{a}t Karlsruhe, 76128 Karlsruhe, Germany} \\
\r {17} {\eightit National Laboratory for High Energy Physics (KEK), Tsukuba, 
Ibaraki 305, Japan} \\
\r {18} {\eightit Ernest Orlando Lawrence Berkeley National Laboratory, 
Berkeley, California 94720} \\
\r {19} {\eightit Massachusetts Institute of Technology, Cambridge,
Massachusetts  02139} \\   
\r {20} {\eightit University of Michigan, Ann Arbor, Michigan 48109} \\
\r {21} {\eightit Michigan State University, East Lansing, Michigan  48824} \\
\r {22} {\eightit University of New Mexico, Albuquerque, New Mexico 87131} \\
\r {23} {\eightit The Ohio State University, Columbus, OH 43210} \\
\r {24} {\eightit Osaka City University, Osaka 588, Japan} \\
\r {25} {\eightit Universita di Padova, Istituto Nazionale di Fisica 
          Nucleare, Sezione di Padova, I-35131 Padova, Italy} \\
\r {26} {\eightit University of Pennsylvania, Philadelphia, 
        Pennsylvania 19104} \\   
\r {27} {\eightit Istituto Nazionale di Fisica Nucleare, University and Scuola
               Normale Superiore of Pisa, I-56100 Pisa, Italy} \\
\r {28} {\eightit University of Pittsburgh, Pittsburgh, Pennsylvania 15260} \\
\r {29} {\eightit Purdue University, West Lafayette, Indiana 47907} \\
\r {30} {\eightit University of Rochester, Rochester, New York 14627} \\
\r {31} {\eightit Rockefeller University, New York, New York 10021} \\
\r {32} {\eightit Rutgers University, Piscataway, New Jersey 08855} \\
\r {33} {\eightit Academia Sinica, Taipei, Taiwan 11530, Republic of China} \\
\r {34} {\eightit Texas A\&M University, College Station, Texas 77843} \\
\r {35} {\eightit Texas Tech University, Lubbock, Texas 79409} \\
\r {36} {\eightit Istituto Nazionale di Fisica Nucleare, University of Trieste/
Udine, Italy} \\
\r {37} {\eightit University of Tsukuba, Tsukuba, Ibaraki 315, Japan} \\
\r {38} {\eightit Tufts University, Medford, Massachusetts 02155} \\
\r {39} {\eightit Waseda University, Tokyo 169, Japan} \\
\r {40} {\eightit University of Wisconsin, Madison, Wisconsin 53706} \\
\r {41} {\eightit Yale University, New Haven, Connecticut 06520} \\
\end{center}
}
% ==================================================

\date{\today}

\maketitle
\begin{abstract}

%----------------- \input{abstract} ----------------
We report the observation of  bottom-charmed mesons $B_c$\ 
in 1.8~TeV \PbarP\ collisions using the CDF detector 
at the Fermilab Tevatron.
The $B_c$ mesons were found through their semileptonic decays, 
$B_c^{\pm} \rightarrow J/\psi \, \ell^{\pm} X$.
A fit to the \Jpsil\ mass distribution yielded 
$20.4^{+6.2}_{-5.5}$\ events from $B_c$ mesons.
A test of the null hypothesis, $i.e.$\ an attempt to fit
the data with background alone, was rejected at the level
of 4.8 standard deviations.
By studying the quality of the fit as a function of the 
assumed \Bc\ mass, we determined
$M(B_c) = 
6.40 \pm 0.39\,{\rm (stat.)} \pm 0.13\,{\rm (syst.)}$\ GeV/$c^2$.
From the distribution of 
trilepton intersection points in the plane
transverse to the beam direction we measured the \Bc\ lifetime to be
$\tau (B_c) = \tauBc$.  
We also measured the ratio of production cross section times
branching fraction for 
$B_c^+ \rightarrow J/\psi \ell^+ \nu$\ relative to that for 
$B^+ \rightarrow J/\psi K^+$\ to be:
\begin{displaymath}
	\frac{\sigma(B_c) \cdot BR(B_c \rightarrow J/\psi \, \ell \nu)}
		   {\sigma(B) \cdot BR(B \rightarrow J/\psi \, K)} 
	= \sigmaBr
\end{displaymath}
% ===============================================

\end{abstract}

\narrowtext

\newpage
\section{Introduction}
\label{sec:intro}
% ------------------- \input{intro} ----------------------
The \Bcp\ meson is the lowest-mass bound state of a
charm quark and a bottom anti-quark.%
\footnote
	{ References to a specific state imply the
	  charge-conjugate state as well. }
It is the pseudoscalar ground state of the third 
family of quarkonium states.
Since the \Bc\ has non-zero flavor, it  
has no strong or electromagnetic decay channels,
and it is the last such meson predicted by the Standard Model.
Its weak decay is expected to yield 
a large branching fraction 
to final states containing a 
\Jpsi~\cite{Lusignoli_decay,isgw2,isgw,Chang_decay},
a useful experimental signature.
 
Non-relativistic potential models are appropriate for the \Bc, 
and they predict its mass.
Kwong and Rosner\cite{Kwong} estimate $M$(\Bc) to be in the 
range 6.194--6.292 GeV/$c^2$.  
Eichten and Quigg\cite{Eichten} discuss four potentials 
that yield values in the range 6.248--6.266 GeV/$c^2$.
In these models, the \cc\ and \bbar\ are tightly bound 
in a very compact system.
These authors describe a rich spectroscopy of 
excited states, which make this the ``hydrogen atom''
or, perhaps, ``the mu-mesic atom'' of QCD.

We expect the full decay width of the \Bc\ to consist of three
major contributions,  $\Gamma = \Gamma_b + \Gamma_c + \Gamma_{bc}$,
which are, respectively,
\begin{itemize}
 \item	$\overline{b} \rightarrow \overline{c} W^+$ 
	with the \cc\ as a spectator, leading to
	final states like $(J/\psi \, \pi)$, $(J/\psi \, \ell \nu)$;
 \item	$c \rightarrow s W^+$,
	with the \bbar\ as spectator, leading to
	final states like $(B_s \, \pi)$, $(B_s \, \ell \nu)$;
 \item	$c \overline{b} \rightarrow W^+$, annihilation leading 
	to final states like $(D^* \, K)$, $(\tau \, \nu_{\tau})$
	or multiple pions.
\end{itemize}
Since these processes lead 
to different final states, their amplitudes do not interfere. 
In the simplest view, the \cc\ and \bbar\ are free, so annihilation is 
suppressed, and the total width is just the sum of the \cc\ and \bbar\ 
total widths, with \cc -decay dominating.  Approximating this by 
$\Gamma(B_c) = \Gamma(D^0) + \Gamma(B^0)$ yields 
$\tau(B_c) \approx 0.3$\ ps~\cite{Bigi}.
When annihilation, phase space considerations 
(which reduce the relative importance of the \cc\ contribution)
and other effects are included, the predictions increase
to the range 0.4--0.9
ps~\cite{Lusignoli_decay,Bigi,Beneke,Gershtein,Colangelo}. 
Quigg~\cite{Quigg} emphasizes the relatively large ratio of
the binding energy to charm-quark mass and the effect 
on $\Gamma_{bc}$\ of the 
compact size of the \cc\bbar\ system, where
the pseudo-scalar decay 
constant is expected to be $f_{B_c} \approx 500$\ MeV.
He predicts lifetimes in the range 1.1--1.4~ps, with $\Gamma_b$ 
as the largest contribution.  
Thus, a \Bc\ lifetime measurement is a test of the different
assumptions made in the various calculations.
Several authors have also calculated the \Bc\ partial decay rates 
to semileptonic final 
states~\cite{Lusignoli_decay,isgw2,isgw,Chang_decay,Choi}.

In perturbative QCD calculations of \Bc\ production using the 
fragmentation approximation, the dominant process 
is that in which a \bbar\ is produced by gluon fusion in the hard
collision and 
fragmentation provides the 
\cc~\cite{Lusignoli_prod,Braaten,Chang_prod,Oakes,Masetti}.
A full $\alpha_s^4$\ calculation shows that fragmentation dominates 
only for transverse momenta large compared to the \Bc\ mass,
$i.e.$\ $p_T \gg M_{B_c}c$~\cite{Oakes}.
This calculation
provides inclusive production cross sections along with distributions in 
transverse momentum \Pt\ and other kinematic variables.

There have been several experimental searches for the \Bc\ meson.
In $e^+e^-$\ collisions at the $Z$ resonance at LEP,
90\% confidence level (C.L.) upper limits have been placed 
on various branching-fraction products by 
the DELPHI collaboration~\cite{BcDELPHI},
the OPAL collaboration~\cite{BcOPAL}, 
and the ALEPH collaboration~\cite{BcALEPH}.
In Sec.~\ref{sec:cross}, we compare these limits with 
our result.
OPAL reported one event in the semileptonic channel 
where the background was estimated to be $(0.82 \pm 0.19)$\ event,
along with two 
$B_c^{\pm} \rightarrow J/\psi \, \pi^{\pm}$\ 
candidates with an estimated 
background of $(0.63 \pm 0.20)$\ events. 
The mean mass of the latter two candidates is 
$(6.32 \pm 0.06)$\ GeV/$c^2$.
ALEPH~\cite{BcALEPH} reported one candidate for 
$B_c^+ \rightarrow J/\psi \mu^+ \nu_{\mu}$, 
with a low background probability and a \Jpsimu\ mass too high to 
be explained by a light $B$ meson.
A prior CDF search placed a limit on the production and 
decay of the \Bc\ to \Jpsi\ and a charged pion 
\cite{BcCDF}.

We report here the observation of \Bc\ mesons produced in 
a 110 \ipb\ sample of 1.8 TeV \pbarp\ 
collisions at the Fermilab Tevatron collider using the CDF detector.
We searched for the decay channels \BJpsimuX\ and \BJpsieX\ 
with the \Jpsi\ decaying to muon pairs.%
\footnote
{Because of the large partial widths for \BJpsilnu%
~\cite{Lusignoli_decay,isgw,Choi},
we assume that these modes dominate \BJpsilX\ , and we 
often refer to them simply as \BJpsil\ or \Jpsil . 
In Sections~\ref{sec:life} and \ref{sec:cross} 
we discuss this further.}
Even the lowest prediction for the \Bc\ lifetime~\cite{Bigi} 
implies that a significant fraction of \Jpsi\ daughters from \Bc\ 
would have decay points (secondary vertices) 
displaced from the beam centroid (primary vertex) 
by detectable amounts.
The existence of an additional identified lepton track
that passes through the same displaced vertex completes the 
signature for a candidate event.  
We have identified 37 events with $J/\psi \, \ell$\ 
mass between 3.35 GeV/$c^2$ and 11.0 GeV/$c^2$.  
Of these, 31 events lie in a signal region 
$4.0$\ GeV/$c^2 < M(J/\psi \, \ell) < 6.0$\ GeV/$c^2$.

The most crucial and demanding step in the analysis is understanding
the backgrounds that can populate the mass 
distribution~\cite{suzuki_thesis,singh_thesis}.
We attribute any excess over expected background to 
production of the \Bc, 
the only particle yielding a displaced-vertex, 
three-lepton final state with a mass in this region. 
The bulk of the background arises from real \Jpsi\ mesons accompanied 
by hadrons that erroneously satisfy our selection criteria for 
an electron or a muon 
or by leptons that have tracks accidentally passing through 
the displaced \Jpsi\ vertex.

In the sections that follow, we begin with a very brief discussion
in Sec.~\ref{sec:detect} 
of some parts of the CDF detector, particle identification, and 
identificaton of \Jpsi\ through its decay to a muon pair.
Following this,  
we describe our selection criteria for tri-lepton events 
(Sec.~\ref{sec:select}), 
our calculation of the number of background events in the signal 
region (Sec.~\ref{sec:background}), 
and the validation procedures 
to establish the accuracy of that calculation (App.~\ref{app:valid}).

Section~\ref{sec:fit} describes the procedures we used to 
establish the existence of the \Bc\ contribution to our
sample of candidates. 
The background calculations and the mass distribution of the 
\Jpsil\ data sample were subjected to a statistical
analysis from which we calculated the \Bc\ contribution to the 
signal region.
We describe first a simple ``counting experiment'' calculation
for events in this region.
However, we base our claim for the existence of the \Bc\ 
on a likelihood fit that exploits information about 
the shape of the signal and background distributions in the 
mass range 3.35--11.0 GeV/$c^2$, which we call the fitting region.
The \Bc\ contribution to these data is $20.4^{+6.2}_{-5.5}$\ events.
The null hypothesis is rejected at a level of 4.8 standard deviations,
$i.e.$\ the probability that the background could fluctuate high enough to 
explain this excess is less than $0.63 \times 10^{-6}$.

In Sec.~\ref{sec:mass},
by studying the quality of the fit as we varied the assumed \Bc\ 
mass, we obtained an estimate of $M(B_c)$.
In Sec.~\ref{sec:life}, 
we describe our measurement of the \Bc\ lifetime, 
and in Sec.~\ref{sec:cross} we describe 
our measurement of the cross-section times branching-fraction ratio:
\begin{displaymath}
\frac{\sigma(B_c^+) \cdot Br(B_c^+ \rightarrow J/\psi \, \ell^+ \nu)}%
{\sigma(B_u^+) \cdot Br(B_u^+ \rightarrow J/\psi \, K^+)}
\end{displaymath}
We chose this form because many of the uncertainties
cancel in the ratio.

% ========================================================

\newpage
\section{Detector and Particle Identification }
\label{sec:detect}
%------------- \input{detect}-----------------------
We collected the data used in this analysis at the Fermilab Tevatron Collider 
with the Collider Detector at Fermilab (CDF)
during the 1992--1995 run.
The integrated luminosity was  110 pb$^{-1}$ of
$\overline{p}p$ collisions at $\sqrt{s}=1.8$ TeV.
We have described the CDF detector in detail elsewhere~\cite{CDF1,CDF2}.
We describe only those components that are important for this
report.

The events we sought, \BJpsil\ where $J/\psi \rightarrow \mu^+ \mu^-$,
have a very simple topology:  three 
charged particle tracks emerging from a decay point displaced from the primary
interaction point.  For each track, the momentum must be known, 
along with its identity, $\mu$ or $e$.  
Below we describe the charged-particle tracking system,
the electron identification system, the muon identification system, 
the real-time triggers, and \Jpsi\ identification.

\subsection{Charged Particles}
Our  cylindrical coordinate system defines 
the $z$ axis to be the proton beam direction, with $\phi$ 
as the azimuthal angle and $r$ as the transverse distance.
Three tracking subsystems detect charged particles as they pass
through a 1.4 T solenoidal magnetic field.  
We discuss them in order of 
increasing distance from the beam axis.
\begin{itemize}
 \item	The silicon vertex detector, SVX, provides $r$--$\phi$ 
information with good resolution close to the interaction vertex. 
It consists of four approximately cylindrical layers of silicon 
strip detectors outside the beam vacuum pipe and concentric with 
the beam line.
The active area of silicon is centered within the overall CDF detector
and extends 25.5 cm in each direction along the beam line.  
The four layers of detectors are at 
radii of 3.0, 4.2, 5.7, and 7.9 cm~\cite{CDF3,CDF4}.
The strips are arranged axially, and have a 
pitch of 60 $\mu$m for the three innermost layers and a pitch of 55 
$\mu$m for the outermost layer.  
 \item	A set of time projection chambers 
provided $r$--$z$ information that was used to 
determine the event vertex position in $z$,  
which serves as a seed in the reconstruction of tracks
in the $r$--$z$ view in the drift chamber described next.
 \item	The central tracking chamber (CTC) 
is an 84-layer cylindrical drift chamber,  
which covers the pseudorapidity interval
$|\eta | < 1$ (where $\eta=-\ln[\tan(\theta/2)]$ and 
$\theta$ is the polar angle with respect to the proton beam direction).  
It consists of five superlayers of axial sense wires interleaved with 
four small-angle stereo superlayers at an angle of about \deg{3} with
respect to the axial wires.   
In each axial (stereo) superlayer there are twelve (six) cylindrical 
layers of sense wires.
The efficiency for track reconstruction is about 95\% 
and independent of \Pt\ for tracks with $p_T > 0.5$\ GeV/$c$. 
From the reconstructed tracks, 
we used charge deposition from hits in the outer 54 layers of the CTC 
to measure the specific ionization ($dE/dx$) of particles 
with about 10\% uncertainty.
This enabled us to determine the relative $\pi/K/p$\ contributions 
in background calculations.  Specific ionization was also used 
as one of the electron identification criteria.
\end{itemize}
The combined data from SVX and  CTC,
required for all tracks in this analysis, 
have a momentum resolution 
$\delta p_T/p_T = [(0.0009\times p_T)^2 + (0.0066)^2]^{1/2}$, 
where \Pt\ is in units of GeV/$c$,
and the average track impact parameter resolution is 
%$(16^2 + (28/p_T)^2)^{1/2}~\mu$m 
$(13 + (40/p_T))~\mu$m 
relative to the origin of the coordinate system 
in the plane transverse to the beam~\cite{CDF3}.

\subsection{Electron Identification}
Electrons were identifed by the association of a charged-particle track 
with $p_T>2$\,GeV$/c$
and an electromagnetic shower in the calorimeter~\cite{CDF2}.
The central ($|\eta|<1.1$) calorimeter is
divided into towers that subtend \deg{15} in azimuth and 0.11 units
of pseudorapidity.
Each tower has two depth segments, 
a nineteen-radiation-length electromagnetic compartment (CEM)
and a hadronic compartment.  

The track must project sufficiently far from a tower boundary that the
energy deposition by an electromagnetic shower would be largely 
contained within a single tower.  
The energy $E$ observed in the CEM tower must be roughly consistent
with the momentum $p$ of the track, 
{\it viz.}, $0.7 < E/pc < 1.5$, and we require that the energy in 
the hadron compartment of this tower be less than 10\% of that
found in the CEM.

Information from other detectors further improves electron identification.
The value of $dE/dx$ measured in the CTC
must be consistent with that expected for an electron.  
Pre-radiator chambers located between the magnet coil 
(one radiation length thick) and the CEM must show a signal
equivalent to at least four minimum-ionizing particles. 
Proportional chambers with both wire and cathode-strip
readout are located in the CEM 
at a depth of six radiation lengths. 
The shower profile observed in orthogonal views in these chambers 
must be consistent in pulse height, 
shape, and position with those found for
electrons.

Real electrons can arise from photon conversions to $e^+e^-$ pairs, 
including internal conversions in $\pi^0\rightarrow\gamma e^+e^-$.
These can be identified and rejected 
when the candidate electron, paired with 
an oppositely charged track in the event, is 
kinematically consistent with the hypothesis 
$\gamma \rightarrow e^+e^-$.
However, such tracks were useful in direct measurements of 
our electron identification efficiency.

\subsection{Muon Identification}
Muons from \Jpsi\ decay were identifed by matching a 
charged-particle track with $p_T>2$\,GeV$/c$
to a track segment found in the  muon drift
chambers  that lie outside the central calorimeter.
The calorimeter presents five interaction lengths for 
$|\eta|<0.6$\ (CMU detector) and six to nine interaction lengths 
for $0.6<|\eta|<1.0$\ (CMX detector). 
Within the uncertainty introduced by multiple Coulomb scattering, 
we required the charged-particle tracks found in the CTC and SVX 
to project to the track segments in these drift chambers
within three standard deviations.

We refer to the muon produced directly in the $B_c$ semileptonic
decay as the ``third muon,'' and we apply stricter requirements 
to identify it~\cite{CDF2}.
The transverse momentum of the third muon was required to exceed 3\,GeV/$c$.
A third muon must project to a track
segment in the CMU, and 
for further suppression of backgrounds   
must pass through 
an additional three interaction lengths of steel to produce 
a track segment in a second set of drift chambers (CMP detector).
These chambers cover about two-thirds of the solid angle for 
$|\eta|<0.6$.  Above 3 GeV/$c$, the efficiency for a muon track to 
match track segments in both the CMU and the CMP 
is independent of $p_T$.%

\subsection{\Jpsi\ Selection}
The CDF detector includes a three-level, real-time trigger system with 
options that can be used to select events appropriate for a wide
range of physics topics.  
In order to ensure consistent treatment for  
$B_c^+\rightarrow J/\psi \, \mu^+ \, X$ decays,
$B_c^+\rightarrow J/\psi \,  e^+ \, X$ decays, and the
$B^+  \rightarrow J/\psi \, K^+$ decays used for the cross-section
normalization, we required that the muons from
the $J/\psi$ decay satisfy the di-muon trigger selection requirements.

The Level-1 trigger identified muon-chamber candidates by requiring a
coincidence between two radially aligned muon chambers.  
Our di-muon trigger required two such coincidences.

The Level-2 di-muon trigger combined the muon 
candidates with information from a fast track processor
that identified tracks from CTC data~\cite{CFT}.
For the first 19.4 \ipb\ of data collected, we required a single match 
between a muon chamber coincidence and a CTC track with 
\Pt\ $> 3$ GeV/$c$.  
The upgraded trigger system used for the remaining 
data required two such matches 
for tracks with \Pt\ $> 2$ GeV/$c$.  
Curves of the \Pt\ thresholds for the fast track processor and for the
muon chambers can be found in Ref.~\cite{B-lifetimes}.

The Level-3 di-muon trigger was a preliminary
event reconstruction in which we required charged
muon candidate pairs with a mass, determined from CTC
information only, between 2.8 and 3.4\,GeV$/c^2$.

Subsequent offline processing performed a comprehensive search for 
all muon candidates in the event.
For consistent treatment of the several decay modes described above,
we required that the muons used to search for \Jpsi\ 
candidates were identical to 
those that triggered the event.
We also required that both muons pass through the SVX.

\begin{figure}
\epsfxsize=6.0in
\centerline{\epsffile{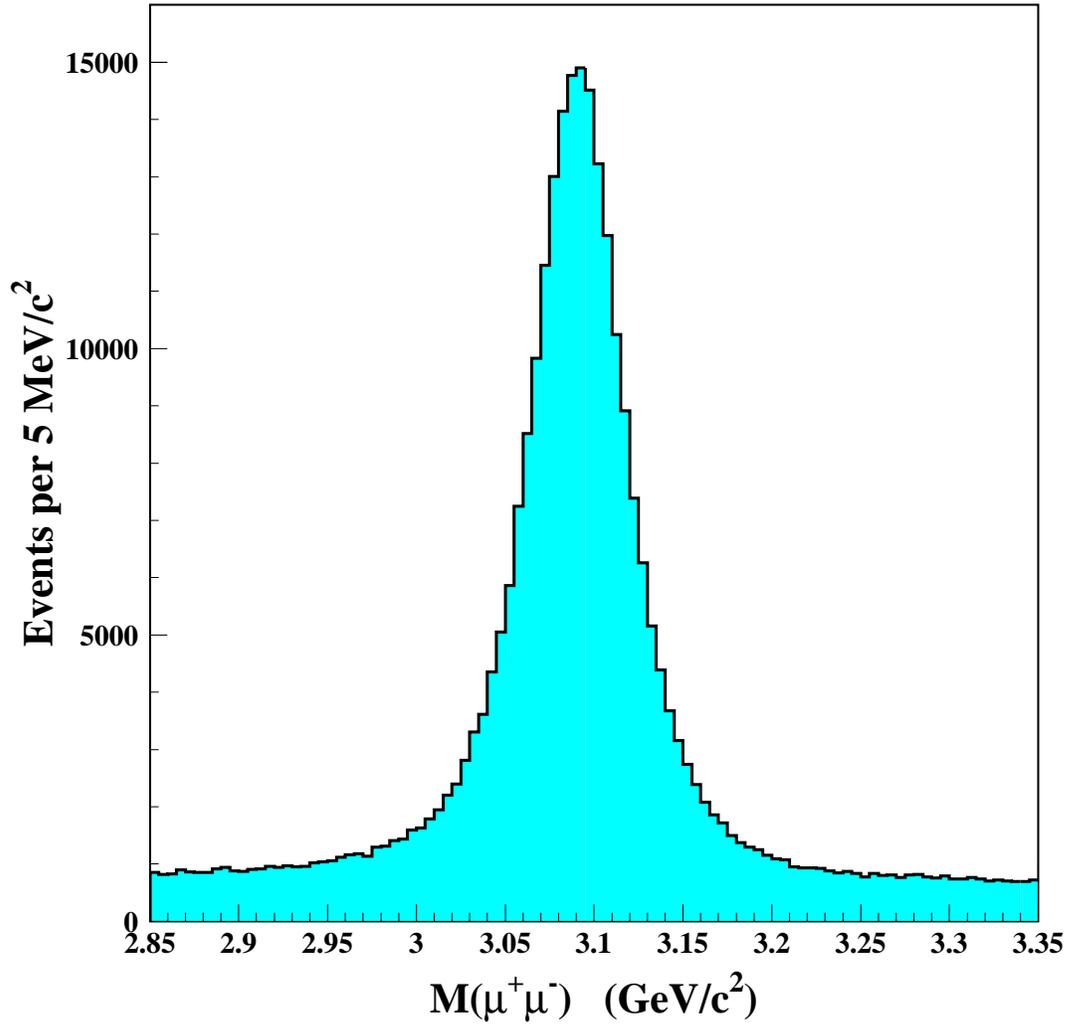}}
\caption{
The distribution of $\mu^+ \mu^-$\ masses.
The data used for further analysis lie between 3.047 and 3.147
GeV/$c^2$\  
and contain $196,000\protect\pm 500$ \Jpsi\ 
events above a background continuum of
$20,000\protect\pm 150$ events under the \Jpsi\ peak.
}
\label{fig:jpsi_mass}
\end{figure}

We performed a $\chi^2$\ fit to the track parameters for pairs of  
oppositely charged muons subject to the constraint that they had
a common origin~\cite{B-lifetimes}.
The di-muon mass was unconstrained.
We required the \chisq\ probability of the fit 
to exceed 1\%. 
The resulting di-muon mass distribution is shown in 
Fig.~\ref{fig:jpsi_mass}.  
The mean mass resolution is 16 MeV$/c^2$.
We required di-muon candidates selected by the offline programs for the \Bc\
analysis to be within 50 MeV/$c^2$\ of the 
world average \Jpsi\ mass of 3096.9 MeV/$c^2$~\cite{Particledata}.

% ========================================================

\newpage
\section{Event Selection}
\label{sec:select}
% ------------------ \input{select} -----------------------
To identify \Bc\ candidates, we searched for events with 
a third track that originated at the $J/\psi$ decay point.
We subjected the three tracks to a \chisq\ fit that 
constrained the two muons to the \Jpsi\ mass and that
constrained all three tracks to orginate from a common point.
We accepted events for which the fit probability satisfied
$P(\chi^2) > 1$\%.
To the resulting samples of \Jpsi\ + track,
we applied further geometric and particle-identification 
criteria for selecting 
\Jpsie\ and \Jpsimu\ events and a kinematic test for 
selecting \JpsiK\ events.%
\footnote
{Differences in the criteria for identifying muons and electrons
yielded different acceptances and backgrounds for the two decay channels.
However, wherever it was possible to adopt common procedures 
for the two channels, we did so.}

The third track for most events was a pion or a kaon.%
\footnote
{Preliminary studies of $dE/dx$\  
for the this sample of tracks showed
the contribution from protons and antiprotons to be negligible 
and it was assumed to be zero thereafter.}
The fitting program corrected individual tracks for ionization 
losses. Consequently, the fit results 
had some slight sensitivity to the mass assumed 
for the third track.
For studies aimed at identifying events with a specific 
third particle ($e^{\pm}$, $\mu^{\pm}$ or $K^{\pm}$) we 
used the appropriate mass.
For generic \Jpsi\ + track studies we used the muon 
mass.

\subsection{ \Jpsi\ + track Decay Vertex Position }
The di-muon fit described in Sec.~\ref{sec:detect} 
constrained the daughter tracks from 
$J/\psi \rightarrow \mu^+ \mu^-$\ 
to come from a common point in space based on information
from the CTC and SVX.
When fitting the two muon tracks of the \Jpsi\ and the 
additional track, we required all three tracks to come 
from the same vertex.
However, the high-resolution information from the SVX provides 
no longitudinal ($z$) coordinate.  Thus, we measured 
the displacement between the beam centroid and the \Jpsi\ 
decay point in the transverse plane.  
The uncertainty in the displacement is typically about 55 \um, 
and the uncertainty in the position of \pbarp\ collision
which produced the \Jpsi\ is 23~\um~\cite{B-lifetimes}.  %Fig 12 in CDF 4227.

$L_{xy}$ is the distance between the beam centroid and the 
decay point of a
\Bc\ candidate projected onto a plane perpendicular to the beam
direction  and projected along the direction of the \Bc\
in that plane.
A measure of the time between production and decay 
of a $B_c$ candidate is the quantity
 \ctstar, defined as
\begin{equation} 
 ct^{\ast} = \frac{M(J/\psi \, \ell) \cdot 
  L_{xy}(J/\psi \, \ell)}{|p_T(J/\psi \,\ell)|} 
\label{eq:ctstar}
\end{equation} 
where $M$(\Jpsil) is the mass of the tri-lepton system and
\Pt(\Jpsil) is its momentum transverse to the beam. 
The average uncertainty in the measurement of \ctstar\ is 
approximately 25 \um.
In order to reduce backgrounds involving prompt \Jpsi\ production,
we required $ct^{\ast} >$\ 60 \um\ 
for all candidates in the analysis of the \Bc\ signal significance.
For the subsequent lifetime analysis (Sec.~\ref{sec:life}), this 
requirement was modified. 

\subsection{\JpsiK\ Identification }
\label{subsec:JpsiK}
The \BJpsiK\ final state has no undetectable particles and 
can be reconstructed fully to calculate the mass of the 
parent $B$\ meson.  
We determined the mass for each \Jpsi\ + track combination
under the hypothesis that the track corresponded to a kaon.

\begin{figure}
\epsfxsize=6.0in
\centerline{\epsffile{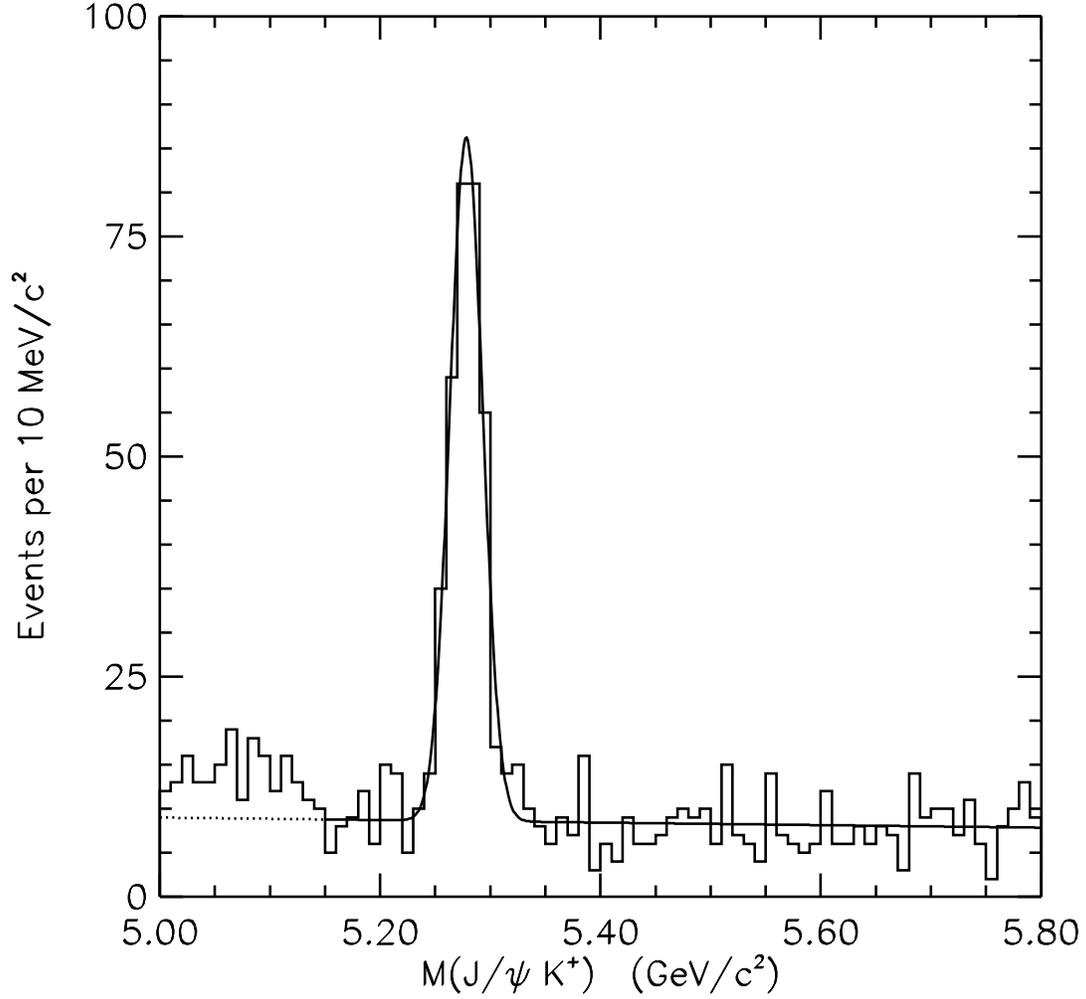}}
\caption{
The distribution of masses of \JpsiKpm\ candidates.  The solid curve represents 
a least squares fit to the data between 5.15 and 5.8 GeV/$c^2$\ 
consisting of a Gaussian signal above a linear background.
The area of the Gaussian contribution is $290 \pm 19$\ events.
}
\label{fig:Jpsi_K}
\end{figure}

Figure~\ref{fig:Jpsi_K} shows the \JpsiK\ 
mass distribution.  The results from this particular data sample 
were used to normalize the 
measurement of the product of the \Bc\ production cross section 
and the \BJpsilnu\ branching fraction
described in Sec.~\ref{sec:cross}.
Events for which $M$(\JpsiK) was within 50 MeV/$c^2$\ 
of $M(B) = 5.2789$\ GeV/$c^2$\ were designated 
as \BJpsiK\ and removed from the sample of candidates 
for \BJpsil.
With different sets of selection criteria, the \JpsiK\ sample
was used to check the calculation 
of the probability for a kaon to be falsely
identified as a muon (Sec.~\ref{subsec:punch}) 
and to  normalize Monte Carlo calculations of backgrounds from
$B \overline{B}$\ pairs (Sec.~\ref{subsec:BBbar}).

\subsection{\Jpsi\ + Lepton Identification }
\label{subsec:leptonid}
\begin{figure}
\epsfxsize=6.0in
\centerline{\epsffile{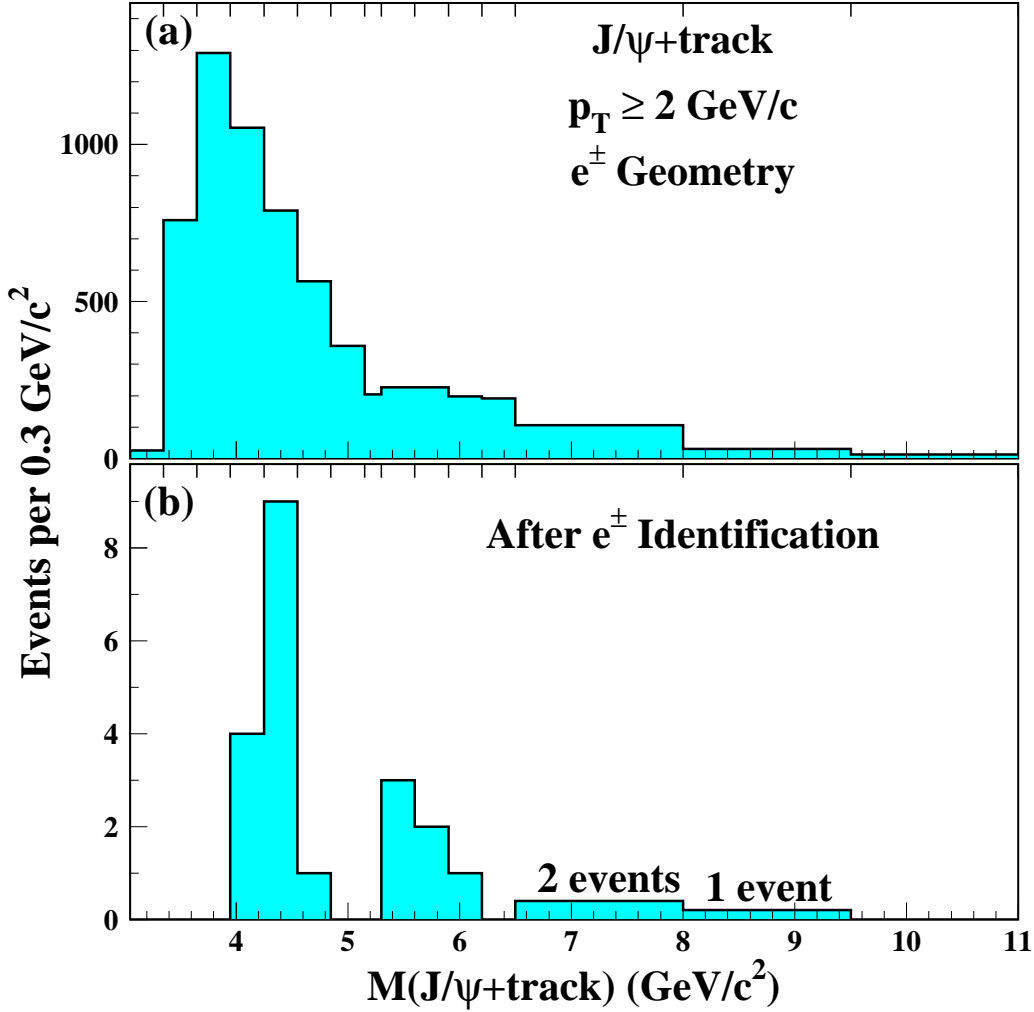}}
\caption{
Histograms of the number of events vs. M(\Jpsi\ + track).
(a) \Jpsie\ candidates.
For the 6530 events in this histogram, we assigned the electron mass to the
third track and required~ $p_T \geq 2.0$\ GeV/$c$. 
We applied the geometric criteria but not the 
particle identification criteria  for electrons.
(b) The 23-event subset of the distribution above 
that satisfies the electron identification criteria.
Note that the bins in  $M$(\Jpsi\ + track) are not uniform in width.
The bin boundaries are indicated by tick marks at the top of the 
figures here and in subsequent mass histograms.
The binning is discussed in the text.
}
\label{fig:jpsi_track_e}
\end{figure}

\begin{figure}
\epsfxsize=6.0in
\centerline{\epsffile{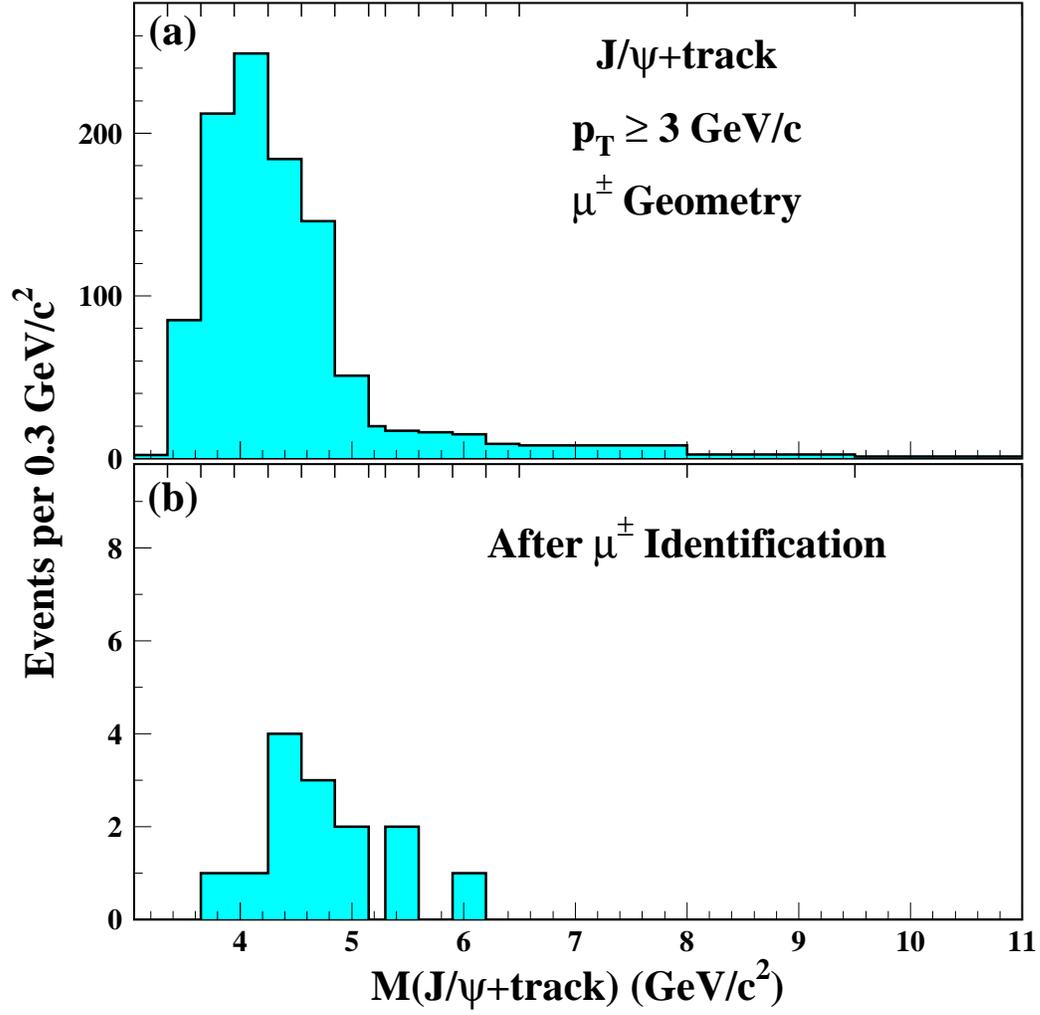}}
\caption{
Histograms of the number of events vs. M(\Jpsi\ + track).
(a) \Jpsimu\ candidates.
For the 1055 events in this histogram, we assigned the muon mass to the
third track and required~ $p_T \geq 3.0$\ GeV/$c$. 
We applied the geometric criteria but not the 
particle identification criteria  for muons.
(b) The 14-event subset from (a) that satisfies the muon
identification criteria.
}
\label{fig:jpsi_track_mu}
\end{figure}

Figures~\ref{fig:jpsi_track_e}(a) and \ref{fig:jpsi_track_mu}(a) 
are histograms of the \Jpsi\ + track   
mass for combinations that  
passed the requirement $P(\chi^2) > 1$\% described above.
We required third tracks to have an opening angle less 
than \deg{90} relative to the \Jpsi\ direction.
This reduced the amount of \BBbar\ background discussed 
in Sec.~\ref{sec:background}.
The \BJpsiK\ events excluded from the \Jpsi\ + track sample  
populate a very narrow region of $M$(\Jpsi+track) in 
Figs.~\ref{fig:jpsi_track_e} and \ref{fig:jpsi_track_mu}.

In the likelihood analysis described 
in Sec.~\ref{sec:fit}, the widths of the mass bins 
are not uniform.
In Fig.~\ref{fig:jpsi_track_e} and in subsequent figures 
containing mass histograms the bin boundaries are indicated
by tick marks at the top of each figure.
Most bins are 0.3 GeV/$c^2$\ wide.
We confined the effects of the excluded events near $M$(\JpsiK) 
to one 0.15 GeV/$c^2$\ bin, which is clearly visible 
in the figures.
We also adopted wider bins at high masses where the 
event population is low.  
We chose the vertical scale so that the number 
of events per 0.3 GeV/$c^2$ is equal to the number
of events per bin for most bins.  
This makes explicit the statistical significance 
for the candidate distributions in 
Figs.~\ref{fig:jpsi_track_e}(b) 
and \ref{fig:jpsi_track_mu}(b).
The event count is displayed for the two
bins in Figs.~\ref{fig:jpsi_track_e}(b) that 
had to be scaled.

With an assumed $B_c$ mass of 6.27 GeV/$c^2$, 
Monte Carlo simulations (App.~\ref{app:monte}) 
reveal that 93\% of the the tri-lepton masses  
reconstructed for \Jpsimu\ and \Jpsie\ decays 
will fall in the range 4.0 to 6.0 GeV/$c^2$.  
We refer to this as the signal region.
When we apply the muon identification criteria 
to events in Fig.~\ref{fig:jpsi_track_mu}(a),
we obtain the mass distribution shown in 
Fig.~\ref{fig:jpsi_track_mu}(b), in which 12 of the 14 events lie 
in the signal region.
When we apply the electron identification criteria described earlier 
to events in Fig.~\ref{fig:jpsi_track_e}(a), 
we obtain the mass distribution shown in 
Fig.~\ref{fig:jpsi_track_e}(b), in which 19 of the 23 events 
lie in the signal region.

The distributions shown in 
Figs.~\ref{fig:jpsi_track_mu}(a) and \ref{fig:jpsi_track_e}(a)
have many events in common because 
most with tracks that satisfy the muon \Pt\ and geometric criteria
also have tracks that 
satisfy the electron \Pt\ and geometric criteria. 
Figures~\ref{fig:jpsi_track_e}(b) and \ref{fig:jpsi_track_mu}(b)
have no events in common.

The two candidate mass distributions  
contain irreducible backgrounds from various
sources over the entire mass range.  
There are 37 candidates, of which 31 lie in the signal region.
Our principal task was to understand the 
shape and normalization of the backgrounds  
over the whole range of masses.  
We then determined their contributions to the signal region
and established the size and significance of a \Bc\ contribution
to that region.

\subsection{Efficiencies }
\label{subsec:eff}
The analyses described in the following sections required the
relative values for the following efficiencies:  
$\varepsilon_{e}	
	\equiv \varepsilon(B_c  \rightarrow J/\psi \, e    \, X)$, 
$\varepsilon_{\mu}
	\equiv \varepsilon(B_c  \rightarrow J/\psi \, \mu  \, X )$\ and 
$\varepsilon_{K}
	\equiv \varepsilon(B^{\pm}  \rightarrow J/\psi \, K^{\pm} )$.
We used a Monte Carlo program (App.~\ref{app:monte}) to study the 
response of our detector and reconstruction programs to each of these
processes.  All Monte Carlo events were subjected to the same 
requirements as the data.
Among these requirements we emphasize $ct^{\ast} > 60$\ \um\ 
and $M(J/\psi \, \ell)$\ in the range 3.35 to 11.0 GeV/$c^2$.
In order to eliminate shared systematic uncertainties,
such as those associated with \Jpsi\ detection, triggering and 
reconstruction, we used only the ratios of these efficiencies:
\begin{eqnarray}
  R^{\varepsilon} & \equiv &
	\frac{\varepsilon_{e} }{ \varepsilon_{e} + \varepsilon_{\mu} } 
	= 0.58 \pm 0.04 
\label{eq:Reps}
	\\
  R^{K} & \equiv & \frac{ \varepsilon_{e} }{ \varepsilon_{K} } = 0.244 \pm 0.033.
\label{eq:RK}
\end{eqnarray}

The principal differences between the efficiencies for \Jpsie\ and 
\Jpsimu\ are 
the larger geometric acceptance for the electron 
identification relative to that for muon identification,
electron isolation requirements in the calorimeter,  and
the different \Pt\ thresholds: 2.0 GeV/$c$\ for electrons and
3.0 GeV/$c$\ for muons.

\begin{figure}
\epsfxsize=6.0in
\centerline{\epsffile{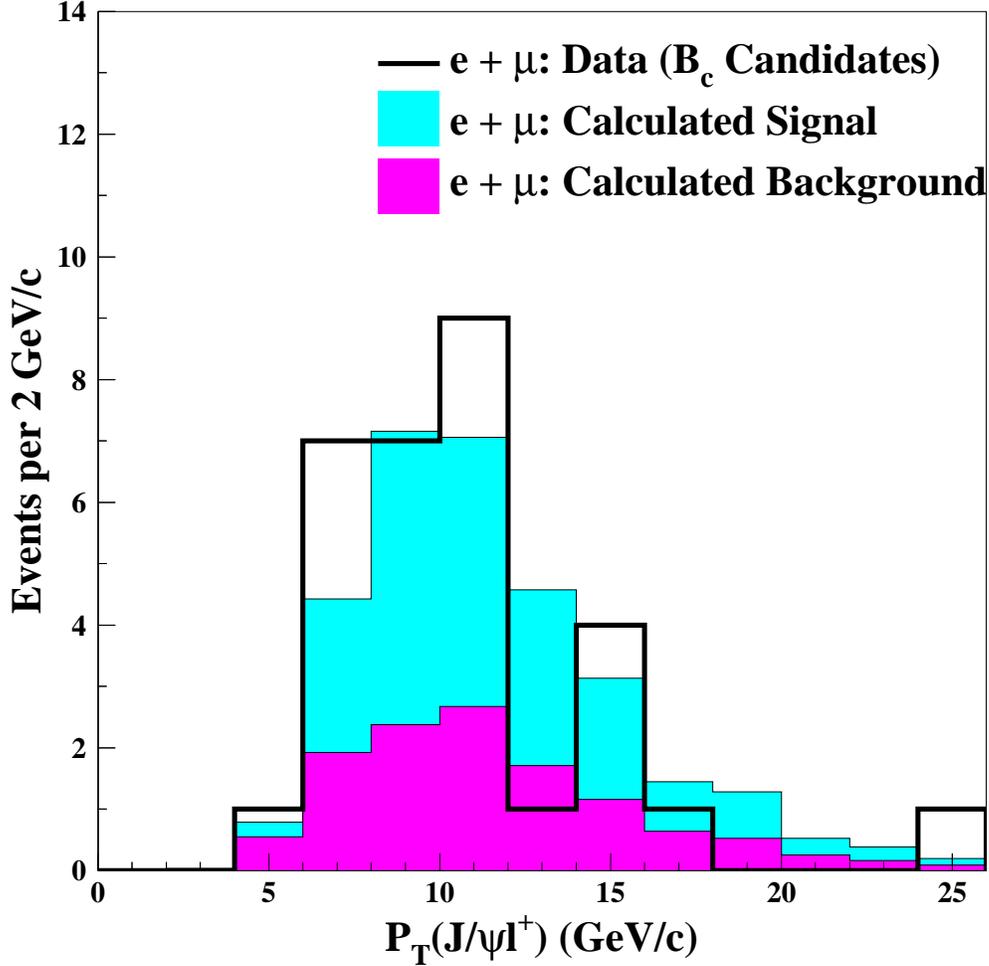}}
\caption{ 
	The transverse-momentum distribution for the \Jpsil\ system
	in \Bc\ candidates (line). 
	It is compared with the normalized \Pt\ 
        distribution for all backgrounds (dark shading) 
	and with the \Pt\ distribution for \BJpsil\ events generated 
	by Monte Carlo calculations (light shading).
	The latter is normalized to the fitted number of \Bc\ events 
	determined in Sec.~\ref{sec:fit}.
	}
\label{fig:pt_sum}
\end{figure}

The uncertainties in $\varepsilon_{e}$\ and $\varepsilon_{\mu}$\
that do not cancel in $R^{\varepsilon}$\ come from  
differing particle identification procedures\cite{CDF2} for 
electrons (10\%) and muons (5\%), 
uncertainty in the Monte Carlo
calculation (10\%), and model dependence (App.~\ref{app:monte}) 
due to the differing \Pt\ thresholds for muons and electrons (5\%). 
This model dependence arises from uncertainty in
the \Pt\ spectrum for \Bc\ production. 
As a check of our \Bc\ production model, 
we show in Fig.~\ref{fig:pt_sum} the tri-lepton \Pt\ distribution for 
the 31 candidate events in the signal region compared to those for
simulated \Bc\ events and for 
calculated backgrounds (Sec.~\ref{sec:background}).
There are no major differences in shape among the three distributions.

The uncertainties in $R^{K}$\ come from Monte Carlo statistics (4\%),
uncertainties in the model (App.~\ref{app:monte})
for production \Pt\ spectra (5\%) 
and in the fragmentation parameter (2.3\%),
uncertainties in the detector (5\%) and
trigger (4\%) simulations, and 
uncertainty in the electron identification (10\%). 

We calculated the efficiencies for \Bc\ decays assuming a 
\Bc\ lifetime \ctau\ = 120 \um.  
Lifetime effects cancel in $R^{\varepsilon}$
but not in $R^{K}$.
$R^{K}$ scales as the number of \Bc\ that survive the
60 \um\ threshold in \ctstar, $i.e.$\  
\begin{eqnarray}
  R^{K}(c\tau) & = & R^K(120 \mu {\rm m})
	\frac
	{\exp{ \left( -\frac{60 \, \mu {\rm m} }
	{ \langle 1/K \rangle c \tau } \right)}}
%	\
	{\exp{ \left( -\frac{60 \, \mu {\rm m}}
	{ \langle 1/K \rangle 120 \, \mu {\rm m}} \right)}}
\label{eq:RK_corr}
\end{eqnarray}
where $\langle 1/K \rangle c \tau$\ is the effective mean
decay length, and the average correction factor is $\langle 1/K \rangle =
0.88 \pm 0.02$.  (See Sec.~\ref{sec:life}.)

% ========================================================

\newpage
\section{Background Determination}
\label{sec:background}
% ------------- \input{background} -----------------------
Backgrounds in the sample of \Bc\ candidates can arise 
from misidentification of hadron tracks as leptons 
($i.e$. false leptons),
from random combinations of real leptons with \Jpsi\ mesons,  
and from incorrectly identified 
\Jpsi\ candidates~\cite{suzuki_thesis,singh_thesis}.

We describe three sources of false lepton identification.
\begin{itemize}
 \item	The third track is a kaon or pion that has 
passed through the muon detectors without being absorbed.
We call this ``punch-through background.''
 \item	The third track is a kaon or pion that has
decayed in flight into a muon in advance of entering the muon
detectors.  We call this ``decay-in-flight background.''
 \item	The third track is a kaon or pion that has
been falsely identified as an electron.
We call this ``false electron background.''%
\footnote{
As stated in Sec.~\ref{sec:select}, 
we made the conservative assumption that the hadron tracks
are all from mesons.
Protons do not decay in flight.  They have an  
interaction cross section higher than that for mesons 
and, therefore, a lower punch-through probability.  
Abandoning this assumption would lower our estimate 
of false muon backgrounds by a fraction of an event.
The assumption does not apply to 
our procedure for estimating false electron backgrounds,
which was validated with jet data containing a mix of mesons and 
baryons (App.~\ref{app:valid}).}
\end{itemize}

Random combinations arise from the following sources:
\begin{itemize}
 \item	External or internal conversions, $i.e.$\  
electrons from photon pair-production in the material around 
the beam line or from Dalitz decay of $\pi^0$.
Electrons from these sources that escape identification as 
conversions are called ``conversion background.''
 \item	A $B$ that has decayed into a $J/\psi$ 
and an associated $\overline{B}$ that has decayed
semileptonically (or through semileptonic decays of its daughter
hadrons) into a muon or an electron.  
The displaced $J/\psi$ and the lepton can accidentally 
appear to originate from a common point. 
We call this ``\BBbar\ background.''
\end{itemize}

Table~\ref{tab:count} (Table~\ref{tab:like}) 
summarizes the results of the data and background 
for the muon and electron channels in the signal (fitting) region
defined in Sec.~\ref{sec:intro}.
The procedures used to obtain these results are 
described in the remainder of this section.
We have also conducted studies to verify the accuracy of our background
calculations, applying them to independent data samples where 
they can be checked against direct measurements.  These studies are 
described in App.~\ref{app:valid}. 

\subsection{False Muon Backgrounds}
\label{subsec:false_mu}
\subsubsection{Punch-Through Background}
\label{subsec:punch}
One of the backgrounds that can mimic a \BJpsimuX\
event results when a $\pi^{\pm}$\ or $K^{\pm}$\  
or one of the particles in the resulting shower
is not completely contained in the 
calorimeter and CMP steel.  This can cause the original 
track to be misidentified as a muon.
Although the probability for this is about 1 in 500, 
a large number of events have 
tracks that meet the fiducial requirements, which offsets
the low punch-through probability.
Such tracks can be reconstructed with a \Jpsi\ to mimic a \Bc\ decay. 

We used a model of the distribution of material in the CDF detector 
and the absorption cross sections for $\pi^{\pm}$\ and $K^{\pm}$\ as
functions of energy~\cite{Particledata}
to calculate the total number of nuclear interaction lengths 
traversed by a particle.
The particle type, its energy and corrections to its momentum for 
energy loss through ionization were included.  
Given this information and the particle
trajectory, we obtained the probability of punching through 
the absorbing material and producing track segments in the muon chamber.

\begin{figure}
\epsfxsize=6.0in
\centerline{\epsffile{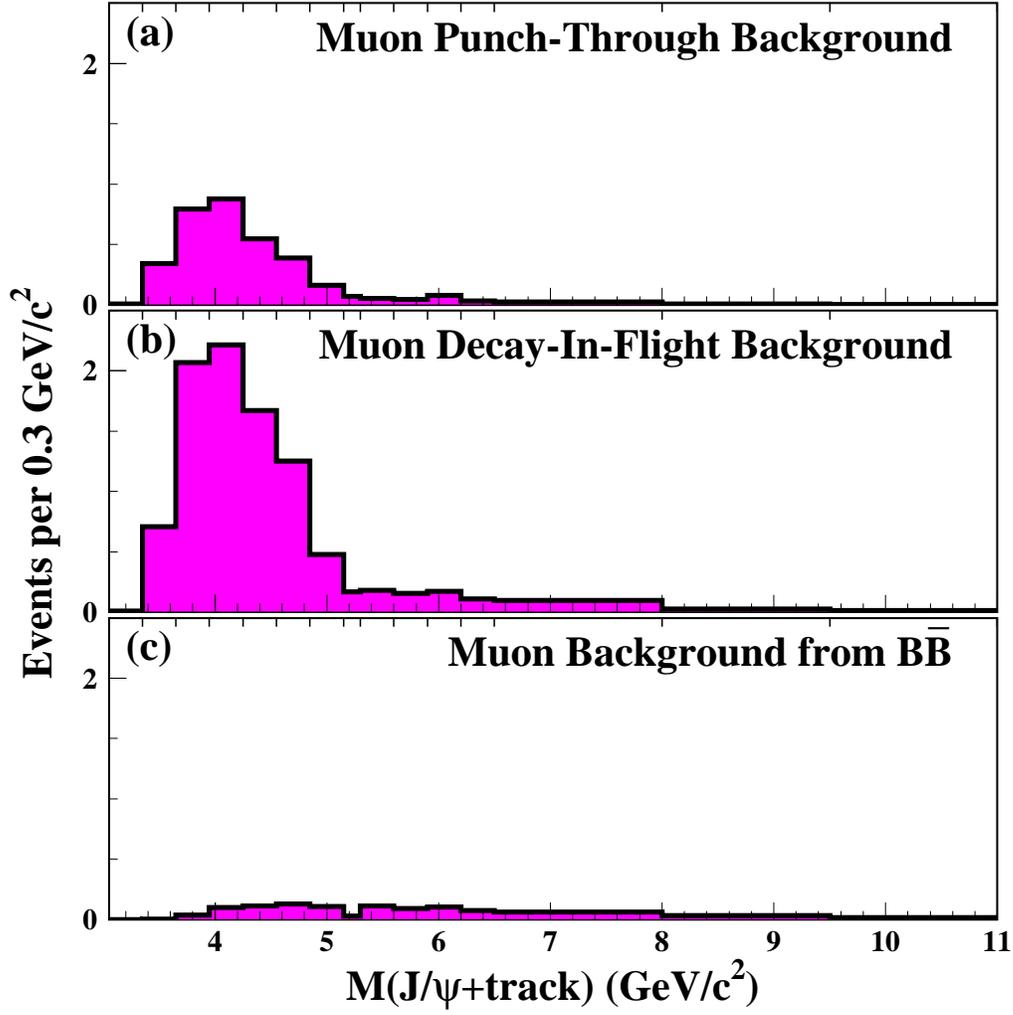}}
\caption{ Mass histograms for backgrounds from hadrons
misidentified as muons. 
(a)~The sum of punch-through background contributions from $\pi^{\pm}$, 
$K^{+}$\ and $K^{-}$.  
The dominant contribution to the punch-through background is from
$K^{+}$\ because of its lower interaction cross section.
(b)~The sum of decay-in-flight background contributions from $\pi^{\pm}$ 
and $K^{\pm}$.
(c)~The contribution from \BBbar\ background. 
These plots are normalized by their calculated contribution to
the candidate distribution in Fig.~\ref{fig:jpsi_track_mu}.
}
\label{fig:fake_mu}
\end{figure}

With the events in Fig.~\ref{fig:jpsi_track_mu}(a) that project to
the CMU and CMP chambers, we assumed 
the third particle to be a pion and calculated its punch-through probability.
We did similar calculations for $K^+$ and $K^-$.
Using $dE/dx$ information from the CTC, 
we determined that $(56.0\pm 3.4)\%$\ of the third tracks are pions,
where the uncertainty is purely statistical based on a fit.
We assume charge symmetry for the relative numbers of $K^+$ and $K^-$.
The shapes of the mass histograms from all these calculations are 
nearly identical to each other and their sum is shown in
Fig.~\ref{fig:fake_mu}(a).
The dominant contribution to the punch-through background 
is from $K^+$ because of its lower absorption cross section.

As a check, we used this procedure to compute 
the number of $K^+$\ and $K^-$\ punch-throughs 
from  \BJpsiK\ events and compared it with the
actual number of punch-throughs in the data.
For $K^+$\ we predict $3.36 \pm 0.46$\ events and observe 
2 events.  For $K^-$\ we predict $0.65 \pm 0.08$\ events 
and observe 1 event. 
With such small samples, it is difficult to evaluate
the systematic uncertainty and we arbitrarily assigned
it a value comparable to these differences between the expected 
and observed number of \JpsiK\ events.

We estimate 
$0.88 \pm 0.13 \, ({\rm stat.}) \pm 0.33 \, ({\rm syst.})$\ 
events in the signal region
due to hadron punch-through.

\subsubsection{ Decay-in-Flight }
Pion or kaon decay-in-flight can contribute background to \BJpsimu\ 
when a daughter muon from a meson decay is reconstructed as a  
track that projects to the \Jpsi\ decay point.

We estimated this background from the 
events in the \Jpsi\ + track mass distribution
shown in Fig.~\ref{fig:jpsi_track_mu}(a).
We assumed the third track to be a pion or a kaon and added it 
to a histogram with a weight that was the product of the 
following factors:
\begin{itemize}
 \item	the probability that it would decay before reaching the
	muon chambers,
 \item	the probability that the data from the tracking system
	would be reconstructed as a track that points to the
	\Jpsi\ decay vertex.
\end{itemize}
The decay probability is a simple calculation for each track.
The probability for reconstruction and vertex-pointing was 
calculated with a Monte Carlo program described 
in App.~\ref{app:monte}.
For the decay channels containing a \Jpsi , the program 
forced pion or kaon daughters of a $B$ to decay into a
muon in the region upstream of the CMU chambers.  
It then traced the particles through the detector.
This study included cases where the track did not originate
at the \Jpsi\ decay vertex, but decayed in a way that allowed 
a perturbed reconstruction which accidentally satisfied 
the vertex requirement.

\begin{figure}
\epsfxsize=6.0in
\centerline{\epsffile{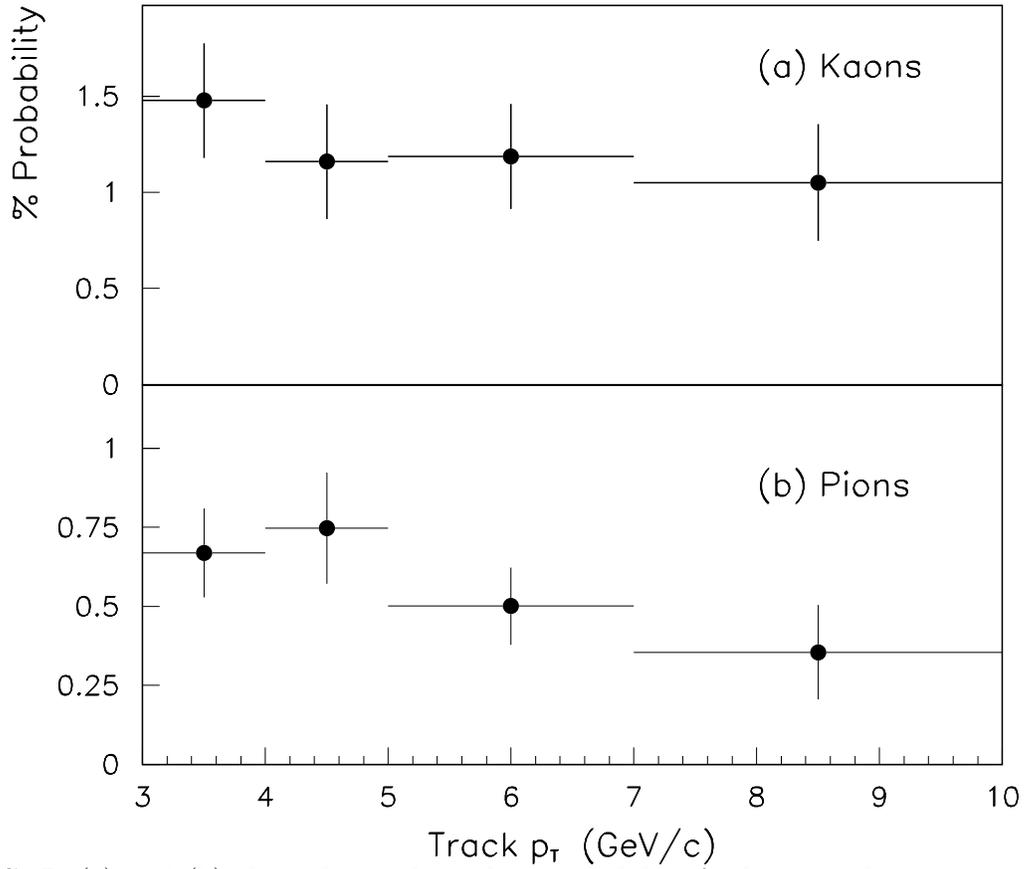}}
\caption{(a) and (b) show the $p_T$-dependent probability for
kaons and pions, respectively, to decay in flight and be misidentified
as muons. The specific ionization $dE/dx$ was used to determine
the correct proportion of pions and kaons in the data.}
\label{fig:DIF_pt}
\end{figure}

The events thus simulated were analyzed to determine the fraction of 
events for which the hadron and subsequent decay muon 
satisfied the muon identification criteria
with a reconstructed track
that projected to the \Jpsi\ decay point.
The fraction depends only on the type of particle and on \Pt.
The results of the calculation are shown in Figure~\ref{fig:DIF_pt} 
for kaons and pions.  
The $\pi/K$ ratio was determined from $dE/dx$\ as  
described in Sec.~\ref{subsec:punch}. 
The appropriate fractions of the 
distributions for pions and kaons were added to yield  
the background mass distribution in Fig.~\ref{fig:fake_mu}(b).

The systematic uncertainty in the number of decay-in-flight 
background events arises from several sources:
\begin{itemize}
 \item	uncertainties in the Monte Carlo calculation
	(12\%),
 \item	uncertainties in the reconstruction efficiency for tracks
	from mesons that decay in the CTC (17\%),
 \item	uncertainty in the $\pi/K$ ratio (10\%).
\end{itemize}
We estimate 
$5.5 \pm 0.5 \, ({\rm stat.})\pm 1.3 \, ({\rm syst.})$\ 
events in the signal region due to the decay-in-flight background.

\subsubsection{Total False Muon Background}
The mass distributions for punch-through and decay-in-flight
backgrounds are statistically indistinguishable in shape, 
and we have combined 
them for the likelihood analysis discussed in Sec.~\ref{sec:fit}.
In the fitting region (3.35--11.0 GeV/$c^2$) we estimate 
$11.4 \pm 2.4$\ (stat. $\oplus$\ syst.) false muon events
of which $6.4 \pm 1.4$ are in the signal region (4.0--6.0 GeV/$c^2$).

\subsection{False Electron Background}
Because of the requirement that the third lepton originates 
from the \Jpsi\ decay point, the main source of false-electron events
among our \Bc\ candidates is $B \rightarrow J/\psi \, + $\ hadrons
where one of the hadrons is misidentified as an electron.

To determine the probability that a hadron was misidentified as 
an electron, we studied two independent sets of events 
deliberately chosen because they contain few real electrons:  
a dataset based on an inclusive jet trigger 
with a threshold transverse energy of 20~GeV (JET20) and 
minimum bias dataset based on a trigger 
that sampled beam crossings with no physics requirements (MB).

The probability of misidentification of a track 
as an electron can depend on its transverse momentum
and on the presence of nearby tracks.  
Therefore, we express this probability as a function 
of \Pt\  and an isolation parameter $I$, 
defined to be the scalar sum of 
the momenta of particles within a cone $\Delta R < 0.2$, 
divided by the  momentum of the track under consideration.
$\Delta R = \sqrt{ (\Delta \eta)^2 + (\Delta \phi)^2 }$ 
is the radius of a cone in \etaphi\ centered on that track.
In this definition of isolation, a smaller $I$\ means
more isolated.  

\begin{figure}
\epsfxsize=6.0in
\centerline{\epsffile{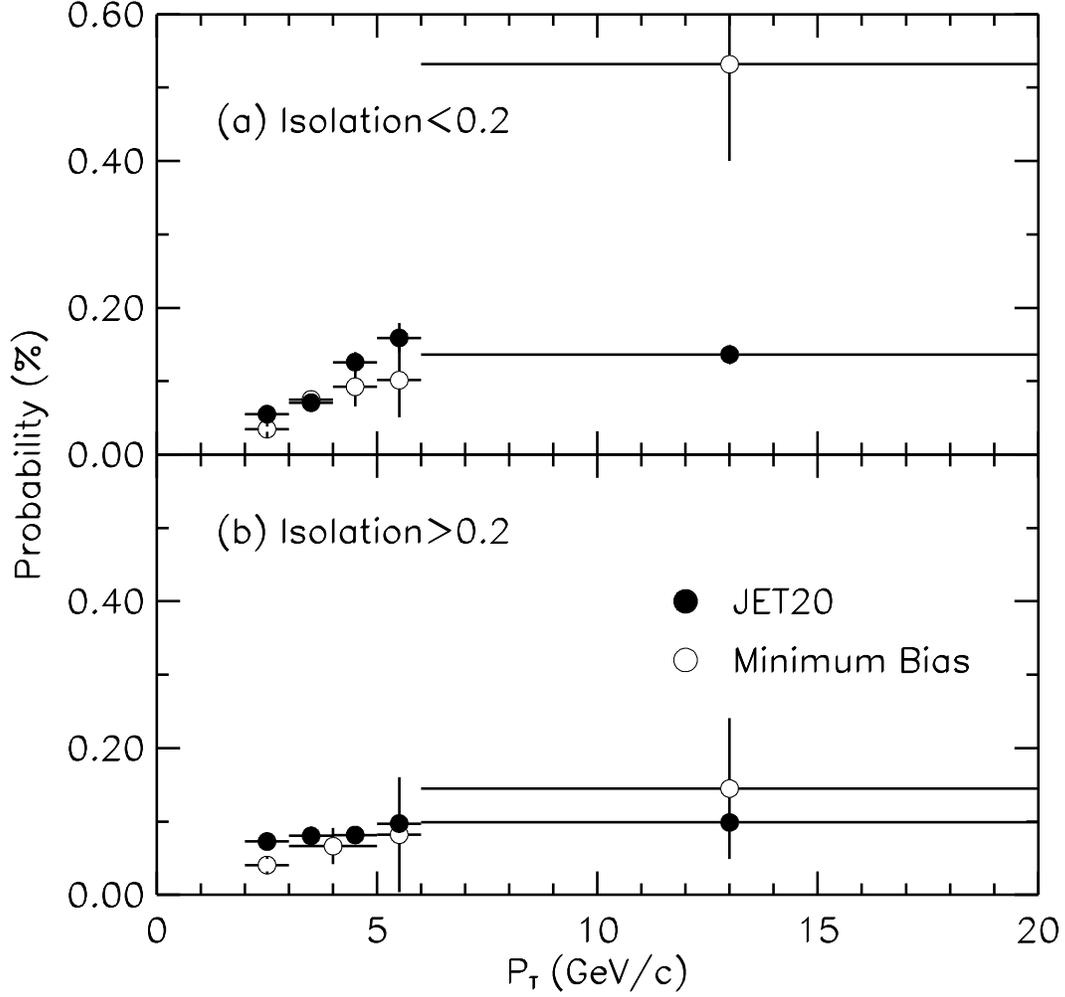}}
\caption{The probability of incorrectly identifying a hadron as an electron
as a function of $p_T$.  Tracks
from both the JET20 sample and minimum bias sample were used 
(App.~\ref{app:valid}). 
(a) and (b) show the data 
for the isolation parameter $I < 0.2$ and $I > 0.2$, respectively.
The probability averaged over the third-track momentum spectrum for the 
events in Fig.~\ref{fig:jpsi_track_e}a is $(0.066 \pm 0.006)$\%.}
\label{fig:fake_ele}
\end{figure}

The data in the JET20~and MB triggers contain a number of
real electrons. 
In order to calculate the false electron probability for hadrons,
the electrons  were removed statistically from the sample 
using $dE/dx$\ measurements.
We computed the fraction $f_m$\ of hadrons 
wrongly identified as electrons from the ratio of
$N^e$, the number of tracks satisfying all electron criteria,
to $N^t$, the number of tracks satisfying the purely geometric criteria.
However, a fraction $f_e$\ of the tracks 
passing all electron criteria were, in fact,
real electrons from heavy-flavor 
decays and from conversions, $i.e$., pair production by photons 
and Dalitz pairs as discussed in Sec.~\ref{subsec:conv}.  
From $dE/dx$ measurements we found $f_e$ to be $0.74 \pm 0.02$ in the 
JET20 data and $0.64 \pm 0.07$ in the MB data.  Thus,
\begin{equation}
f_m =  \frac{N^e}{N^t} \times (1 - f_e).  \nonumber
\end{equation}
Figure~\ref{fig:fake_ele} shows $f_m$\ as a function of \Pt\ for
the two data sets and for two ranges of the isolation parameter.
The results from the MB data differ from those of the JET20 data 
by 10\%, and we adopted this as a measure of the systematic uncertainty
in this calculation.

\begin{figure}
\epsfxsize=6.0in
\centerline{\epsffile{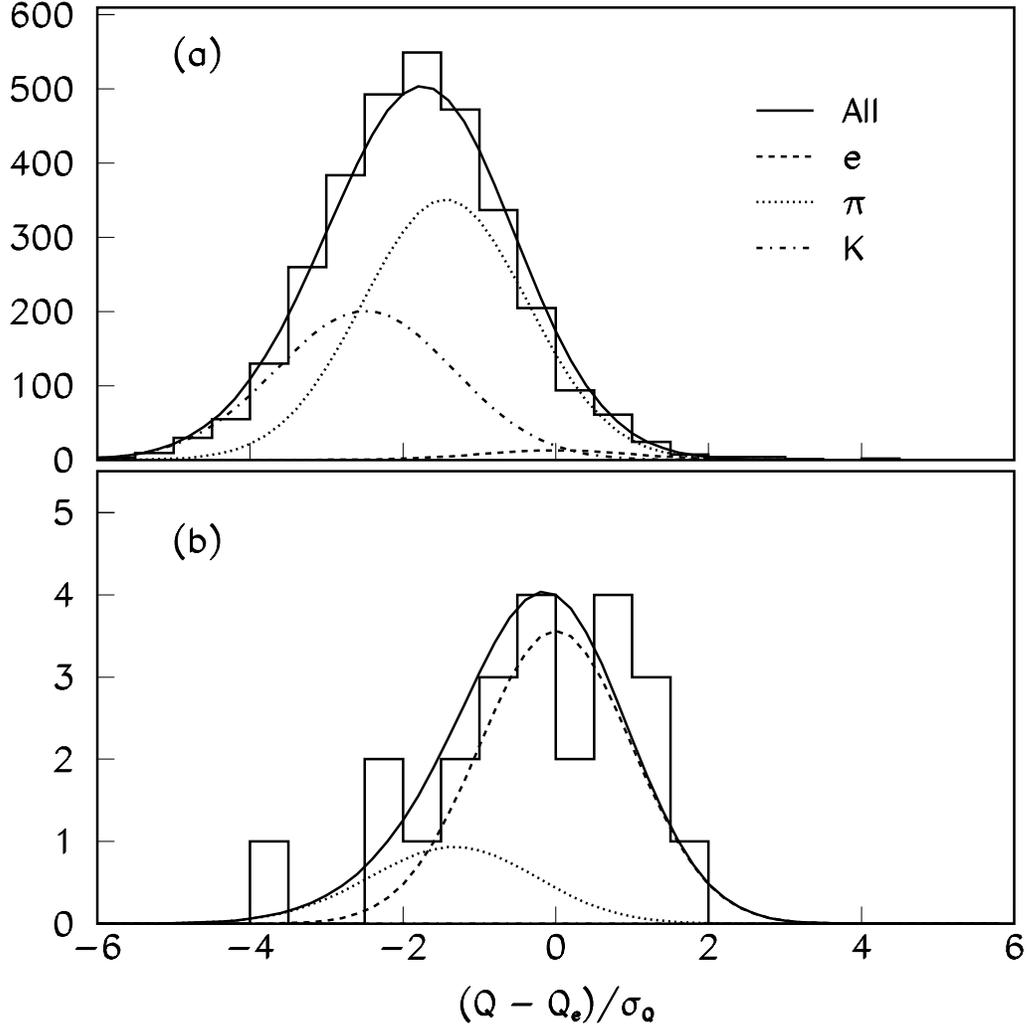}}
\caption{ 
The difference in $dE/dx$\ observed for the third track in
\Jpsi\ + track events and that expected for an electron.
For electrons, $Q$\ has a mean of $Q_e$\ and a standard
deviation of $\sigma_Q$. 
We scaled the difference to yield a distribution with 
zero mean and unit standard deviation for a pure sample of electrons. 
(a) The same events shown in Fig.~\ref{fig:jpsi_track_e}(a),
where  we assigned the electron mass to thethird track, required $p_T > 2.0$\ GeV/$c$, 
and applied the geometric criteria, but not the 
particle identification criteria  for electrons.
(b) The subset of the distribution above 
that satisfy all the electron
identification criteria except $dE/dx$. 
}
\label{fig:dedx_psi}
\end{figure}

We calculated the number of  background events due to misidentified 
hadrons in \Jpsie\ (Fig.~\ref{fig:jpsi_track_e}(b))
by selecting $J/\psi+{\rm track}$ events (Fig.~\ref{fig:jpsi_track_e}(a))
in which the third track is required to satisfy the purely geometric 
criteria for electron identification.
For each such track, we calculated $I$ and 
weighted its contribution by the probability $f_m(p_T,I)$ 
determined in the JET20 studies.
A mass histogram of the weighted sum is given in 
Fig.~\ref{fig:fake_e_conv}(a).
The number of hadronic background events determined
with this technique was consistent with that expected from
the $dE/dx$\ distribution data prior to the application 
of the $dE/dx$\ requirement. 
Figure~\ref{fig:dedx_psi}(a) shows the results of a $dE/dx$\
calculation applied to the third track for events in 
Fig.~\ref{fig:jpsi_track_e}(a).  Most are hadrons.
These tracks were then required to satisfy all the 
electron identification criteria {\it except} the 
$dE/dx$\ requirement.  
Results of the $dE/dx$\ calculation for the surviving 
events are shown in Fig.~\ref{fig:dedx_psi}(b).
For most of the surviving events, the third track is
an electron.

We estimate $2.6\pm 0.3$\ (stat.~$\oplus$\ syst.) 
events in the signal region
due to false electrons 
and $4.2 \pm 0.4$\ such events in the fitting region.

\subsection{ Conversion Background}
\label{subsec:conv}
Photon pair production in material around the beam 
and Dalitz decays both produce $e^+ e^-$\ pairs.
The reconstructed track for 
one member of a pair can pass through the \Jpsi\ decay point and 
be selected as a candidate for \BJpsiX.
After applying other electron identification criteria 
and the vertex constraint
(Sec.~\ref{sec:select}), 
we found and rejected two such ``conversion'' events 
by searching for the partner track in the \Jpsi\ + track sample 
with \ctstar~$>$~60 \um.
However, a track can contribute to the background in 
the \Jpsie\ events
if its partner track has low momentum and escapes detection.

\begin{figure}
\epsfxsize=6.0in
\centerline{\epsffile{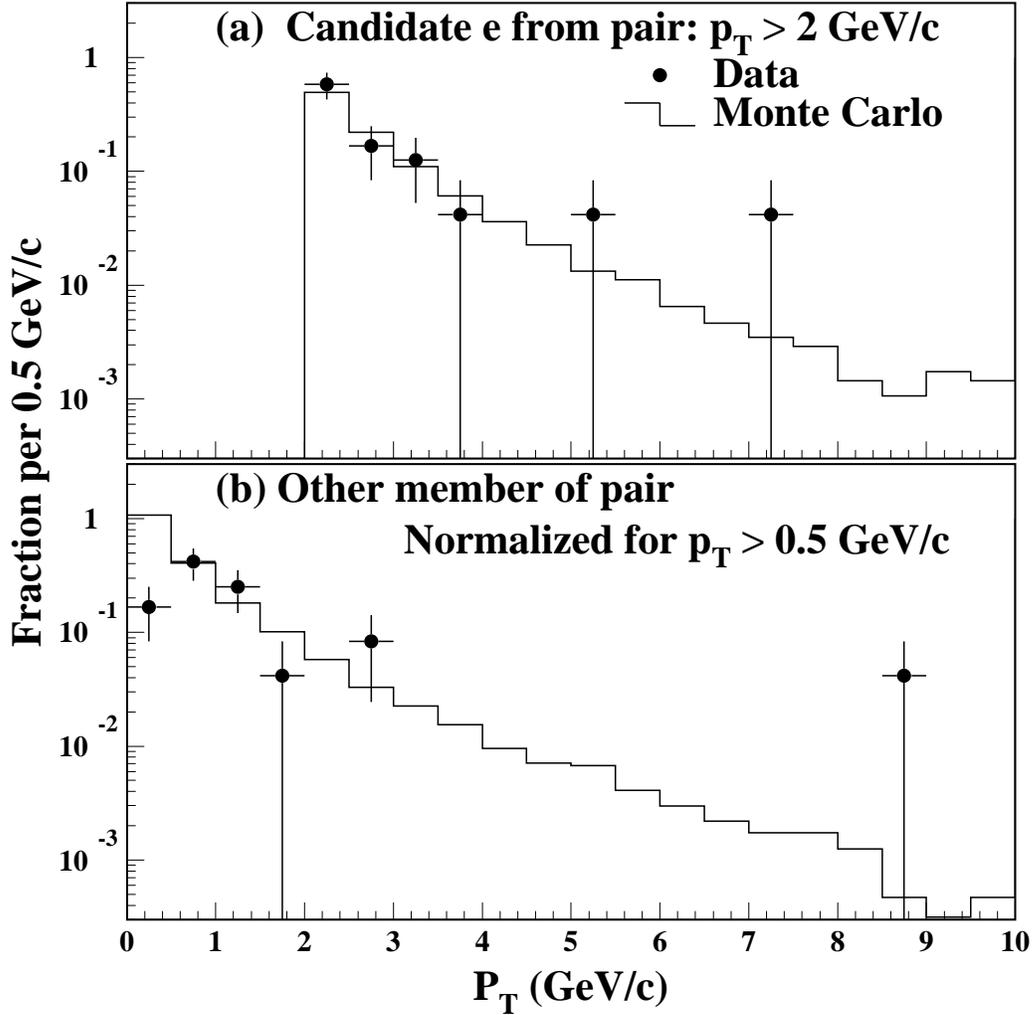}}
\caption{ 
Conversion Background.  
(a) The momentum spectrum for 
the track from an electron-positron pair 
that fulfilled the requirements for the third lepton 
and (b) the momentum spectrum for the other member of the pair.
To obtain a larger data sample, we removed the requirements
on \ctstar\ and $P(\chi^2)$\ for the vertex fit.
In both (a) and (b) the data distributions are normalized
to unit area.  The Monte Carlo distributions are normalized
to the data for \Pt\ $> 0.5$\ GeV/$c^2$.
In the lowest bin in (b), the difference between data and 
Monte Carlo arises from the dropoff in track reconstruction
efficiency for \Pt\ $< 0.5$\ GeV/$c$.  This is the reason
for the undetected conversion background.
}
\label{fig:psi_cv_pt}
\end{figure}

To estimate the magnitude and shape of this background 
in the $M(J/\psi \, e)$\ distribution, 
we performed a hybrid Monte Carlo calculation based on the 
\Jpsi\ + track events.  
The Monte Carlo program replaced the third track in the event by a $\pi^0$.  
It forced 1.2\% of the $\pi^0$'s to decay through
the Dalitz channel and the rest through two-photon final states.  
The program propagated the photons through the surrounding material with 
tabulated probabilities for $e^+e^-$\ production, 
and it propagated the resultant charged particles through the
detector simulation.  
We used each event 100 times, rotating its azimuth by 
a random angle to sample all parts of the detector.
Figure~\ref{fig:psi_cv_pt} shows the momentum spectrum for 
the track which fulfilled the requirements for the third 
lepton and the spectrum for the other member of the pair.
These hybrid events were subjected to the \Bc\ analysis procedures.
Roughly half, ($48.6 \pm 1.9$)\%, of 
these ``conversion'' background events 
were rejected because the partner was detected.  
Thus the ratio of undetected or 
residual conversions to detected conversions is 
$R^{ce} = 1.06 \pm 0.08$\ (stat.).

\begin{figure}
\epsfxsize=6.0in
\centerline{\epsffile{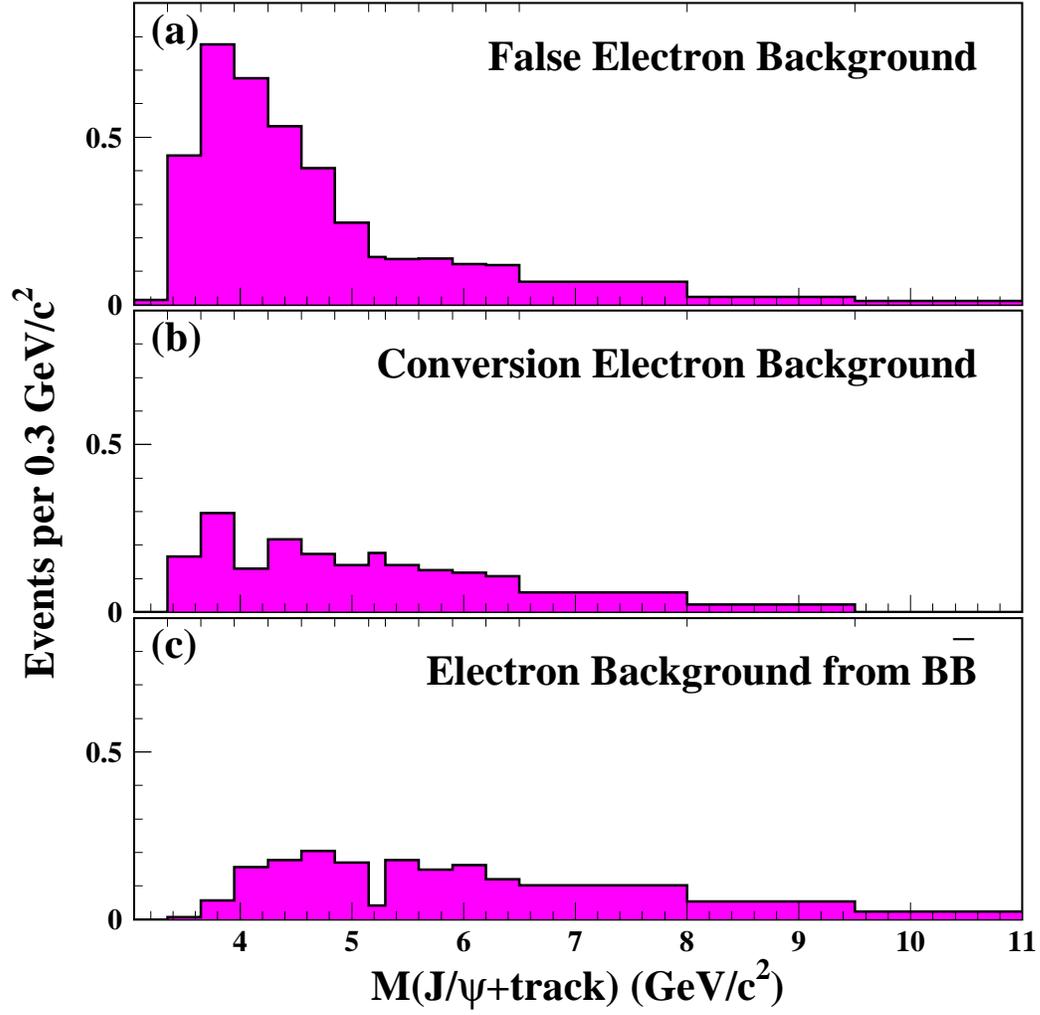}}
\caption{ The \Jpsie\ mass distribution  
(a) for background events resulting from misidentified electrons.
(b) For events in which the electron originated 
from a $\gamma$ conversion or Dalitz decay that was not 
identified as such.
(c) For \BBbar\ events in which the \Jpsi\ came from one 
parent and the electron from another.
These plots are normalized by their expected contribution to
the candidate distribution in Fig.~\ref{fig:jpsi_track_e}.
}
\label{fig:fake_e_conv}
\end{figure}

In the simulation,
the \Jpsie\ mass distributions arising from detected and undetected 
conversions have the same shape. 
Fig.~\ref{fig:fake_e_conv}(b) shows this shape normalized to an area
equal to the expected 2.1 undetected conversion background events.

Systematic uncertainties arise from statistical uncertainty in the efficiency 
for finding the conversion partner (28\%), from uncertainty in the 
shape of the \Jpsi\ + track mass distribution for these events (9\%),
and from differences in \Pt\ distributions between the data and 
the sample used to calculate this background (13\%).  
Combined, they are 32\%.

The statistical uncertainty from two events is the largest 
contribution to the overall uncertainty in the conversion background,
and we quote the Gaussian approximation of the 
uncertainty here.  
In the likelihood analysis of Sec.~\ref{sec:fit}, the two detected events,
$N^{ce} = 2$, enter as a Poisson term.  The systematic uncertainties
are incorporated in the ratio of undetected to detected conversions,
$R^{ce} = 1.06 \pm 0.36$\ (stat.~$\oplus$\ syst.).  
The residual background is the product $N^{ce}R^{cv}$.

The mass distribution for the conversion background distribution 
in Fig.~\ref{fig:fake_e_conv}(b) contains 
$2.1 \pm 1.7$\ (stat.~$\oplus$\ syst.) events 
in the fitting region.
Of these $1.2 \pm 0.9$  events are in the signal region.

\subsection{\BBbar\ Background}
\label{subsec:BBbar}
\BBbar\ pairs produced during the $p\overline{p}$ collision can
mimic the \BJpsil\ signature when a $B$ decays into a $J/\psi$ and
its associated $\overline{B}$\ 
or any of its daughters decays into a lepton.  
If the lepton track projects through the \Jpsi\ vertex,
the event may not
be distinguishable from a $B_c$ decay and would be a part of the
irreducible background.

The \BBbar\ background was determined by a Monte Carlo simulation 
(App.~\ref{app:monte}).
One  $B$ was required to decay into a final state containing 
a \Jpsi , and 
the other $B$\ was allowed to decay through all channels.  
We simulated the detector response,
and we required the simulated events to pass the di-muon trigger criteria.
To avoid double-counting false-lepton backgrounds, 
we eliminated candidates where the third track was a hadron.
We then performed the \Bc\ analysis on these events.
We used \BJpsiK\ events to normalize the Monte Carlo simulation 
to the data. 
The resulting mass distributions are shown in 
Figs.~\ref{fig:fake_mu}(c) and \ref{fig:fake_e_conv}(c).

The systematic uncertainties in the estimate of this background include the
trigger simulation (5\%), the uncertainty in the branching ratio 
$B^+ \rightarrow J/\psi \, K^+$\ (10\%), and Monte Carlo statistics (11\%).

We estimate $0.7\pm 0.3$\ (stat.~$\oplus$\ syst.) 
\Jpsimu\ events and $1.2\pm 0.5$ \Jpsie\ events in the signal region 
due to \BBbar\ background.
The corresponding numbers in the fitting region are
$1.44 \pm 0.25$\ \Jpsimu\ events and $2.3 \pm 0.9$\ \Jpsie\ events.

\subsection{Other Backgrounds}
We have considered three additional potential sources 
of background to the decay \BJpsil.  They are
\begin{itemize}
\item false $J/\psi$ candidates from the continuum background 
of the di-lepton spectrum,
\item $J/\psi + c\overline{c}$ production in which the charm decays 
semileptonically, and
\item decays of as yet undiscovered baryonic $bc$ states such as the
$\Xi_{bc}$.
\end {itemize}
We estimate that these make negligible contribution to our background.

The false \Jpsi\ background is very small after mass and vertex constraints
are applied to the data.  We selected two side bands in the $J/\psi$ 
mass distribution.  In each we substituted the central mass for the 
side band in our fitting procedures.  We found 3 ``\Jpsi'' + track 
events that satisfied our track selection criteria.
In none of these did the third track satisfy 
our criteria for muons or electrons.  
The dominant source of false \Jpsi\ candidates is $B$\ decay to a real muon 
along with a hadron falsely identified as a muon because of 
punch-through or decay-in-flight (Sec.~\ref{subsec:false_mu}).
Either the associated $\overline{B}$\ or a daughter $D$\ has  
a branching fraction of roughly 0.1 for yielding a third lepton.
The probability for another hadron falsely identified as the third lepton
is even lower, roughly 0.01.
Our background estimate is $3 \times 0.1 \times 0.5 = 0.15$, where the
factor 0.5 is the ratio of widths for the central peak vs. two side
bands.  The 90\% confidence upper limit on 3 events is 6.7 events,
which yields an upper limit of 0.34 events.
We neglect this source of background.

It is possible for additional charm to be produced along with 
prompt \Jpsi\ mesons with production mechanisms similar 
to those for \Bc\ production.  
Several of our selection requirements suppress background
from such events in which the additional charm decays semileptonically.
As is the case with the \BBbar\ background,
the prompt \Jpsi\ + charm  background is suppressed  
because the \Jpsi\ and lepton do not generally form a common vertex.  
Additional suppression of charm-daughter leptons results from 
the isolation cut in the electron
channel and the high transverse momentum requirements in both channels.  
Finally, since these events are prompt, they mostly fail the \ctstar\  
requirement.  For the lifetime meaurement discussed in 
Sec.~\ref{sec:life}, we studied the \ctstar\ dependence of the
signal and various backgrounds.  
They account for the distribution of candidate events 
at low \ctstar , and there is no evidence for additional
background from  \Jpsi\ + charm.  Therefore, we neglect it.

The as yet undiscovered hyperon $\Xi_{bc}$ can decay into a tri-lepton
topology, $e.g$. $\Xi_{bc} \rightarrow J/\psi \, \Xi_c$ followed
by $\Xi_c \rightarrow \Xi \, \ell \, \nu$.
The production cross section for such a particle is likely to
be significantly less than that for the \Bc.  
Alternate standard-model decay modes for $\Xi_{bc}$  
fail our \Bc\ identification criteria.
The same observation can be made for
other baryons containing a $b$ quark.  
We assumed no background from these particles.

% ========================================================

\newpage
\section{The Magnitude and Significance of the \Bc\ Signal}
\label{sec:fit}
\begin{figure}
\epsfxsize=6.0in
\centerline{\epsffile{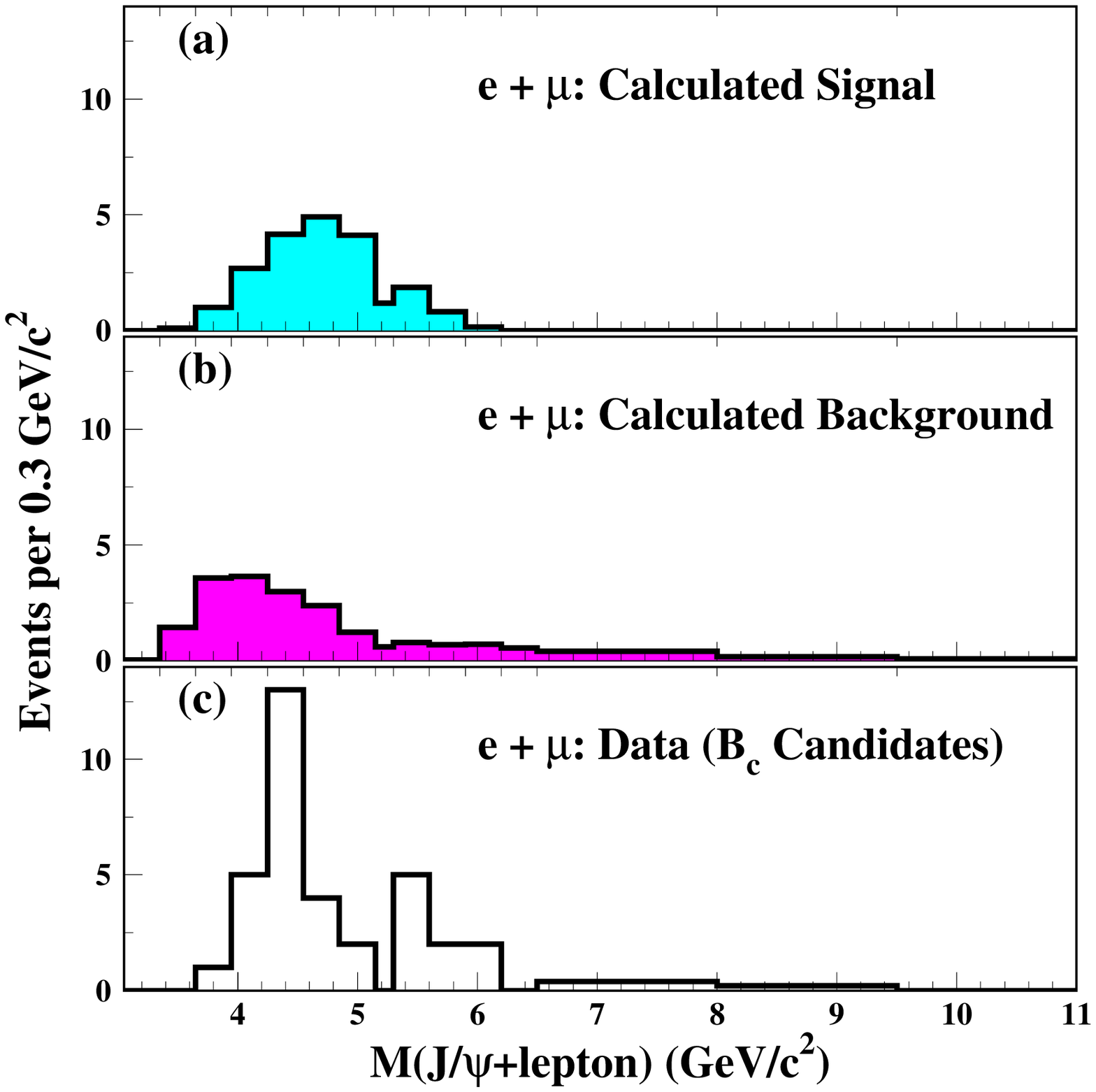}}
\caption{ 
	(a) A tri-lepton mass distribution for \BJpsil\ based on Monte Carlo
	calculations.  It is normalized to the fitted number of 
	\Bc\ events.
        The distribution was generated under the assumption that
	the mass of the \Bc\ is 6.27 GeV/$c^2$.  
	There are negligible differences between
	the shapes for \BJpsimu\ and \BJpsie.
	Note that ($93.0 \pm 0.6$)\% of the area falls in the
	signal region 4.0--6.0 GeV/$c^2$.
	(b) The normalized mass distribution for all backgrounds
	for both muon and electron channels.
	(c) The mass distribution for \Bc\ candidates in the
	data for both muon and electron channels. 
	Note that each of these is a summary histogram, {\it i.e.} 
	the sum of several individual histograms presented earlier.  
	We emphasize that the fitting procedures use the full
	information from individual distributions rather than the
	sums.  
	}
\label{fig:plot_sum}
\end{figure}

% -------------- \input{fit} ---------------------------
Monte Carlo calculations (App.~\ref{app:monte}) for \BJpsimu\ and \BJpsie\ 
with the \Bc\ mass assumed to be 6.27 GeV/$c^2$\ yielded the 
tri-lepton mass distribution shown in Fig.~\ref{fig:plot_sum}(a).
The normalization anticipates the results of the fit described below.
The electron and muon mass distributions are used
seperately, but the figure shows the combined
distribution since the differences are small.
We assume equal branching fractions for the two decay modes, 
and we expect the ratio of \Jpsie\ to total \BJpsil\ events 
to be given by the efficiency ratio 
$R^{\varepsilon} =  0.58 \pm 0.04$\ discussed in Sec.~\ref{subsec:eff}. 
The mass distribution for the sum of the normalized backgrounds for
muons and electrons is shown in Fig.~\ref{fig:plot_sum}(b).
The mass distribution for all \Bc\ candidates is shown 
in Fig.~\ref{fig:plot_sum}(c).

The expected background is unable to account for the 
observed data distribution. 
In order to test this statistically and to determine the 
magnitude of the signal needed to account for the excess,
we adopted two approaches.  The first was a simple ``counting
experiment'' based on the number of events in the \Jpsi\ + lepton
mass range from 4.0 to 6.0 GeV/$c^2$.
However, this ignores additional information in the 
shapes of the distributions and the yield in the 
extended mass range populated by backgrounds but not 
by signal.
Our second approach employed a binned likelihood fitting 
procedure that includes the shape of the 
distributions over the full mass 
range, 3.35 to 11.0 GeV/$c^2$.  
To account for the excess in the data over expected background,  
the fit varied the normalization of the signal shape of 
Fig.~\ref{fig:plot_sum}(a) and calculated its uncertainty.
The bins are those shown in
Figs.~\ref{fig:jpsi_track_e} and \ref{fig:jpsi_track_mu} 
except that the lowest bin in the figures, 
3.05 to 3.35 GeV/$c^2$, was not used in the fit.

In both approaches, we computed the probability that a random
fluctuation of the background is sufficient to account for the 
observed data in the absence of a \Bc\ contribution. 
This is the ``null hypothesis.''  

We also performed an unbinned likelihood analysis using 
spline fits to the parent distributions.
The results are completely consistent with the 
binned likelihood analysis.  We also varied the assumed \Bc\
mass from 5.28 to 7.52 GeV/$c^2$.
Within the range 6.1-6.5 GeV/$c^2$, which embraces 
all the theoretical predictions,
we found the fitted number of \Bc\ events 
to be insensitive to the assumed mass
These issues are discussed in Sec.~\ref{sec:mass}.

\subsection{ The Counting Experiment }
In the signal region of \Jpsil\ mass, we observe 19 
\Jpsie\ candidates and 12 \Jpsimu\ candidates.
Table~\ref{tab:count} summarizes the backgrounds from 
the various sources of background 
discussed in Sec.~\ref{sec:background}.
The expected total backgrounds  are
$5.0 \pm 1.1$\ events for \Jpsie\ and $7.1 \pm 1.5$\ events 
for \Jpsimu, 
leading to a combined signal of $18.9 \pm 5.6$\ events.  
From these results, we tested the null hypothesis by folding 
the Gaussian uncertainties in the estimated mean number 
of background counts with their Poisson fluctuations. 
This allowed us to determine the probability that the 
backgound would fluctuate up to the observed number of events.
The null hypothesis probabilities are 
$2.1 \times 10^{-5}$\ for the \Jpsie\ sample and 
0.084 for the \Jpsimu\ sample.

\subsection{ Likelihood Analysis:  Fit to the \Bc\ Signal }
We used a normalized log-likelihood function for testing and fitting
our data and background estimates.
It used the shapes of the distributions over the mass range
3.35 to 11.0 GeV/$c^2$, and it included as input all the information  
on the tri-lepton mass distributions for signal and for background 
discussed in earlier sections.
The likelihood function has a necessary and sufficient set of
parameters to fit these distributions to the observed data.
It also included constraints such as the expected fractions
of events in the two decay channels.

In  Appendix~\ref{app:like}, we discuss 
the normalized log-likelihood function 
$\xi^2\equiv-2\log({\cal L}/{\cal L}_0)$\ used to fit our data,
where ${\cal L}$\ is the likelihood function and 
${\cal L}_0$\ is its value for a perfect fit. 
Maximum likelihood is equivalent to minimum $\xi^2$ which 
has properties similar to those of $\chi^2$.
The only unconstrained parameter in the fit is 
$n^{\prime \ell}$, 
the total number \BJpsimu\ and \BJpsie\ events
in the fitting region, 
$i.e.$ in the \Jpsil\ mass range 3.35--11.0 GeV/$c^2$.
All other parameters in the fit are constrained by 
externally derived information.

At the minimum in $\xi^2$, the number of \Jpsil\ events in the 
fitting region is
\begin{equation}
	n^{\prime \ell} = (n^{\prime \mu} + n^{\prime e}) 
			= 20.4^{+6.2}_{-5.5}\: {\rm events}
\label{eq:evts_fit_region}
\end{equation}
with $\xi^2/N_{d.o.f.} = 38.1/26$,
where $N_{d.o.f.}$\ is the number of degrees of freedom in the fit.
In the Monte Carlo signal distribution in Fig.~\ref{fig:plot_sum}, 
($93.0 \pm 0.6$)\% of the events fall in the signal region
(4.0--6.0 GeV/$c^2$).  We scale 20.4 events by this value to calculate
\begin{equation}
	n^{\ell} = (n^{\mu} + n^{e}) 
			= 19.0^{+5.8}_{-5.1}\: {\rm events}
\label{eq:evts_sig_region}
\end{equation}
in the signal region.
This is in excellent agreement with the counting experiment result.

\begin{figure}
\epsfxsize=6.0in
\centerline{\epsffile{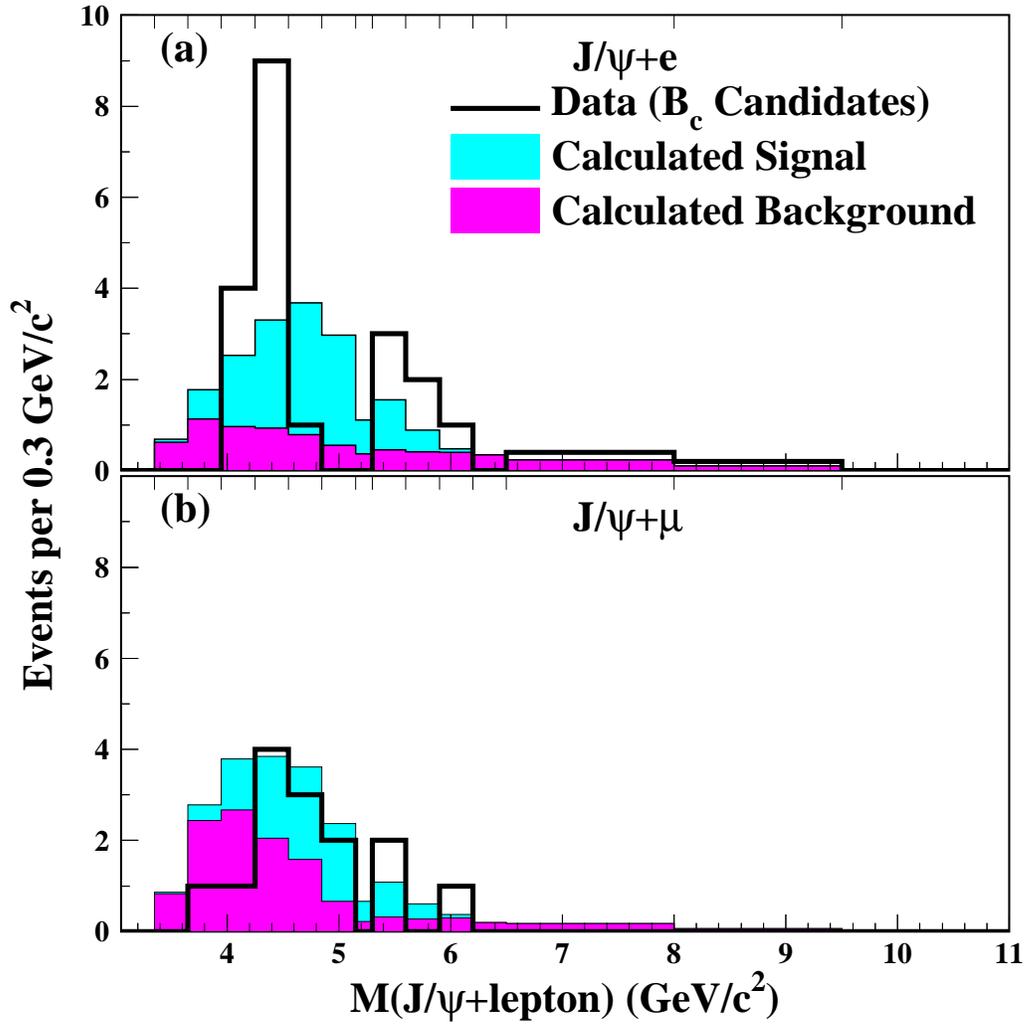}}
\caption{ 
Histograms of the \Jpsil\ mass that compare 
the signal and background contributions determined in the likelihood fit to 
(a) the data for \Jpsie\ and (b) the data for \Jpsimu.
Note that the mass bins vary in width.
}
\label{fig:fit_emu}
\end{figure}

\begin{figure}
\epsfxsize=6.0in
\centerline{\epsffile{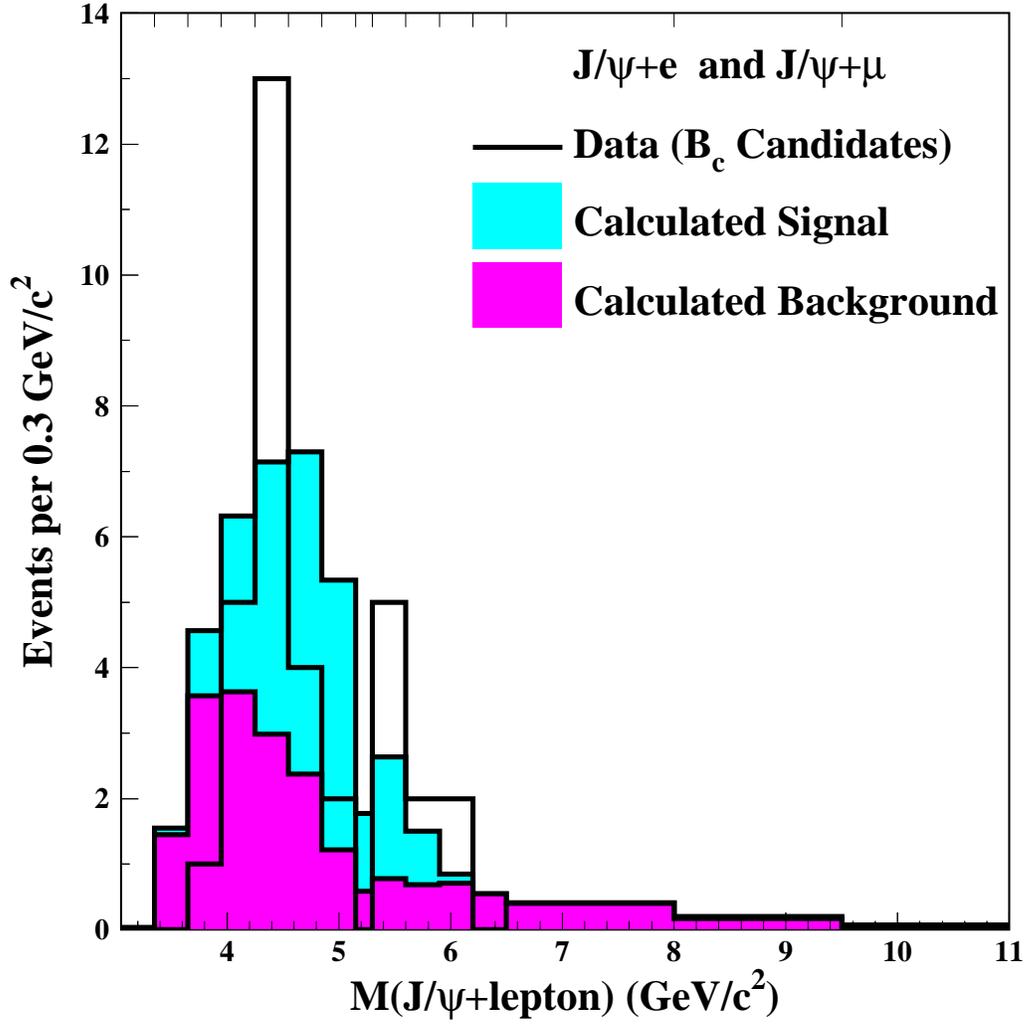}}
\caption{
Histogram of the \Jpsil\  mass that compare 
the signal and background contributions determined in the likelihood fit to 
the combined data for \Jpsie\ and \Jpsimu.
Note that the mass bins vary in width.
The total $B_c$\ contribution is $20.4^{+6.2}_{-5.5}$ events. }
\label{fig:fit_comb}
\end{figure}

Figure~\ref{fig:fit_emu} 
shows the contributions to the
background and signal for \BJpsieX\ and \BJpsimuX\ 
separately resulting from the binned likelihood fit, and 
Fig.~\ref{fig:fit_comb} shows the combined data.

\begin{figure}
\epsfxsize=6.0in
\centerline{\epsffile{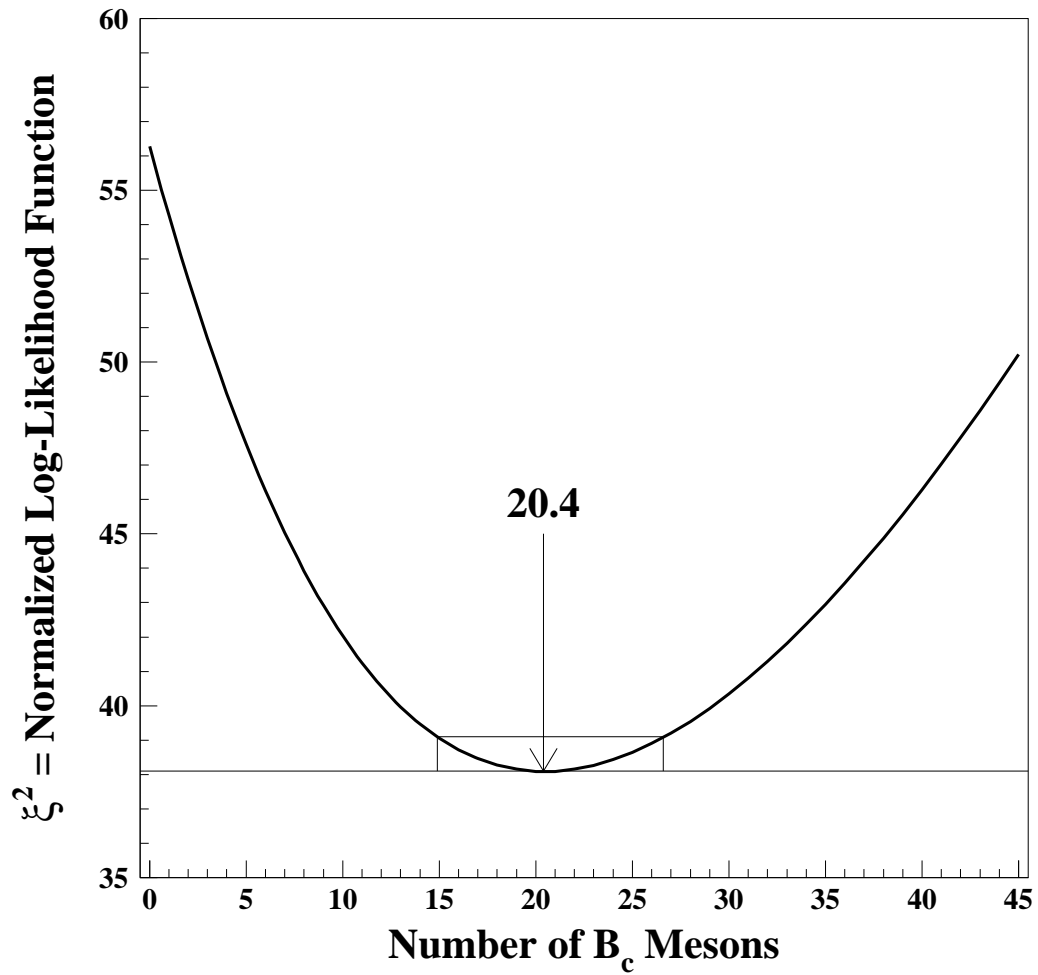}}
\caption{The variation of  
$\xi^2 = -2\ln{( \cal L/ L_{\rm 0})}$ as a function of the number
of $B_c$ mesons.  For each fixed value of $N(B_c)$\ all other parameters
were adjusted for the best fit. 
We find $N(B_c) = 20.4^{+6.2}_{-5.5}$ at the minimum.}
\label{fig:likelihood}
\end{figure}

Figure~\ref{fig:likelihood} shows $\xi^2$\ 
plotted as a function of the
assumed number of $B_c$ mesons in the data sample.
For each value of $n^{\prime \ell}$,  $\xi^2$\ was minimized 
as a function of the other parameters.
Table~\ref{tab:like} shows the input constraints 
and fitted values for the background normalizations and 
for other parameters.

\begin{figure}
\epsfxsize=6.0in
\centerline{\epsffile{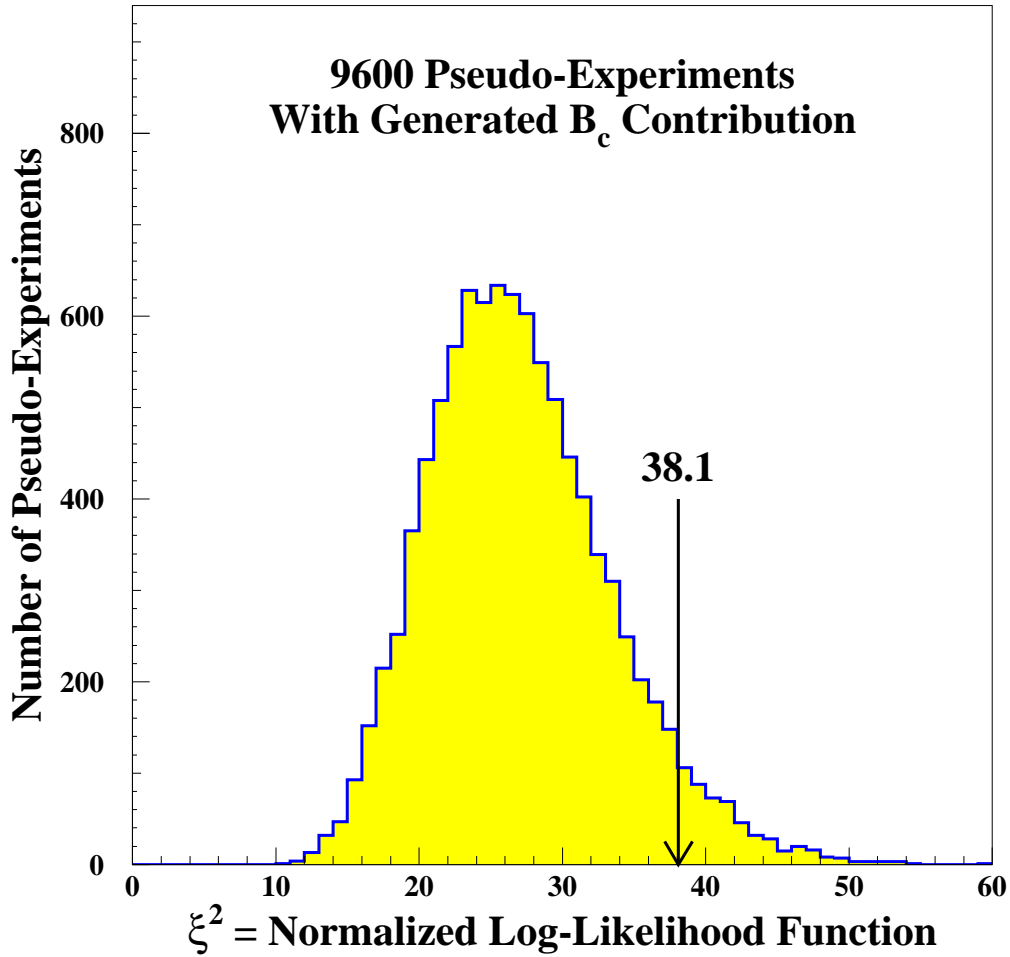}}
\caption{Each entry in this histogram is the result of a fit to a
Monte Carlo pseudo-experiment that simulated the statistical
properties of our data.
The backgrounds were generated
with the measured means and varied using Poisson or Gaussian statistics
as appropriate. $B_c$ events were included with statistical 
fluctuations from the mean of 20.4 and bin-by-bin fluctuations. 
The resulting muon and electron events were fit as with the data.
The values of $\xi^2$\ are histogrammed here and compared 
with the value found for the experimental data.}
\label{fig:bc_like}
\end{figure}

To evaluate the quality of the fit, we observe that,
to the extent that $\xi^2$\ behaves like $\chi^2$, $P(\xi^2) = 8$\%.  
We made a more reliable estimate of this probability by 
generating a large number of Monte Carlo ``pseudo-experiments.''
First, we generated random backgrounds 
with Gaussian-distributed uncertainties based on the shapes and
normalizations determined in Sec.~\ref{sec:background}.
To this we added a signal contribution with the fitted magnitude 
varied according to the uncertainty from the fit.
Bin-by-bin, the signal plus background value served as the 
mean for a number of events randomly generated according to 
a Poisson distribution.  This constituted a pseudo-experiment 
with a \Bc\ signal. 
We ran the fitting program on each pseudo-experiment.
The $\xi^2$\ distribution
for these is shown in Fig.~\ref{fig:bc_like}.  The probability
of finding $\xi^2 \geq 38.1$\ is 5.9\%. 

Only two assumptions about the \Bc\ signal distribution 
were used in the fit: the \Bc\ mass and the relative 
contributions to the electron and muon channels.  
The choice of 6.27 GeV/$c^2$\ for the 
mass will be considered in Sec.~\ref{sec:mass}.
As a test we fit the data with the electron fraction 
$r^{\varepsilon}$\ allowed
to vary freely, not constrained to  
$R^{\varepsilon} = 0.583 \pm 0.043$.
The results of this fit were: 
$\xi^2 = 37.7$;
the number of signal events was 20.3, 
and the fitted electron fraction was
$r^{\varepsilon} = 0.65 \pm 0.14$, 
consistent with $R^{\varepsilon}$.

\subsection{ Likelihood Analysis:  The Null Hypothesis }
\begin{figure}
\epsfxsize=5.0in
\centerline{\epsffile{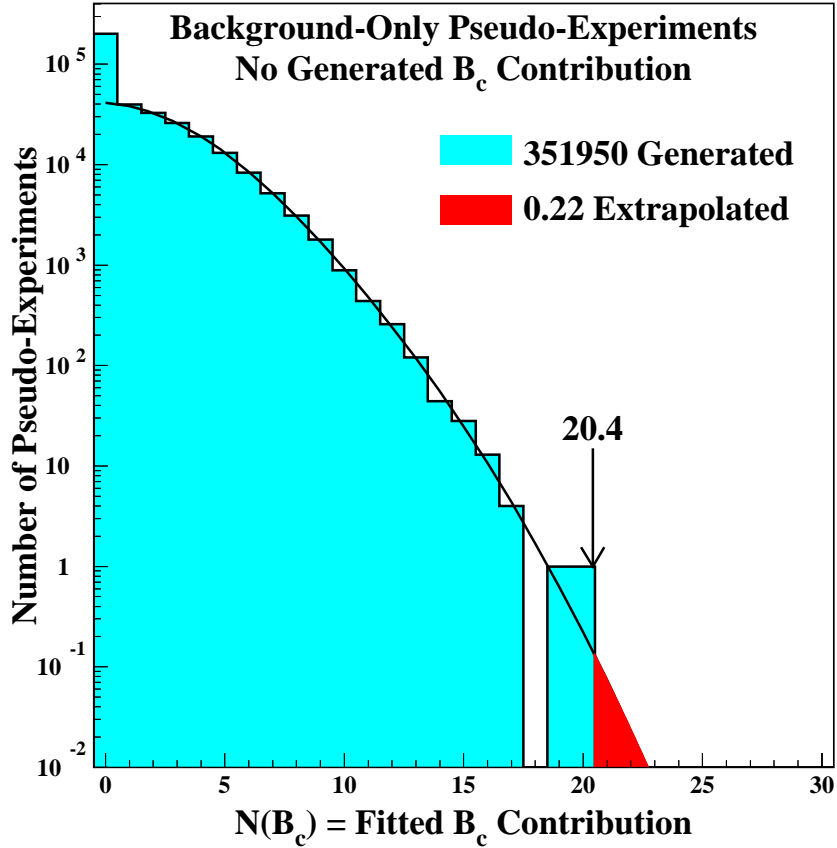}}
\caption{Each entry in this histogram is the result of a fit to a
Monte Carlo simulation of the statistical properties of this
experiment.  
We generated the backgrounds randomly according
to the measured means and varied using Poisson or Gaussian statistics
as appropriate.  The $B_c$ contribution was set to zero in 
generating the distribution. 
We then fit the resulting numbers of muon and electron events 
using the likelihood function.  The fitting function included 
a $B_c$\ contribution.  The histogram above is a measure of the 
probability of finding a false $B_c$\ contribution of size $N(B_c)$\ 
where none exists.
Upward and downward fluctuations of the generated
samples can require both positive and negative solutions
for  $N(B_c)$.  We chose to collect all negative solutions
in the lowest bin in this figure where these 
events produce a prominent excess.
The smooth curve represents a fit of a convenient 
extrapolation function (the sum of two Gaussians) to estimate the 
area beyond 20.4 events.
}
\label{fig:bc_sig}
\end{figure}

The null hypothesis is the postulate 
that there is no \Bc\ signal and that a
statistical fluctuation in the background is responsible for the
apparent excess in the data.
In order to test this, we again computed the results for a large
number of ``pseudo-experiments'' or trials in the manner described
above, except that we omitted the signal contribution.
With $n^{\prime \ell}$\ allowed to vary, 
we ran the fitting program to return the fitted 
number of \Bc\ events in a distribution devoid of real signal.
Figure~\ref{fig:bc_sig} shows a histogram of $n^{\prime \ell}$\ for 
351,950 pseudo-experiments.  
The fitted signal tends to 
compensate for statistical fluctuations, positive or negative, 
from the correct background shape.
The peak at zero events includes those trials consistent with
a negative contribution from the signal distribution.
No pseudo-experiments gave 
values of $n^{\prime \ell}$\ exceeding 20.4.
We extrapolated the fitted shape of the distribution and estimate
its area above 20.4 to be $0.22^{+0.10}_{-0.06}$\ out of 351,950 trials.
Thus, the probability that a random fluctuation of the 
background could produce the observed data distribution 
is $0.22/351,950 = 0.63 \times 10^{-6}$.
This is equivalent to 4.8 standard deviations in significance.

In the following sections, we assume that the 
excess events are due to the existence of the 
\Bc\ meson. 
We describe measurements of its mass, 
its lifetime and 
its relative cross section times branching 
fraction, all of which are consistent with 
values expected for the \Bc .

% ========================================================

\newpage
\section{The \Bc\ Mass}
\label{sec:mass}
% ------------- \input{mass} -----------------------------
In order to check the stability of the \Bc\ signal, 
we varied the value assumed for the  \Bc\ mass.
With the procedures described in Sec.~\ref{sec:fit} 
and App.~\ref{app:monte},
we generated Monte Carlo samples of \BJpsil\  with various  
values of  $M(B_c)$\ from 5.52 to 7.52 GeV/$c^2$.    
For each of these samples, we propagated the \BJpsil\ 
final-state particles through the detector simulation programs 
to obtain the tri-lepton mass spectrum, $i.e.$\ a signal template.
The signal template for each value of $M(B_c)$\ together 
with the background mass distributions was used 
to fit the mass spectrum for the data.  
The best-fit log-likelihood value shows a rough parabolic
dependence on the assumed \Bc\ mass,
and this yields a measurement of $M(B_c)$.

We performed this analysis with the binned log-likelihood analysis
described in Sec.~\ref{sec:fit} and with an un-binned log-likelihood
analysis.  The two methods yielded nearly identical results, but 
the binned method exhibited slightly more scatter about a smooth dependence 
on mass.  We present the unbinned results here because this 
method is not sensitive to binning fluctuations.

\begin{figure}
\epsfxsize=6.0in
\centerline{\epsffile{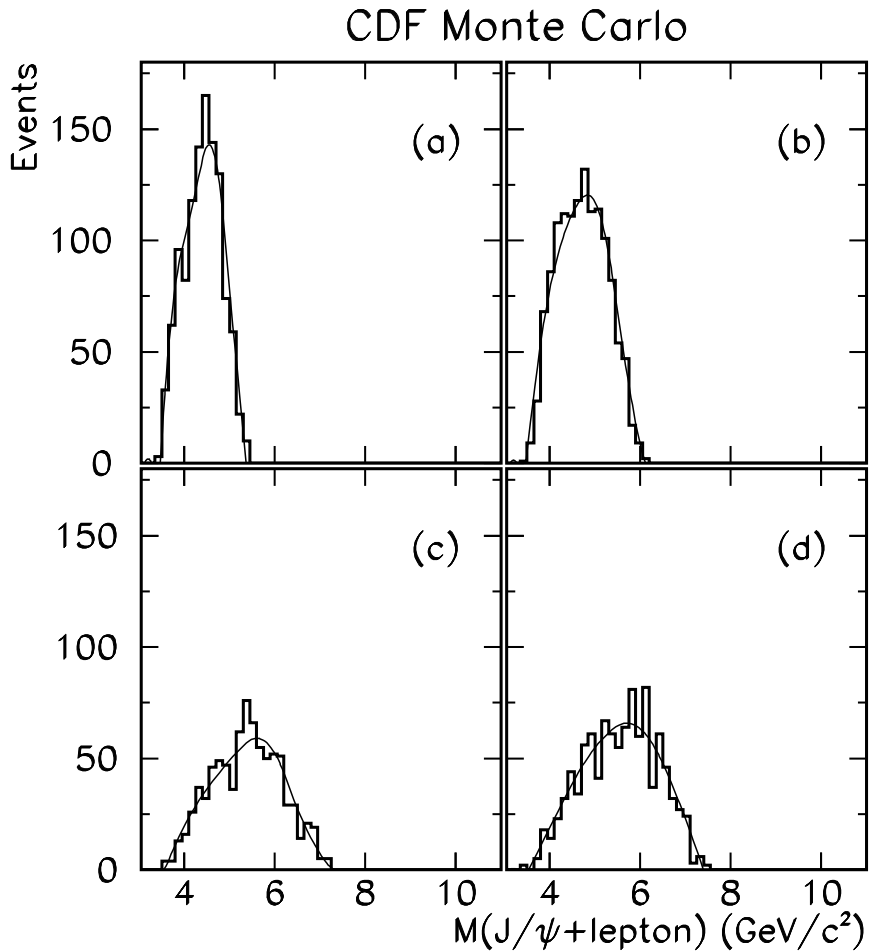}}
\caption{ Templates used to determine the quality of the 
fit to the mass spectrum for various assumed values of 
the \Bc\ mass.  Of the 11 values used, four templates are shown 
here for the following values of $M(B_c)$: 
(a) $5.52$\ GeV/$c^2$, (b) $6.27$\ GeV/$c^2$, 
(c) $7.27$\ GeV/$c^2$, and (d) $7.52$\ GeV/$c^2$.
In each case 
the histogram displays the binned results of the Monte
Carlo calculation and the smooth curve is a spline fit
to the histogram. 
}
\label{fig:bc_templates}
\end{figure}

For each assumed \Bc\ mass, a signal template was formed 
with a smooth spline fit to the Monte Carlo distribution.
Figure~\ref{fig:bc_templates} shows the generated distributions
and spline fits for a sample of the templates used in this
study.
Background templates formed in the same way were independent
of the assumed \Bc\ mass.  
Most contributions to the unbinned log-likelihood function 
were the same as those in Sec.~\ref{sec:fit} 
and App.~\ref{subapp:bin_m} for the binned fit.
However, the sum over bins of Poisson terms was 
replaced by of sum over events of log-probabilities.
This is discussed in App.~\ref{subapp:unbin_m}.  
In this analysis we compare the log-likelihood to 
its value at minimum ${\cal L}_{min}$,
and we define the relative log-likelihood function $\xi^2_m$\
as a function of $M(B_c)$.
\begin{equation}
	\xi^2_{m} \equiv 
	-2 \ln \left( \frac{ \cal{L}}{ {\cal L}_{min}} \right).
\label{eq:xi2_min}
\end{equation}
At each assumed value of $M(B_c)$, several Monte Carlo samples and 
corresponding signal templates were generated in order to determine
the sensitivity of the fit to statistical fluctuations in the 
Monte Carlo simulation.  This provides us with an uncertainty 
on the values of $ \xi^2_m$.

\begin{figure}
\epsfxsize=6.0in
\centerline{\epsffile{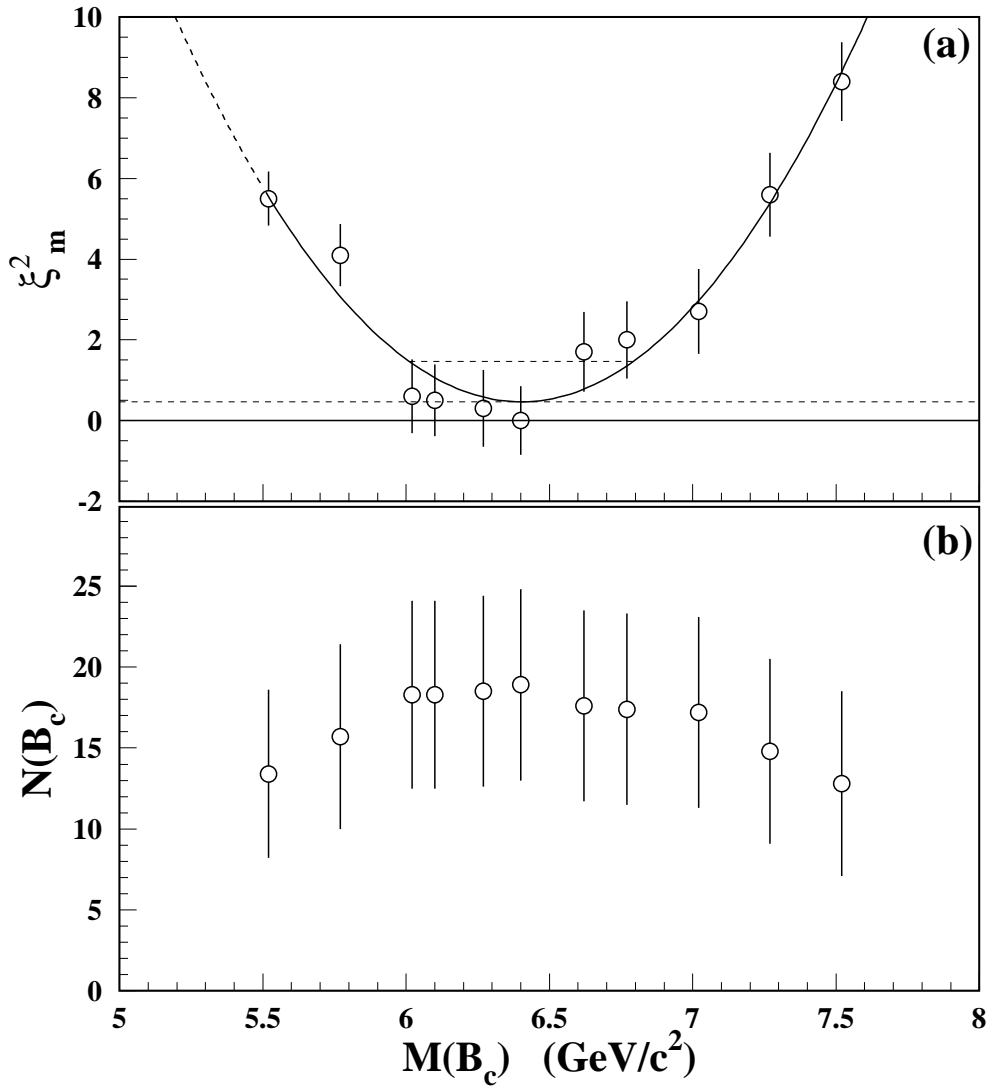}}
\caption{(a) The relative log-likelihood function $\xi^2_m$\ from
fits to the data for various values of the assumed mass of the \Bc.
Error bars on $\xi^2_m$\ represent its fluctuations with
different Monte Carlo samples of \Bc\ events at the same mass. 
The parabolic curve is a fit to the plotted points with
$\chi^2/n_{d.o.f.} = 4.3/8$.
A horizontal line is drawn through the parabola's 
minimum which occurs at $M(B_c) = 6.40$\ GeV/$c^2$.
Another line one unit above its minimum indicates the 
one-standard-deviation uncertainties of $\pm 0.39$\ GeV/$c^2$.
(b) The fitted number of \Bc\ events vs. $M(B_c)$.
It is stable over the range of theoretical 
predictions for $M(B_c)$, 6.1 to 6.5 GeV/$c^2$.
}
\label{fig:bcmass}
\end{figure}

 Figure~\ref{fig:bcmass}(a) shows the dependence of 
$\xi^2_m$\ on $M(B_c)$.
The figure includes a parabolic fit to $\xi^2_m$.
The parabolic fit yields a best fit value of 6.40 GeV/$c^2$\ with
a statistical uncertainty of 0.39 GeV/$c^2$.  

As in Sec.~\ref{sec:fit} we generated a sample of pseudo-experiments
based on the fitted results with the assumed mass of 6.27 GeV/$c^2$.
The distributions of $M(B_c)$, its uncertainty, the number of \Bc\
and its uncertainty were consistent with the results 
in the experimental data.
This provides some confidence that the model used to fit the 
data is adequate to the task.
The comparison between the unbinned log-likelihood function for the 
experimental data and that for the pseudo-experiments was closely 
similar in shape and width to that for the binned likelihood analysis 
(Fig.~\ref{fig:bc_like}).

We considered a number of sources of systematic uncertainty
in this measurement: 
\begin{itemize}
 \item	distortion of the signal mass distribution arising from
	decay to higher-mass $c\overline{c}$\ states 
	rather than \Jpsi\ (0.09 GeV/$c^2$).
 \item	fitting procedures, estimated from the difference between
	binned and unbinned analyses (0.08 GeV/$c^2$), 
 \item	finite Monte Carlo statistics in the signal template 
	(0.04 GeV/$c^2$),
 \item	variations in the \Bc\ mass distribution due to $b$-quark 
	production spectrum (0.02 GeV/$c^2$),
 \item	Monte Carlo simulation of the CDF trigger (0.02 GeV/$c^2$), 
\end{itemize}
These uncertainties are small in comparison with the 
statistical uncertainty.  In quadrature, they sum to
0.13 GeV/$c^2$.

Figure~\ref{fig:bcmass}(b) shows 
that the magnitude of the \Bc\ signal is stable over the
range of theoretical predictions for $M(B_c)$, and our 
experimental measurement of the mass is 
$M(B_c) = 6.40 \pm 0.39\,{\rm (stat.)} \pm 0.13\,{\rm (syst.)}$\
GeV/$c^2$.

% ========================================================

\newpage
\section{The \Bc\ Lifetime}
\label{sec:life}
\begin{figure}
\epsfxsize=6.0in
\centerline{\epsffile{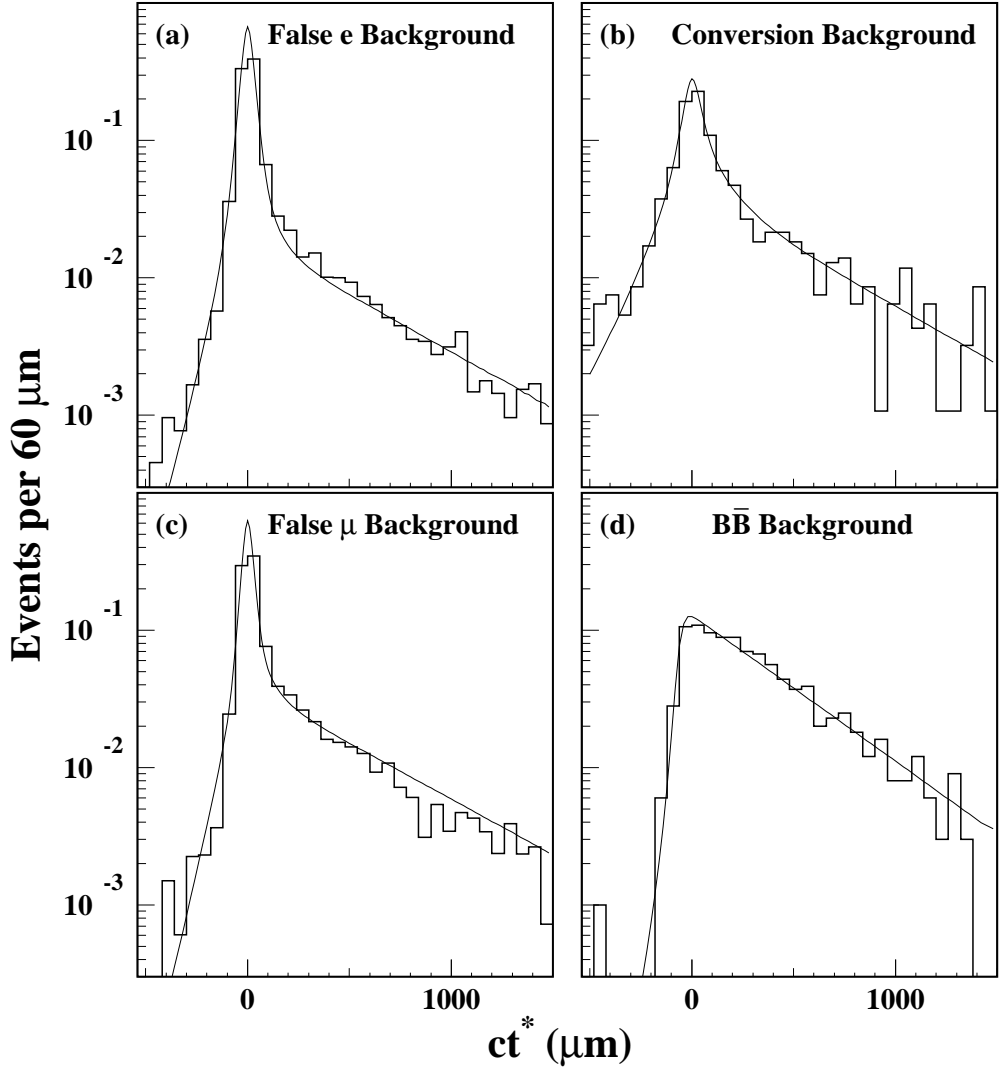}}
\caption{
Pseudo-proper decay length distributions 
for the background distributions.  
(a) \Jpsie\ background from false electrons.
(b) \Jpsie\ background from conversion electrons.
(c) \Jpsimu\ background from false muons.
(d) \BBbar\ background. Its shape is the same for
both \Jpsie\ and \Jpsimu .
}
\label{fig:bkg_shape}
\end{figure}

% ----------------- \input{life} -------------------------
We extended our analysis to obtain a best estimate of the mean proper
decay length \ctau\ and hence the lifetime $\tau$\ of the
$B_c$\ meson.  The information to do this is contained in the 
distribution of \ctstar\ which is defined in Eq.~\protect\ref{eq:ctstar}.  
We changed the threshold requirement on 
\ctstar\ from $c t^* > 60$\ \um\ to $c t^* > -100$\ \um.  
This yielded a sample of 71 events, 42 \Jpsie\ and 29 \Jpsimu.  
We determined a functional form for the shapes in \ctstar\ for each
of the backgrounds (Fig.~\ref{fig:bkg_shape}).  To these, we added a 
resolution-smeared exponential decay
distribution for a \Bc\ contribution, parametrized by its mean decay
length \ctau.  Finally, we incorporated the data from each of the 
candidate events in an unbinned likelihood fit to determine the 
best-fit value of \ctau.

Since the neutrino in \BJpsil\ carries away undetected momentum,
the true proper time for the decay of each event cannot be 
calculated from \ctstar .
The relationship between \ctstar\ and \ct\ is:
\begin{equation}
	ct^{\ast} = \frac{ct}{K}
\end{equation}
where $K$\ for an event is given by 
\begin{equation}
 K = 
\frac{M(B_c)}{M(J/\psi \ell)} \times 
\frac{p_T(J/\psi \ell)}{p_T(B_c)}.
\label{eq:K}
\end{equation}
We assume $M(B_c) = 6.27$\ GeV/$c^2$, but $p_T(B_c)$\ is 
unknown for single events, and therefore, we cannot correct
for $K$\ event-by-event.
In an ideal data sample with no background and a known \Pt(\Bc)
distribution, one finds  
$\langle ct^{\ast}\rangle = \langle ct \rangle \langle 1/K \rangle
 = c\tau \langle 1/K \rangle$,
where $\langle ct^{\ast}\rangle$\ is the average over the data,
and $\langle 1/K \rangle$\ is the average over \Pt(\Bc) and
\Pt(\Jpsil).

\begin{figure}
\epsfxsize=6.0in
\centerline{\epsffile{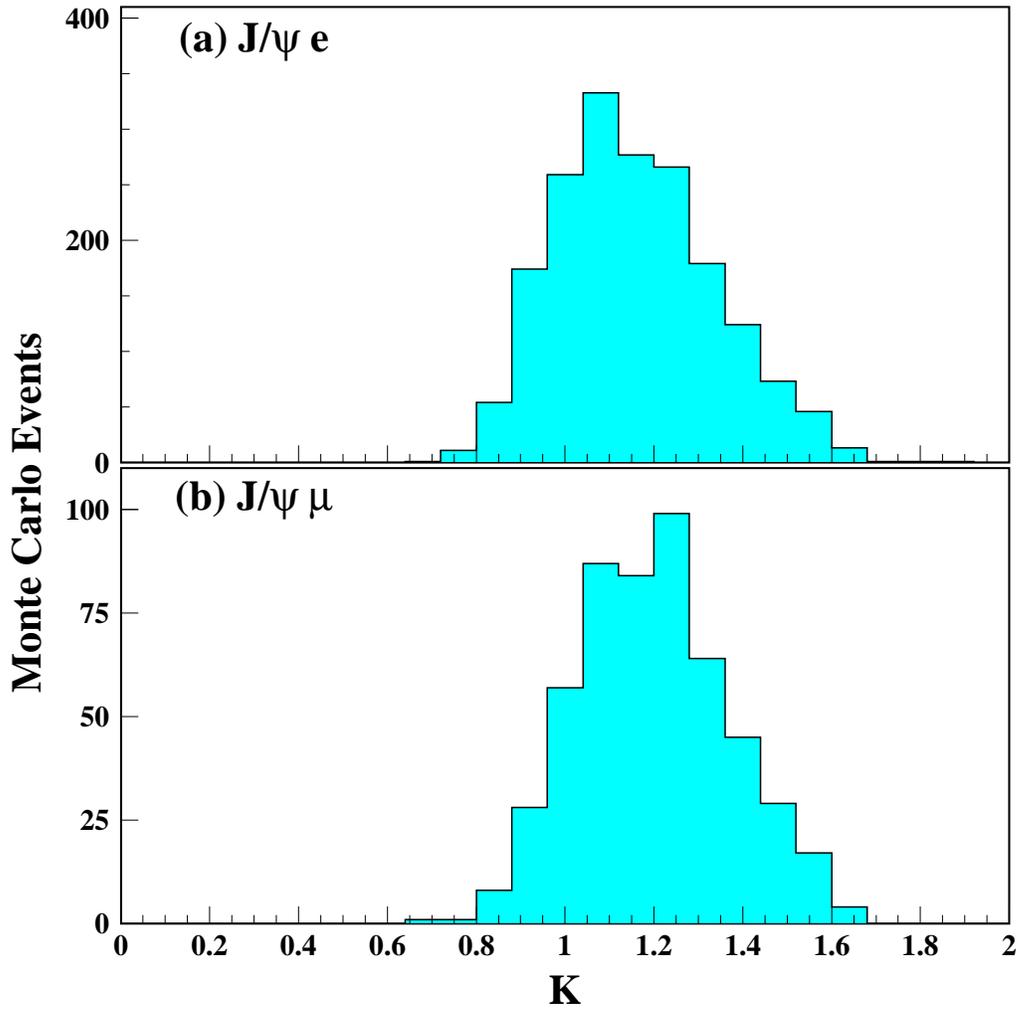}}
\caption{
$ K=(M(B_c)/p_T(B_c)/(M(J/\psi \, \ell)/p_T(J/\psi \, \ell))$ 
distribution using Monte Carlo simulation  
(a) for the electron channel and (b) for the muon channel.
}
\label{fig:kfig}
\end{figure}

For \BJpsie\ and \BJpsimu, 
we obtained the $K$\ distributions $H(K)$\ by Monte Carlo methods.
Figure~\ref{fig:kfig} shows the results of these calculations 
for the kinematic criteria 
$p_{T}(e) > 2$ GeV/$c$ or $p_{T}(\mu) > 3$ GeV/$c$, and
4 GeV/$c^{2} < M(J/\psi \ell) < 6$ GeV/$c^{2}$.
Since the criteria differ for the electron and muon, 
the $K$-factor distributions for these
channels were determined separately.  
For the exponential dependence of $R_K$\ on (1/\ctau) 
(Sec.~\ref{subsec:eff}),
the distributions in Fig.~\ref{fig:kfig} can be adequately
represented by $\langle 1/K \rangle = 0.88 \pm 0.02$, where 
we have adopted the difference between the two distributions 
as the uncertainty.

\begin{figure}
\epsfxsize=6.0in
\centerline{\epsffile{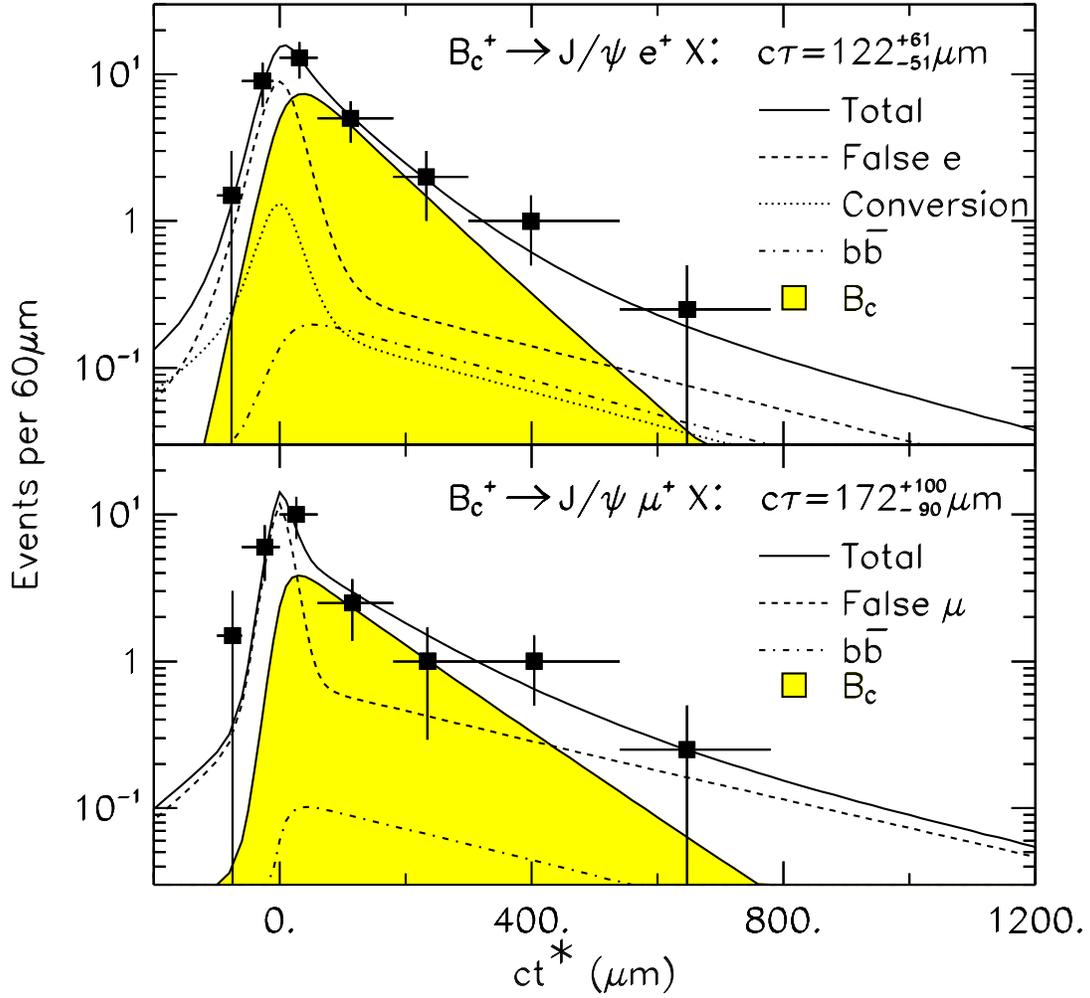}}
\caption{
Pseudo-proper decay length distributions for data with 
the fitted curve and the contributions from backgrounds 
(a) for the electron channel and (b) the muon channel.
}
\label{fig:ct_fitk1}
\end{figure}

\begin{figure}
\epsfxsize=6.0in
\centerline{\epsffile{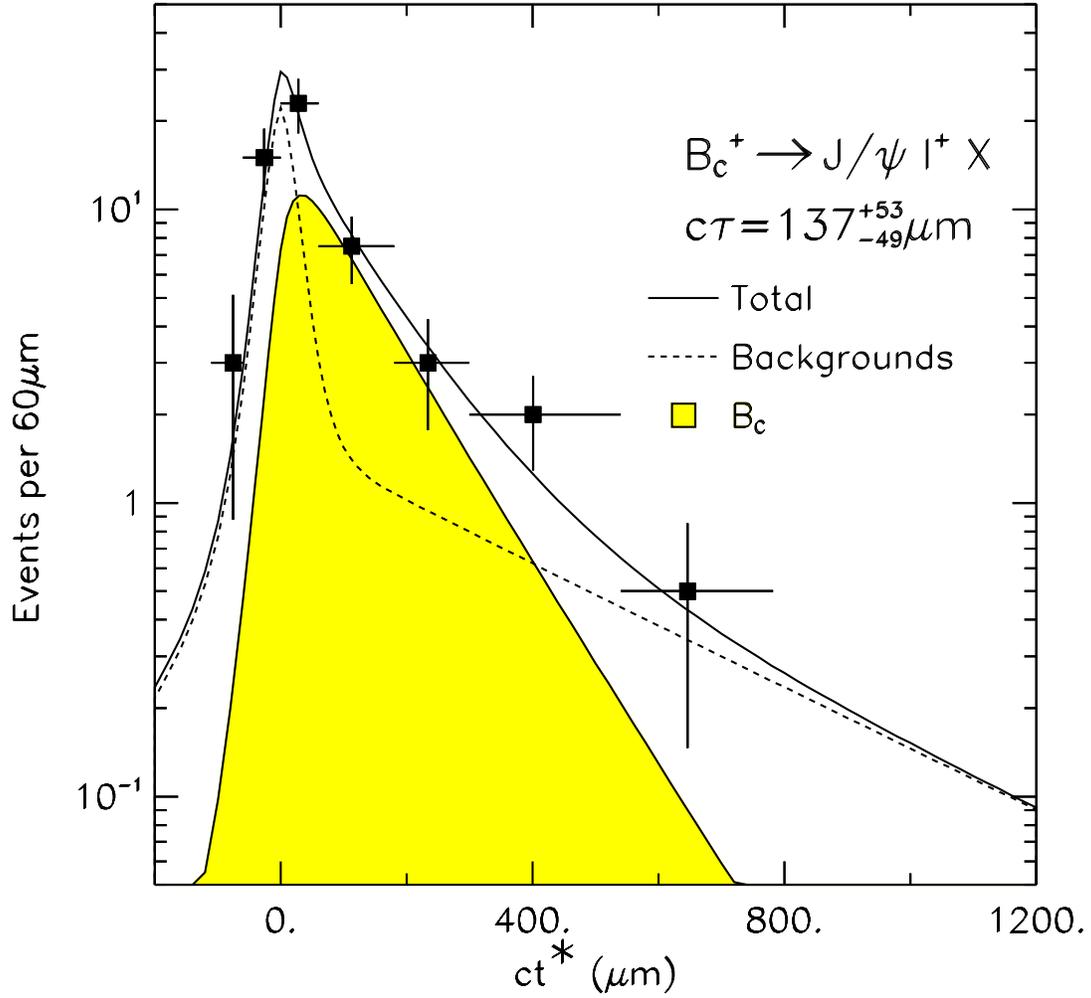}}
\caption{Pseudo-proper decay length distribution 
for the combined \Jpsimu\ and \Jpsie\ data along with 
the fitted curve and contributions to it from signal and background 
}
\label{fig:ct_fitk2}
\end{figure}

The quantity $c t^{\ast}$ was determined
for each event by the relation given in Eq.~\protect\ref{eq:ctstar}.
The points with uncertainties in Fig.~\ref{fig:ct_fitk1} 
show the binned \ctstar\ distributions for the 
$J/\psi e$ and $J/\psi \mu$ data.  The two decay channels are 
combined in Fig.~\ref{fig:ct_fitk2}.

\subsection{Background and Signal Distributions in \ctstar}
We used a procedure similar 
to that described in detail in Ref.~\cite{B-lifetimes} to
account for backgrounds.
We constructed functions to represent the \ctstar\ distributions,
for signal and backgrounds and 
convoluted them with a Gaussian resolution function.

The evaluation of backgrounds for events with \ctstar\
greater than 60 \um\ was described in Sec.~\ref{sec:background}.
The same procedures were used independently for events with 
\ctstar\ between -100 \um\ and 60 \um\ 
which have ``prompt'' contributions from direct charmonium 
production.
 
We obtained the best fit to the \ctstar\ distributions for each 
of the backgrounds using the same methods discussed in
Sec.~\ref{sec:background} for the background rate determinations.
The general shape in $x =$\ \ctstar\ 
used for each of the backgrounds was a sum of three terms: 
\begin{itemize}
 \item	A right-side (\ctstar$>0$) exponential dominated
	by the decay of ordinary $B$s in the background.
	Its fractional contribution is $f^j_{+}$\ and
	its exponential slope is $\lambda^j_{+}$.
 \item	A left-side (\ctstar$<0$) exponential to account for 
	an observed low level background from daughters of 
	$B$\ decay incorrectly associated with particles
	from the primary intraction vertex.
	Its fractional contribution is $f^j_{-}$\ and
	its exponential slope is $\lambda^j_{-}$.
 \item	A central Gaussian to account for prompt decays.
	Its fractional contribution is $(1 - f^j_{+} - f^j_{-})$.	
\end{itemize}
The index $j$\ stands for the various background contributions from 
false muons ($j = f \mu$), 
false electrons ($j = fe$) 
and undetected conversion electrons  ($j = ce$).
For the \BBbar\ backgrounds  ($j = B \mu$, $Be$), 
the central Gaussian term in Eq.~\protect\ref{eq:ct_bgshape} 
was not needed, $i.e.$\ $f^{B\ell}_+ + f^{B\ell}_- = 1$.
The exponentials were convoluted with a Gaussian resolution function.
This sum can be written
\begin{eqnarray}
{\cal F}^{j}(x) & = & (1-f^j_{+}-f^j_{-})G(x;s^j\sigma) \nonumber \\
& & + \frac{\displaystyle{f^j_{+}}}{\displaystyle{\lambda^j_{+}}}
\theta(x)
\exp \left(-\frac{\displaystyle{x}}{\displaystyle{\lambda^j_{+}}} \right)
\otimes G(x;s^j\sigma) \nonumber \\
& & +\frac{\displaystyle{f^j_{-}}}{\displaystyle{\lambda^j_{-}}}
\theta(-x)
\exp \left(+\frac{\displaystyle{x}}{\displaystyle{\lambda^j_{-}}} \right)
\otimes G(x;s^j\sigma).
\label{eq:ct_bgshape}
\end{eqnarray}
where the Heaviside function $\theta(x)$ is defined as 
$\theta(x) = 1$ for $x$ $\geq$ 0 and $\theta(x) = 0$ for $x$ $<$ 0.
The product $s^j \sigma$\ is the one-standard-deviation width
of the Gaussian distribution, where $\sigma$\ is the measurement
uncertainty on $x$\ for each event and $s^j$\ is a fitted 
scale factor.  
In all background fits,
the $s^j$\ were consistent with a common value of $s = 1.4$.
Therefore, $s$\ was fixed at that value.
Figure~\ref{fig:bkg_shape} shows the distributions and 
fitted functions for the backgrounds.
Table~\ref{tab:ct_bg} shows the fitted shape parameters
for each background.
The values of $\lambda^j_+$\ suggest that the backgrounds 
are dominated by partially reconstructed $B$\ mesons.  
Table~\ref{tab:ct_bg} also shows 
the numbers of events for each background.
These differ from the corresponding numbers in
Tables~\ref{tab:count} and \ref{tab:like} because of 
differences in the selection criteria for 
\ctstar\ and tri-lepton mass used here.
For this reason, we adopt a double-prime notation
for this analysis, $e.g.$\ $n^{\prime \prime f \mu}$\ for 
the number of false muon events with $M(J/\psi \, \ell)$\ 
in the range 4.0 to 6.0 GeV/$c^2$\ and 
with $-100 \, \mu m < ct^* < 1500 \, \mu m$.

Our fitting procedure accounted for a difference
between the relative pion and kaon fractions contributing  
to the prompt background and 
that contributing to background 
in the $B$-like region with $ct^* > 60$\ \um.
The fit also allowed variation in the relative 
probability for pions and kaons to be falsely
identified as electrons or muons.  These considerations 
allow additional variation of the values of $f^j_{\pm}$\ in 
Table~\ref{tab:ct_bg} and are discussed in
App.~\ref{subapp:unbin_ct}.

We assumed an exponential decay for the contribution from \Bc,
but we convoluted it with the $K$\ distribution and a Gaussian 
distribution to account for measurement uncertainty.
\begin{eqnarray}
{\cal F}_{sig}^{\ell}(x,c\tau) & = & \int \left[ H(K) 
\left( \frac{K}{c \tau} \right)
e^{\left(- \frac{K x}{c \tau} \right)}
\otimes G(x;s^{\ell}\sigma)  \right] dK
\label{eq:ct_sigshape}
\end{eqnarray}
where $\ell = \mu, \, e$.
The weighted sums of signal and background probability
distributions are defined in App.~\ref{subapp:unbin_ct}.

\subsection{ Unbinned Likelihood Fit for \ctau}
We used an unbinned likelihood method to obtain a best estimate of 
\ctau\ for each decay channel individually and for the combined 
dataset.  A parameter in the fit was assigned to each of the 
quantities in Table~\ref{tab:ct_bg}.  The numbers of events 
in each background were constrained by their measured or 
calculated values as in the previous sections.  
The full covariance matrices from the fits that determined the 
background shape parameters were used to constrain them
in the lifetime fit.  As before, we used the total number of  
events $n^{\prime \prime \ell}$\ and the electron fraction
$r^{\varepsilon}$\ to describe the \Bc\ signal with
$n^{\prime \prime e} =  r^{\varepsilon} n^{\prime \prime \ell}$\ and 
$n^{\prime \prime \mu} =  (1 - r^{\varepsilon}) n^{\prime \prime \ell}$.
The only parameter unconstrained by information beyond the  
candidate events was \ctau , the 
mean decay length for the \Bc\ contribution to the \ctstar\
distribution.  The likelihood function is presented in 
App.~\ref{subapp:unbin_ct}

\begin{figure}
\epsfxsize=6.0in
\centerline{\epsffile{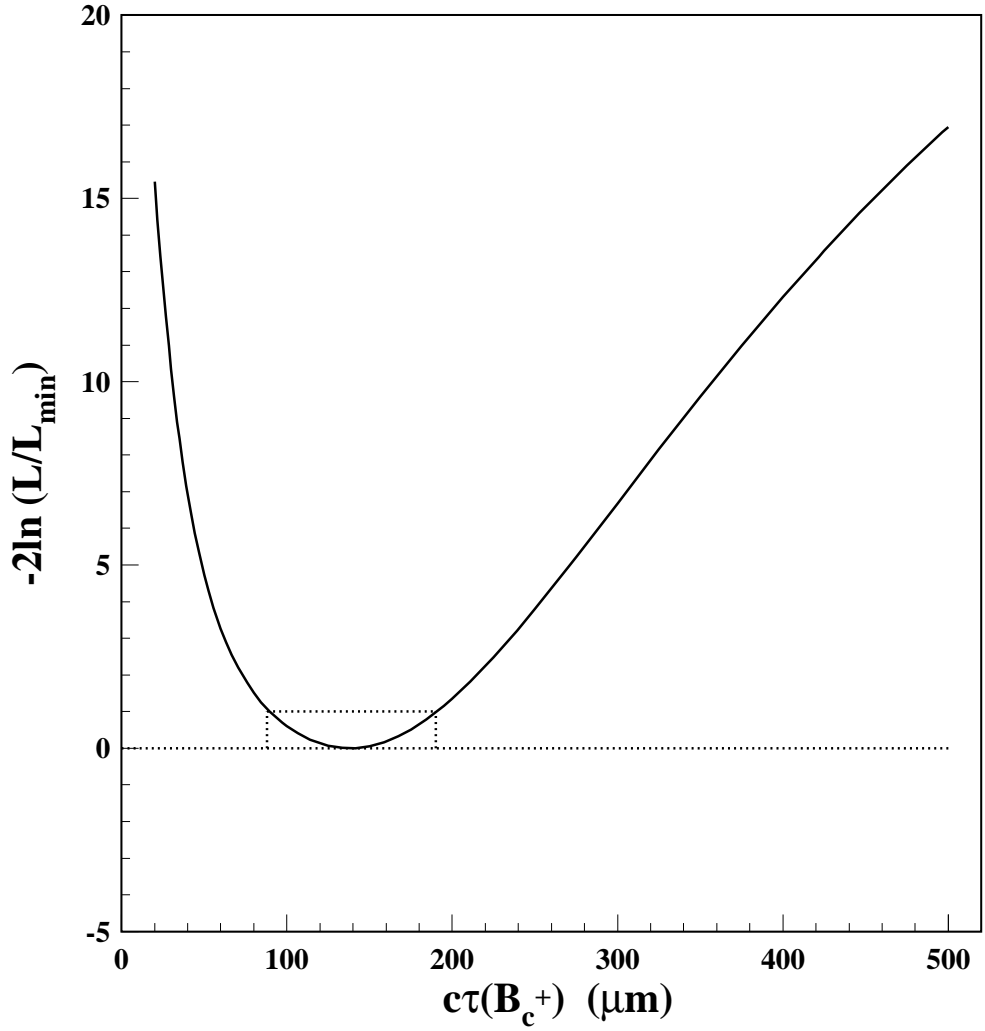}}
\caption
{The change in $-2\ln({\cal L})$\ from its minimum as a function of 
\ctau\ for the fit to the \ctstar\ distribution of \Bc\ candidates.
}
\label{fig:lk}
\end{figure}

The result of the log-likelihood fit to the \ctstar\ distribution
for \Jpsie\ events is 
\begin{eqnarray}
	c \tau = 122^{+61}_{-49} \: \mu{\rm m} 
%\label{eq:ctau_mu}
\end{eqnarray}
For \Jpsimu\ events, the fit yielded
\begin{eqnarray}
	c \tau = 172^{+100}_{-90} \: \mu{\rm m} 
%\label{eq:ctau_e}
\end{eqnarray}
The solution for a simultaneous fit to all events is
\begin{eqnarray}
	c \tau = 137^{+53}_{-49} \: \mu{\rm m} 
\label{eq:ctau}
\end{eqnarray}
\begin{eqnarray}
	  \tau = 0.46^{+0.18}_{-0.16} \: {\rm ps }
%\label{eq:tau}
\end{eqnarray}
The variation of $-2\ln({\cal L})$\ from its minimum as 
a function of \ctau\ is shown in Fig.~\ref{fig:lk}.
The simultaneous fit also determined the number of \Bc\ events to be
\begin{eqnarray}
	n^{\prime \prime \ell} = 34.2^{+8.2}_{-7.5} \:{\rm events} 
\end{eqnarray}
With the mean decay length above, the 
acceptance for \ctstar\ greater than 60 \um\ is 
$0.61^{+0.09}_{-0.15}$, and we can calculate 
\begin{eqnarray}
	n^{\ell} = 20.9^{+5.3}_{-5.5} \:{\rm events} 
\end{eqnarray}
for comparison with Eq.~\protect\ref{eq:evts_sig_region}.  
Clearly there is a large correlation between these two numbers
because of the largely overlapping event samples.
However, the consistency of the size of the \Bc\ signal as
determined from both the tri-lepton mass distribution and 
the \ctstar\ distribution adds confidence to the result.

\subsection{Statistical Tests of the Fit}
In order to test the adequacy of our model for signal
and background, we ran a number of pseudo-experiments
based on the fitted values of 
$R^{\varepsilon}$, $n^{\prime \prime \ell}$,
and the background parameters.  
For each of the pseudo-experiments, we varied these parameters
randomly according to the appropriate 
Poisson or Gaussian uncertainties.
The value of \ctau\ was fixed at 140 \um\ for all
pseudo-experiments.
From these quantities, we 
constructed the \Jpsie\ and \Jpsimu\ probability distributions 
for the independent variable \ctstar.
The dataset for a pseudo-experiment 
consisted of contributions from a signal plus three types of 
background for \Jpsie\ and a signal plus two types of background
for \Jpsimu.  
For each of the five backgrounds the number of events was 
allowed to fluctuate according to Poisson statistics, and 
the value \ctstar\ was chosen randomly according to the 
appropriate probability distribution.
The total number of signal events was chosen according 
to Poisson statistics, and each event was designated 
\Jpsie\ or \Jpsimu\ with probability determined 
by $R^{\varepsilon}$.
These samples were then subjected to the same
fitting procedures as the experimental data.
The comparison between the results for the pseudo-experiments
and those for the data tests the adequacy of
the fitting function to represent the data.

\begin{figure}
\epsfxsize=6.0in
\centerline{\epsffile{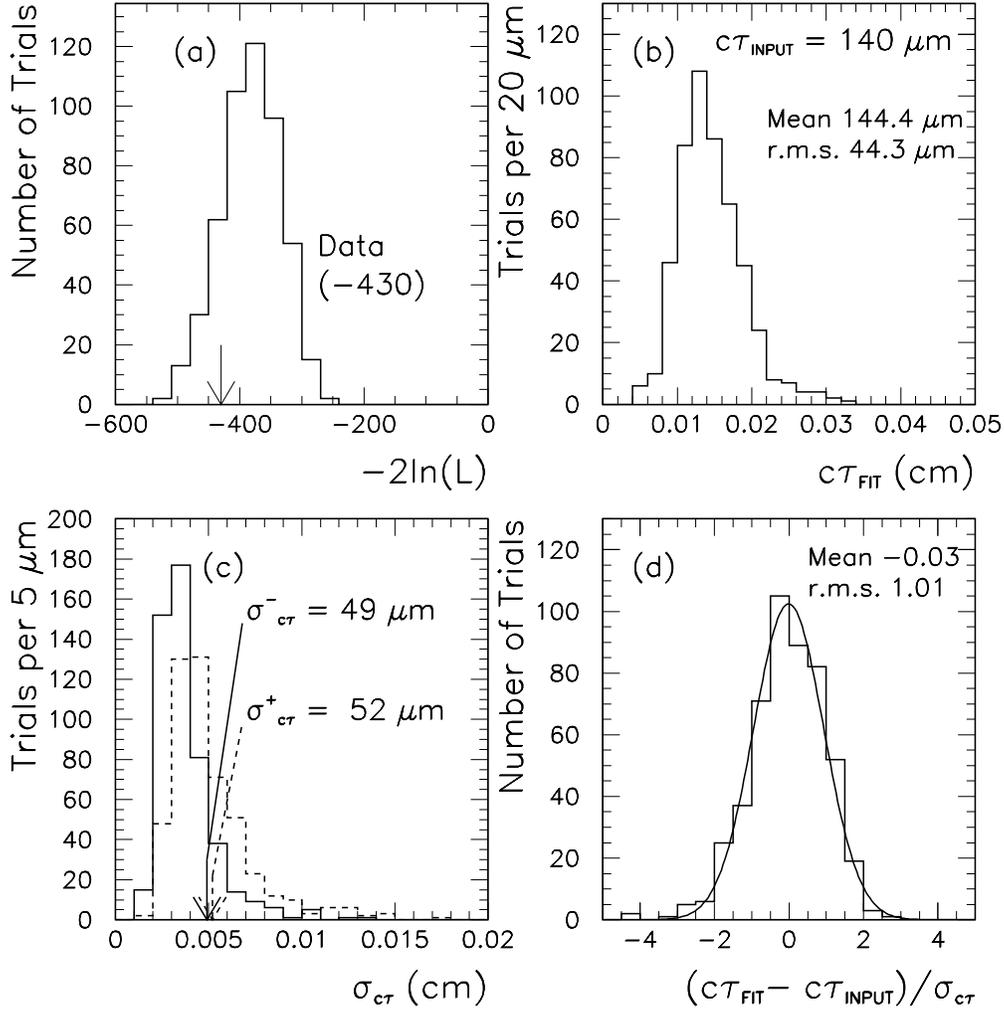}}
\caption
{Results from 500 pseudo-experiments to
simulate the statistics in the  \Bc\ lifetime analysis:
(a)~$-2\ln{\cal L}$; 
(b)~fitted lifetime; 
(c)~solid (dashed) line for the negative (positive) uncertainty;
(d)~($c\tau_{fit}$ $-$ $c\tau_{input}$)/$\sigma_{fit}$. 
In (d), the positive (negative) uncertanty was 
used when the fitted lifetime was smaller (larger)
than the input lifetime.}
\label{fig:ct_fit_test}
\end{figure}

Figure~\ref{fig:ct_fit_test}(a) shows the distribution for 
the log-likelihood with a mean value of $-382$\ and 
an r.m.s. width of 49.  The experiment yielded $-430$,
which corresponds to an 84\% confidence level. 
Figure~\ref{fig:ct_fit_test}(b) shows the distribution of 
fitted values of \ctau.  The mean of the distribution, 144~\um,
agrees closely with the input value of 140 \um, and the 
width is 44~\um, which consistent with the measured uncertainty.
Figure~\ref{fig:ct_fit_test}(c) shows the distributions of 
the upper (solid histogram) and lower (dashed histogram)
uncertainties from the fits.  Arrows indicate the 
corresponding uncertainties from the experimental data.  They 
are in reasonable agreement with the results from 
the pseudo-experiments. 
Figure~\ref{fig:ct_fit_test}(d) shows the distribution 
for deviation of the 
fitted \ctau\ from the input value normalized to the 
uncertainty from each fit.

We conclude that the model used to fit the data is adequate 
and that the resulting log-likelihood value and 
fitting uncertainties are consistent with expectations
based on the uncertainties in the data.

\subsection{Systematic Uncertainties}
The uncertainty reported by our fitting program already
includes some sources of systematic uncertainty 
because of the way we constrained the parameters 
describing the signal and backgrounds.
The fit shows a correlation of $-10\%$ between \ctau\ and 
the prompt electron fraction discussed 
in App.~\ref{subapp:unbin_ct}.
The correlations with all other fitting parameters are less 
than 5\%.  Thus, the \ctau\ value varies only a fraction of 
a standard deviation as other parameters in 
the analysis are varied.  
Refitting with parameters fixed at values different  
from nominal gives results consistent with this.
We estimate the systematic uncertainty included in
the fitting uncertainty to be less than 10 \um.
Thus, the fitting uncertainty is 
overwhelmingly statistical, and we quote it as such.

Below we discuss additional sources of systematic
uncertainty. Combined in quadrature, they amount to
about one-fifth the statistical uncertainty.

The $K$\ distribution (Eq.~\ref{eq:K} and Fig.~\ref{fig:kfig}), 
which was used to compensate for the 
information lost by our inability to detect the 
neutrino, is vulnerable to errors in our model of the 
\Bc\ production spectrum and its decay kinematics.

Figure~\ref{fig:pt_sum} shows that the \Pt\ spectra for 
data and background are very similar to that calculated 
for \Bc\ which was used to generate the $K$\ distribution.
To generate the \Bc\ Monte Carlo events, we used the 
next-to-leading order 
calculation of the $b$ quark spectrum~\cite{Mangano,Nason} 
with the MRSD0 parton distribution functions (PDF)~\cite{MRSD}, 
$m_{b}$ = 4.75 GeV, and the renormalization scale 
$\mu$ = $\mu_{0}$ $\equiv$ $\sqrt{m_b^2 + p_{T}(b)^2}$.
%The spectrum with CTEQ4M is steeper than  
%those with older structure functions.
We also generated a \Bc\ Monte Carlo sample using 
the CTEQ4M PDFs~\cite{CTEQ}  
to obtain a new $K$\ distribution 
and used it to fit the signal sample.
The value of \ctau\ thus obtained differed by 2 \um\ from  
the value in Eq.~\protect\ref{eq:ctau}.
Therefore, we assign $\pm$2 \um\ systematic uncertainty
for the PDFs.

We also refit the data with the assumed \Bc\ mass changed 
by $\pm 150$\ MeV.  This yielded a variation in 
\ctau\ of $\pm 1.6$\ \um.

A \Bc\ can decay to a lepton, a neutrino, and a higher 
mass \ccbar\ state 
that can subsequently decay to \JpsiX.  This would satisfy
the requirements for a candidate event, but would give rise to
a different $K$\ distribution.  Calculations based on the ISGW
model~\cite{isgw} indicate that the largest such contribution
comes from $B_c \rightarrow \psi(2S) \ell \nu$, which
could account for 12\% of the \BJpsil\ candidate sample.
We generated events of this type to obtain a $K$\ distribution
that we used to refit the candidate events. The value 
of \ctau\ changed by 1.9 \um\ which we adopt as a measure
of the systematic uncertainty for this effect.
We also considered the effects of
$B_c \rightarrow J/\psi \, \tau \nu$ ,
$B_c \rightarrow J/\psi \, D_s$ , and 
$B_c \rightarrow J/\psi \, D_s^{\ast}$.
We estimate their contribution to the \Bc\ sample to be 
less than 5\%. We assume that they produce no change
in the lifetime.

Our model for \Bc\ decay~\cite{CLEOMC} uses a $V-A$\ matrix
element.  As alternative, we generated events with the 
ISGW model~\cite{Scora} to obtain a new $K$\ distribution and 
refit the data.  This indicates a possible systematic 
uncertainty of $\pm 2.0$\ \um\

In order to test possible bias in our experimental trigger,
we turned off the trigger simulation in our Monte Carlo 
program and generated a sample of events without it to 
obtain a $K$\ distribution.  We assign $\pm 1$\ \um\ uncertainty
for this effect.

For each event in the lifetime analysis, the raw uncertainty 
in \ctstar\ was multiplied by a scale factor, $s=1.4$ that 
best fits the distributions in our background studies.
We changed this factor by $\pm 0.4$ and re-fit the 
background shapes.  We assign a
systematic uncertainty in \ctau\ of $\pm 8.3$\ \um\ for 
this effect.

In another analysis of $B$\ hadron lifetimes~\cite{B-lifetimes}, 
we studied the effects of detector alignment.  From this work,
we assign an uncertainty on \ctau\ of $\pm 2.0$\ \um.

In quadrature, these uncertainties sum to $\pm 9.4$\ \um, and 
we quote this as our systematic uncertainty 
with the caveat 
that some other sources have already been included in the 
fitting uncertainty which, nevertheless, remains 
predominantly statistical.
Thus, our  result is:
\begin{eqnarray}
	c \tau & = & \ctauBc \\
	  \tau & = & \tauBc
\end{eqnarray}

% ========================================================

\newpage
\section{ \Bc\ Production}
\label{sec:cross}
% ------------------ \input{cross} -----------------
From the event yield of Sec.~\ref{sec:fit}, we calculated the $B_c$\ 
production cross section times the \BJpsil\ branching fraction
\sigBRBcl. 
We express this product relative to that for the 
topologically similar decay \BJpsiK\ 
because the systematic uncertainties arising from the luminosity,
from the $J/\psi$\ trigger efficiency,
and from the CTC track-finding efficiency cancel in the ratio.  
Our Monte Carlo calculations yielded 
the values for the efficiencies that
do not cancel in the ratio.
We assumed that the branching fraction is the same for 
\BJpsie\ and \BJpsimu. 

We use the number of \Bc\ events from Eq.~\protect\ref{eq:evts_fit_region}
and the number of \JpsiK\ events from the fit in Fig.~\ref{fig:Jpsi_K}.
\begin{eqnarray}
	n^{\mu} + n^{e} & = & 20.4^{+6.2}_{-5.5}\: {\rm events} \\
	n^K	& = & 290 \pm 19\: {\rm events}
\end{eqnarray}
In order to be consistent with the efficiency
calculations of Sec.~\ref{subsec:eff},
the \Bc\ event count is that for $M(J/\psi \, \ell)$\ in the range 
3.35 to 11.0 GeV/$c^2$. 
We relate these quantities to the luminosity $\cal{L}$,
to the products of cross section and branching
fraction $\sigma \cdot BR$, and to the efficiencies discussed 
in Sec.~\ref{subsec:eff}.
\begin{eqnarray}
	n^{e}	& = & {\cal L} \cdot \sigma(B_c) 
		\cdot BR(B_c \rightarrow J/\psi \, \ell \nu) 
		\cdot \varepsilon^{e} \\
	n^{\mu}	& = & {\cal L} \cdot \sigma(B_c) 
		\cdot BR(B_c \rightarrow J/\psi \, \ell \nu) 
		\cdot \varepsilon^{\mu} \\
	n^{K}	& = & {\cal L} \cdot \sigma(B) 
		\cdot BR(B \rightarrow J/\psi \, K) 
		\cdot \varepsilon^{K} \\
	\frac{n^{e}+n^{\mu}}{n^{K}} & = &
		\frac{\sigma(B_c) \cdot BR(B_c \rightarrow J/\psi \,
		\ell \nu )}
		   {\sigma(B) \cdot BR(B \rightarrow J/\psi \, K)}
		 \cdot \frac{ \varepsilon^{e} + \varepsilon^{\mu}}
		   {\varepsilon^{K}} \\
	& = & \frac{\sigma(B_c) \cdot BR(B_c \rightarrow J/\psi \,
		\ell \nu)}
		   {\sigma(B) \cdot BR(B \rightarrow J/\psi \, K)}
		 \cdot \frac{ R^K }{ R^{\varepsilon} }
\end{eqnarray}
We used the value of $R^{\varepsilon}$\ from Eq.~\protect\ref{eq:Reps}.
We calculated the efficiency ratio $R^K$\ from Eq.~\protect\ref{eq:RK_corr}
and the lifetime discussed in Sec.~\ref{sec:life} to be
\begin{eqnarray}
	R^K & = & 0.263 \pm 0.035 \, ({\rm syst.}) \, ^{+0.038} _{-0.062} 
		\, ({\rm lifetime}).
\end{eqnarray}
As was discussed in Sec.~\ref{sec:life}, there can a contribution to
our data sample from other decay modes of the \Bc. 
Estimates of partial widths for higher charmonium states~\cite{Scora} 
yield an upper limit of 12\% for their contribution to the signal.
The estimated contributions from final states involving $D_s$, 
$D_s^{\ast}$, and $\tau$\ with subsequent decay to $e$\ or $\mu$\ 
total less than 5\%.  We assume an uncertainty equal to the 
magnitude of the correction $1/1.17 = 0.85 \pm 0.15$.
With these values we find 
\begin{mathletters}
\begin{eqnarray}
	{\cal R}(J/\psi \, \ell \nu) & \equiv & 
	\frac{\sigma(B_c) \cdot BR(B_c \rightarrow J/\psi \, \ell \nu)}
		   {\sigma(B) \cdot BR(B \rightarrow J/\psi \, K)} 
\label{eq:calR_def} \\
	 & = & 0.85 \frac{n^{e}+n^{\mu}}{n^{K}}
	   \cdot  \frac{ R^{\varepsilon} }{ R^K } 
	 \\
	&=& \sigmaBr .
\label{eq:sigB_ratio}
\end{eqnarray}
\end{mathletters}
The statistical uncertainty is from the event counts and the 
systematic uncertainty is from the efficiency ratios and the 
correction for other decay modes. 

%The kinematic selection criteria placed on the \Jpsi\ and third 
%particle for the events used in this study cannot be transformed 
%in a simple way to the transverse momenta and rapidity for the 
%parent $B$\ and \Bc\ in the above ratio.
%However, 
Based on Monte Carlo studies, 
the effective kinematic limits for \Bc\ mesons in this study are:
transverse momenta $p_T > 6.0$\ GeV/$c$\ and rapidity $|y| < 1.0$.
 
\begin{figure}
\epsfxsize=6.0in
\centerline{\epsffile{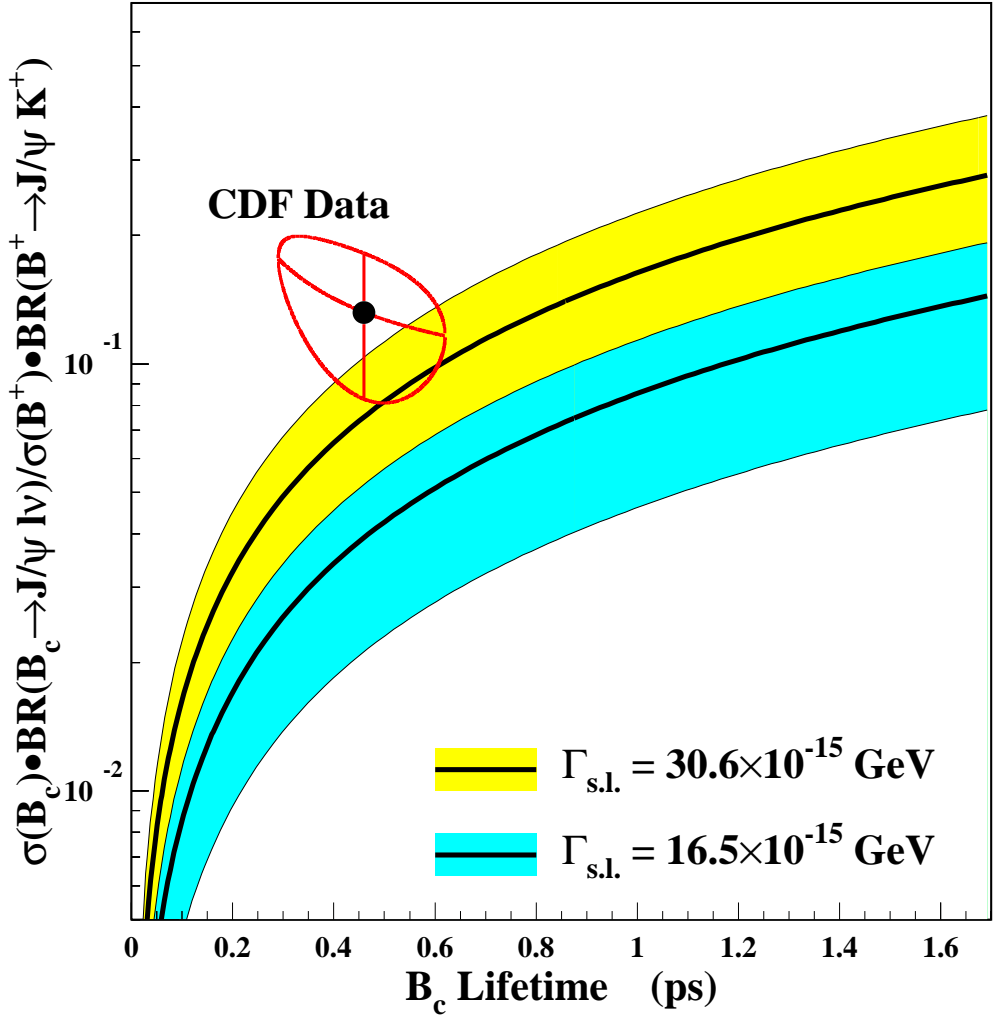}}
\caption{The point with 1-standard-deviation contour 
shows our measured value of the ratio 
$\sigma \cdot BR (B_{c}^{+} \to
J/\psi \; l^{+} X)/\sigma \cdot BR(B^{+} \to J/\psi K^{+})$ 
plotted at the value we measure for the \Bc\ lifetime.
The shaded region represents theoretical predictions 
and their uncertainty corridors for two
different values of the semileptonic width 
$\Gamma_{s.l.}$ based on 
Refs.~\protect\cite{Lusignoli_decay} 
and \protect\cite{isgw}.
The other numbers assumed in the theoretical predictions 
are discussed in the text.
} 
\label{fig:ct_crss1}
\end{figure}

Figure~\ref{fig:ct_crss1} shows theoretical predictions of the ratio 
${\cal R}(J/\psi \, \ell \nu)$\ 
as a function
of the assumed lifetime of the $B_{c}$.
The shaded regions in the figure represents the prediction and 
its uncertainty for two different assumptions about the 
semi-leptonic width 
$\Gamma_{s.l.} = \Gamma(B_{c}^{+} \to J/\psi \, \ell^{+} \nu)$.
Assumed in the theoretical predictions are
\begin{eqnarray}
	V_{bc} 
	&=& 0.041 \pm 0.005
	\hfill \cite{Particledata}, \\
	\Gamma(B_{c}^{+} \to J/\psi \, \ell^{+} \nu)
	&=& (30.6 \pm  16) \times 10^{-15}\, {\rm GeV} 
	\hfill \cite{Lusignoli_decay}, \\
	{\rm or } \, \Gamma(B_{c}^{+} \to J/\psi \, \ell^{+} \nu)
	&=& 16.5 \times 10^{-15}\, {\rm GeV} 
	\hfill \cite{isgw}, \\
	\frac{\sigma(B_{c}^{+})}{\sigma(\overline{b})}
	&=& 1.3 \times 10^{-3}
	\hfill \cite{Lusignoli_prod}, \\
	\frac{\sigma(B^{+})}{\sigma(\overline{b})} 
	&=& 0.378 \pm 0.022
	\hfill \cite{Particledata},  \\
	BR(B^{+} \to J/\psi K^{+}) 
	&=& (1.01 \pm 0.14) \times 10^{-3}
	\hfill \cite{Particledata}.
\end{eqnarray}
Fig.~\ref{fig:ct_crss1} also shows the measured cross section ratio 
(Eq.~\ref{eq:sigB_ratio}) plotted at the measured value of the lifetime.

In Sec.~\ref{sec:intro} we referred to results from previous searches
for the \Bc\ meson through its decay to various final states 
(f.s.) including 
$J/\psi \,  \pi$, 
$J/\psi \,  \pi^+ \pi^- \pi$, 
$J/\psi \,  a_1$\ and
$J/\psi \,  \ell \nu$.
We have converted the upper limits quoted in these searches to 
calculate in each case a corresponding upper limit on 
${\cal R}({\rm f.s.})$\ as defined in Eq.~\protect\ref{eq:calR_def}.
For these conversions, we used 
$BR(Z \rightarrow b \overline{b}) = 0.1546 \pm 0.0014$, 
$BR(Z \rightarrow q \overline{q}) = 0.6990 \pm 0.0015$, 
$BR(\overline{b} \rightarrow B^+) = 0.378 \pm 0.022$,
$BR(B^{+} \to J/\psi \,  K^{+}) = (1.01 \pm 0.14) \times
10^{-3}$\ \cite{Particledata}.
The limits reported for the LEP experiments are for the 
sums of the two charged conjugate modes, and they are 
modified by a factor of 2 for this calculation.
Table~\ref{tab:Br_limits} shows the results of these calculations.

% ========================================================

\newpage
\section{ Summary and Conclusions }
\label{sec:conclude}
% ------------- \input{conclude} ----------------------
This paper reports the observation of $B_c$ mesons.  The decay
mode used for the study was  $B_c \rightarrow J/\psi \, \ell X$
where $\ell$ is either an electron or a muon.  
A total of 31 events for which the mass of \Jpsil\ system was between  
4.0 and 6.0 GeV/$c^2$\ were found.
We performed a detailed study of backgrounds and 
estimate their contribution to this sample to be $12.1 \pm 1.9$\ events.
In the wider mass range 3.35 to 11.0 GeV/$c^2$ we found 37 $B_c$\ 
candidates with an estimated background of $21.4 \pm 3.1$\ events.
We performed a shape-dependent likelihood fit to the mass distribution
and found that it required a $B_c$\ contribution of 
$20.4^{+6.2}_{-5.5}$\ of which $19.0^{+5.8}_{-5.1}$\ have masses 
between 4.0 and 6.0  GeV/$c^2$.  A fit without a $B_c$\ 
contribution was rejected at the level of 4.8 standard deviations.

By repeating the above procedure with a number of assumed masses
between 5.52 GeV/$c^2$\ and 7.52 GeV/$c^2$\ we determined that 
the mass of the \Bc\ meson is 
$M(B_c) = 6.40 \pm 0.39\,{\rm (stat.)} \pm 0.13\,{\rm (syst.)}$\
GeV/$c^2$.

We studied the displacement of the $B_c$\ decay vertex position from
the average beam line, and from it we measured the $B_c$\ lifetime 
to be $\tau (B_c) = \tauBc$ .

Finally, we estimated ratio of 
the product of the production cross section times
branching fraction for $B_c^+ \rightarrow J/\psi \ell^+ \nu$ 
to that for $B^+ \rightarrow J/\psi K^+$\ to be 
\begin{displaymath}
	\frac{\sigma(B_c^+) \cdot BR(B_c^+ \rightarrow J/\psi \, \ell^+ \nu)}
		   {\sigma(B^+) \cdot BR(B^+ \rightarrow J/\psi \, K^+)}
	 =  \sigmaBr .
\end{displaymath}

% ========================================================

\newpage
%\section*{Acknowledgments }
\acknowledgements
% ------------------- \input{acknowledge} ----------------
We thank the Fermilab staff and the technical staff at the 
participating institutions for their essential contributions
to this research.  This work is supported by the 
U.~S.~Department of Energy and the National Science Foundation;
the Natural Sciences and Engineering Research Council of Canada;
the Istituto Nazionale di Fisica Nucleare of Italy;
the Ministry of Educaton, Science and Culture of Japan;
the National Science Council of the Republic of China;
and the A.~P.~Sloan Foundation.
% ========================================================

\newpage
\appendix
\section{Event Simulation}
\label{app:monte}
% ------------ \input{monte} -------------------------
A number of quantities and distributions needed for this work 
could not be measured directly from 
the experimental data.  For these we relied on
Monte Carlo simulations of particle production and decay and of 
our detector's response to final state particles.
The Monte Carlo program consisted of several parts:
\begin{itemize}
 \item	We generated \bbbar\ quark pairs according to the 
	predictions of 
	a next-to-leading order QCD calculation 
	\cite{Mangano,Nason} using the MRSD0 parton distribution
	functions~\cite{MRSD}.  We required $p_T > 5$\ GeV/$c$\
	for a $b$-quark. 
	We assumed the distribution in rapidity $y$\ to be flat
	in the range $|y| < 1.2$.
 \item	We determined the $b$\ quark fragmentation into a $B$\ meson 
	using the Peterson parametrization with the parameter 
	$\epsilon = 0.006$~\cite{Peterson,Chrin}.
 \item	For \Bc\ production we used the fragmentation 
	model of Ref.~\cite{Braaten}.
 \item	We used the CLEO $B$\ decay model~\cite{CLEOMC}, 
	for the decay of the $B$\ meson 
	and its daughter particles.
 \item	We used full simulation of the CDF detector to calculate
	its response to the final state particles. 
\end{itemize}
The resulting Monte Carlo events were processed with the same 
programs used to reconstruct the data.  The processes 
we studied with this program were:
\begin{itemize}
 \item	\BJpsie,
 \item	\BJpsimu,
 \item	\BJpsiK,
 \item	Pairs of $B$\ mesons with \BJpsiX\ accompanied by
	$\overline{B} \rightarrow e$\ or $\mu$\ 
	either directly or through its daughters.%
\end{itemize}
These studies yielded ratios of the detection efficiencies 
$\varepsilon($\BJpsie), $\varepsilon($\BJpsimu) and 
$\varepsilon($\BJpsiK), the \BBbar\ backgrounds described 
in Sec.~\ref{subsec:BBbar}, 
and the $K$\ distributions used in Sec.~\ref{sec:life}.

In addition, we employed hybrid Monte Carlo calculations that 
replaced a real track in a \Jpsi\ + track event by another 
particle to study punch-through, decay-in-flight, 
and photon-conversion backgrounds.
These studies are described in Sec.~\ref{sec:background}.

% ========================================================
\newpage
\section{Validation of Background Estimates}
\label{app:valid}
% ----------------- \input{valid} ------------------------
\subsection{Semileptonic $B$ Decay Sample}

We confirm our ability to determine accurately the various background 
rates to our observation of the $B_c$\ meson by using identical methods 
to determine the background rate for a different process studied in a 
data sample independent of that which yielded the \Jpsi\ + track 
distributions in 
Figs.~\ref{fig:jpsi_track_e} and \ref{fig:jpsi_track_mu}.

In $b$ hadron decays, leptons are produced either directly in the
$b\to c$ decay or in the sequential decay of the daughter charm
hadron.  Pairs of leptons thus arise from events in which there is a
both a prompt and sequential semileptonic decay of a single $B$ or
from \BBbar\ pairs.  The leptons in the sequential decays are
necessarily opposite charge and have a two-particle mass less than
5\,GeV/$c^2$.  Leptons from \BBbar\ pairs may be of the same charge
either because of mixing or where one lepton is direct and the second
is sequential.  The pair-mass, however, tends to be large and is
typically greater than the $B$ mass.  Thus, low-mass, same-charge
pairs of identified leptons in $B$ events form a nearly pure
background sample in which we can test our algorithms.

Our overall strategy for obtaining such a sample was to select 
lepton pairs in which one lepton was responsible for the trigger
and came from a displaced vertex.  
We required the other lepton also to originate 
in a displaced vertex in the same jet cone as the trigger lepton.
This emphasized low mass pairs.  

Our inclusive, high-\Pt\ lepton trigger provides a large sample of
semileptonic $b$ (and $c$) decays.  However, even after strict
identification cuts these events are contaminated by events in which
the lepton is a misidentified hadron.  
Therefore, we need to identify the event as
a $B$ decay by other means.  To do so, we take advantage of the long
$B$ lifetime.  In central electron and muon events with lepton
$P_T>7.5$\,GeV/$c$, we reconstruct jets in the calorimeter using a cone
algorithm \cite{jetcone} with a cone radius of
$R\equiv\sqrt{\eta^2+\phi^2}=0.7$.  We require a jet of $E_T>10$\,GeV
and search for displaced decay vertices using charged particle tracks
that lie inside the jet reconstruction cone.  
We define the impact
parameter significance $s\equiv|d_0|/\sigma_d$ where $d_0$ 
is the impact parameter in the transverse plane  
with respect to the beamline, and $\sigma_d$ is its
measured uncertainty including the known transverse beam width.  We
require either that the lepton and two additional tracks in the cone
satisfy $s>2.5$ or that the lepton and one additional track satisfy
$s>4.0$.  In all cases, we require that the displaced tracks originate
from a common point and that the vertex be
forward of the beamline with $L_{xy}/\sigma_{xy}>2.0$, where
$\sigma_{xy}$ is the uncertainty on $L_{xy}$.

To estimate the purity of this sample, we make use of another property
of semileptonic $B$ decays.  The lepton is typically the leading
particle in the decay.  Further, the lepton spectrum in the $B$
rest-frame is well established\cite{bsemi}.  In the candidate events,
we find the distribution of the momentum of the lepton transverse to
the jet direction $P_{T,rel}$ and fit it to Monte Carlo templates for
direct-$b$ and sequential decays, $c\bar{c}$ production, and false 
leptons from mismeasured prompt jets.  We find a sample composition of
approximately 85\% $b\bar{b}$, 10\% ${c\bar{c}}$, and 
5\% false leptons.  

The tracks in the event, except for the trigger lepton, provide the
parent sample to test the backgrounds to our soft-lepton
identification.
For each track that satisfies our
electron or muon geometric requirements and comes from 
a displaced vertex in the same jet 
cone as the trigger lepton, we find the mass of the
trigger-lepton and candidate track combination.  We weight the mass by
the track's false lepton probability (as determined in section IV) and
histogram the mass for same-charge and opposite-charge combinations.
We compare this to combinations in which the candidate track satisfies
our lepton identification criteria.  Next-to-leading-order processes
can contribute to the low-mass regions with leptons from different
$b$ hadrons.  Therefore, to make an accurate comparison, we find the
distribution of lepton-pair masses in \BBbar\ Monte Carlo simulation
subject to our trigger and identification criteria.  
We used the number of trigger leptons to normalize the \BBbar\ 
Monte Carlo calculation to the experimental results.

\begin{figure}
\epsfxsize=6.0in
\centerline{\epsffile{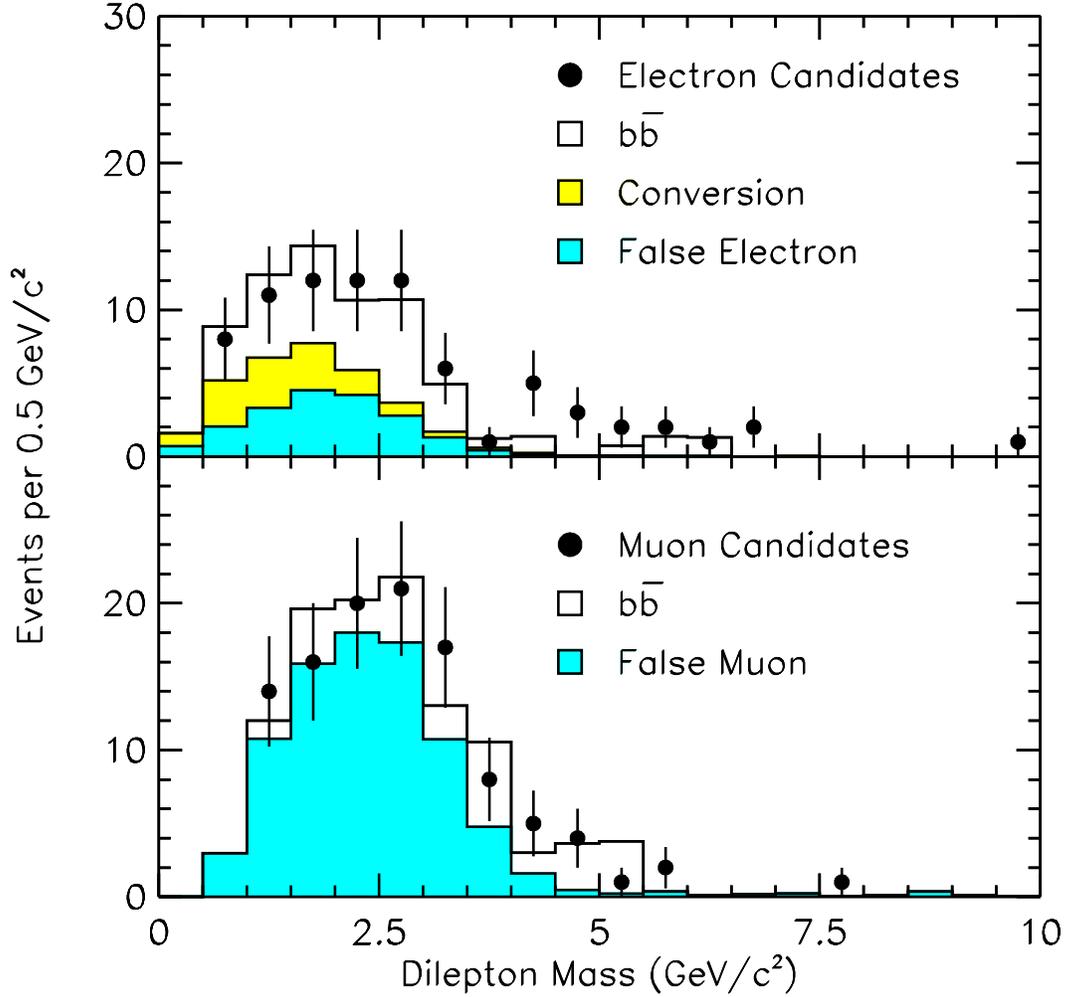}}
\caption{
Same-charge di-lepton mass distributions for a trigger 
lepton and a tagged lepton.  
Both were required to 
come from a displaced vertex and be within the same 
jet cone.
In (a) the tagged 
lepton is an electron, and in (b) the tagged lepton 
is a muon.
In both cases, the data from trigger electrons and 
that from trigger muons are combined.
The points with uncertainties are data, and the 
histograms show the predicted contributions from the various 
backgrounds relevant to the $B_c$ analysis.
} 
\label{fig:secvtx}
\end{figure}

For various combinations of 
electrons and muons identified in the trigger 
and those identified in subsequent analysis (tagged)
Fig.~\ref{fig:secvtx}  
shows the mass distributions of same-sign di-leptons. 
The points with uncertainties are the data, and the histograms
represent the contributions from
the same backgrounds relevant to the $B_c$ analysis.
Table~\ref{tab:secvtx} lists the number of expected and
observed di-lepton pairs for $M_{\ell\ell}<5$\,GeV$/c^2$.
The calculated and observed same-sign di-lepton data are
in reasonable agreement within the statistical uncertainties.
This supports the validity of the background calculation
in the \Bc\ analysis.  

We also removed the requirement that the second lepton come 
from a displaced vertex in the same jet cone as the trigger 
lepton and repeated the analysis with this larger sample.  
In this case, we normalized the
\BBbar\ contribution by requiring that the sum of same- and 
opposite-charge \BBbar\ and false-lepton contributions in the high-mass
($M_{\ell\ell}>5$\,GeV$/c^2$) region be equal to the total number of
di-lepton events.   The two normalization procedures agreed.

\subsection{ Impact Parameter Significance }
\begin{figure}
\epsfxsize=6.0in
\centerline{\epsffile{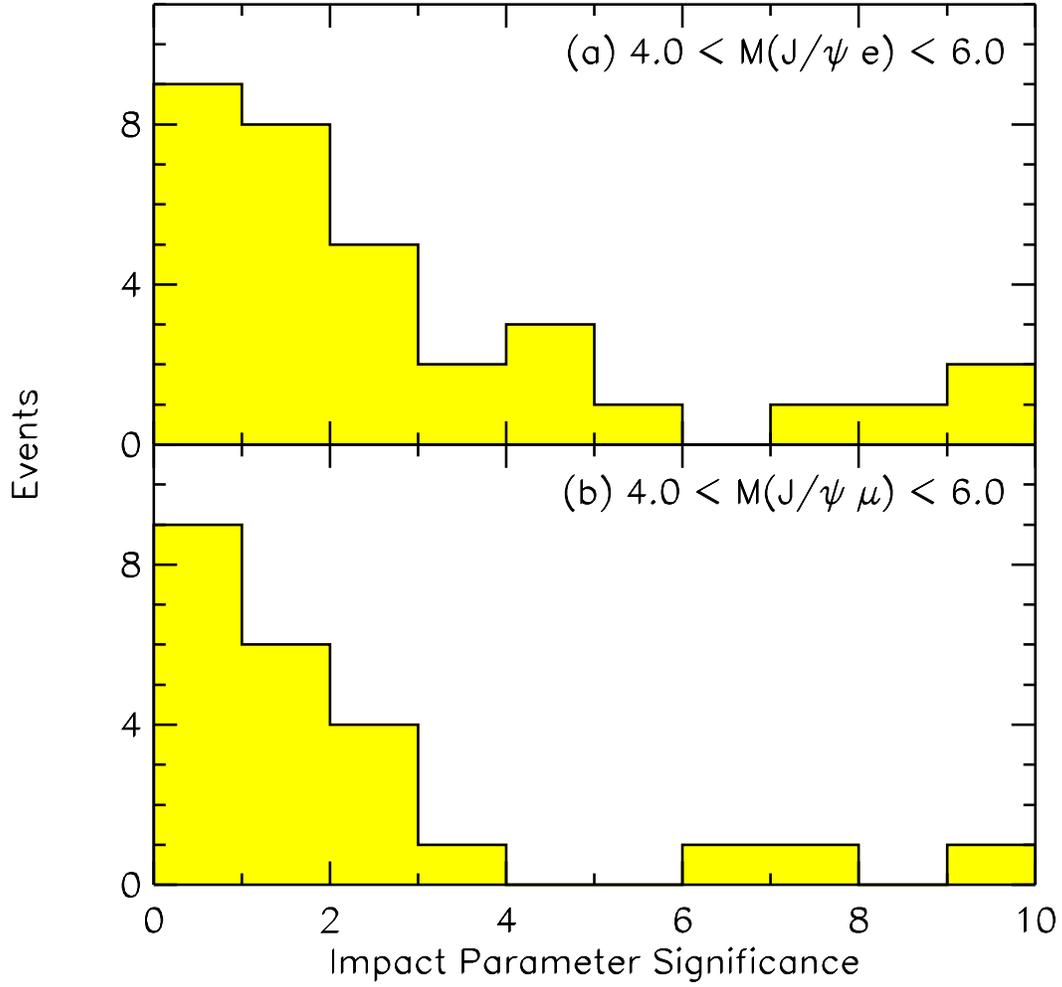}}
\caption{(a) Distribution of the impact parameter significance of the
third track with respect to the $J/\psi$ vertex for \Jpsie\ events.  
(b) The same distribution for the \Jpsimu\ events.
}
\label{fig:ip_bb}
\end{figure}

We present additional evidence that the \BBbar\ background, based 
on a Monte Carlo calculation, is indeed small.
We re-analyzed the \Jpsil\ data with a modified procedure
which relaxed the requirements that the third track come from
the same point as the \Jpsi\ decay vertex.
\begin{itemize}
 \item	We performed a two-track mass and vertex constraint on 
	$J/\psi \rightarrow \mu^+ \mu^-$\ and required the good-fit
	probability to be greater than 1\%.
	This departs from our standard procedure of requiring all
	three leptons to originiate at a common vertex.
 \item	With the third lepton, we calculated the \Jpsil\ mass, \Pt\ and 
	\ctstar\ based on the \Jpsi\ vertex.
 \item	We required \ctstar\ to be greater than 60 \um.
 \item	We calculated the distance of closest approach of the 
	third lepton track to the \Jpsi\ vertex $d$ and its 
	uncertainty $\sigma_d$.  We define 
	the ratio $d/\sigma_d$\ as the impact parameter significance.
\end{itemize}
Figure~\ref{fig:ip_bb}(a) shows the impact parameter significance for 
electrons with respect to a $J/\psi$\ vertex for the data.  
Figure~\ref{fig:ip_bb}(b) shows the same quantity where the third lepton
is a muon.
Backgrounds from \BBbar\ should extend to higher values of the impact
parameter significance because the \Jpsi\ and the third lepton 
come from different vertices.  
\Bc\ events should populate the low impact parameter region because
the \Jpsi\ and the third lepton emerge from a common vertex.
The figure shows that, when this region is included, 
most events have low impact parameters.
Note that the events in Fig.~\ref{fig:ip_bb} are a superset of our final
data sample because of the relaxed vertex requirements.
When we account for the effect of the relaxed requirements 
on these events, the level of events with high impact 
parameters is in good agreement with our predicted levels
of \BBbar\ backgrounds.

% ========================================================

\newpage
\section{The Likelihood Function}
\label{app:like}
% ------------------ \input{like} ------------------------
For the likelihood analysis to test the null hypothesis and 
to estimate the size of the \Bc\ signal we used a normalized 
log-likelihood function.
\begin{equation}
	\xi^2 = -2 \ln \left( \frac{ \cal{L}}{ {\cal L}_0} \right)
\label{eq:xi2}
\end{equation}
where ${\cal L}$\ is the likelihood function, {\it i.e.} the product 
of all the probability distributions in the analysis, and 
${\cal L}_0$\ is its value for a perfect fit.
For purely Gaussian probability distributions, $\xi^2$\ is formally identical
to the commonly used $\chi^2$.
The advantage of $\xi^2$\ for a more general ${\cal L}$\ is that
its properties are quantitatively similar to $\chi^2$.%
\footnote
{ As an example, if ${\mathcal L}$ is a simple product of 
either Binomial or Poisson probabilities, 
it is easy to derive an expression
for the inverse of the co-variance matrix for $\xi^2$\ in the same 
way one does for $\chi^2$.  
This yields the textbook uncertainties in the parameters. 
A Taylor expansion of the logarithmic terms in $\xi^2$\  reveals that
a one-standard-deviation change in a parameter 
from its best-fit value increases $\xi^2$\ by approximately one unit.
}

Below we define the input information and corresponding parameters
along with the  constraints and relationships among them,
and we present the normalized log-likelihood function.  
Upper case letters represent input information,  and 
lower case letters represent parameters of the fit.  
The superscript $\mu$\ ($e$) refers to \Jpsimu\ (\Jpsie).  
We designate background types by additional superscripts, $fe$\ and 
$f\mu$\ for false leptons, $ce$ for conversion electrons,
$B\mu$\ ($Be$) for the \BBbar\ contributions to muon (electron)
backgrounds.
We use $\lambda_i^{\mu}$  ($\lambda_i^e$) 
to represent a function of the parameters corresponding 
to the fitted number of signal plus background events in the $i^{th}$\ bin
for the muon (electron) distribution.
We use primes ($N^{\prime}$, $n^{\prime}$) for the number of events in the
mass range 3.35--11.0 GeV/$c^2$,  and 
elsewhere we use unprimed numbers ($N$, $n$) for the 
subset in the range 4.0--6.0 GeV/$c^2$.

In order to propagate the uncertainties for various measured or 
calculated quantities, each item of input information has 
a corresponding parameter in the fit that we constrained to the 
measured value within its uncertainties.
We include each such constraint as a Gaussian or Poisson 
factor in the likelihood function.
For quantities 
with both Poisson statistical uncertainties and Gaussian systematic 
uncertainties, we adopted a Gaussian approximation of the Poisson
uncertainty and added them in quadrature.

It is important to understand that 
the only freely adjustable parameter in this fit 
is ($a_e + a_{\mu}$), the total number of $B_c$ signal events.
All other parameters are constrained within uncertainties 
by information independent of the \Bc\ candidate mass distribution.

\subsection{ Definitions }
\label{app:like_def}
{\bf Data.}
For the histogram in Fig.~\ref{fig:jpsi_track_mu}(b), we represent the 
number of candidate \BJpsimu\ events in the $i^{th}$\ bin as
$D^{\mu}_i$.  These numbers contribute factors to $\cal{L}$ 
according to the Poisson probabilities 
\begin{equation}
	P_{D^{\mu}_{i}}(\lambda^{\mu}_i) = 
	\frac{(\lambda^{\mu}_i)^{D^{\mu}_i}}{D^{\mu}_i !} 
	e^{-\lambda^{\mu}_i}
\label{eq:Pois}
\end{equation}
where the best estimate for the mean of $D^{\mu}_i$\ is represented by
$\lambda^{\mu}_i$, the function that sums the signal 
and background contributions calculated in the fit.
Each term in the sum is a product of parameters defined below.
In like manner, we symbolize the bin-by-bin numbers of 
candidate \BJpsie\ events 
(Fig.~\ref{fig:jpsi_track_e}(b)) by $D^e_i$\ and the 
functions representing their means by $\lambda^e_i$.

{\bf  \Bc\ Signal.}
The Monte Carlo simulation of \Bc\ production and decay, and response 
of our detector (App.~\ref{app:monte}) yielded 
mass distributions for \Jpsimu\ and \Jpsie\ 
and normalized each to unit area.  
Their values for the $i^{th}$\ bin 
are represented by $S^{\mu}_i$\ and $S^e_i$, respectively.
We symbolize the total number of \Jpsil\ events by
$n^{\prime \ell}$\ and the fraction of these in the \Jpsie\ channel
by $r^{\varepsilon}$.
For convenience, we express the numbers of events in 
the two decay channels as
$n^{\prime \mu} = (1-r^{\varepsilon}) n^{\prime \ell}$\ 
and $n^{\prime e} = r^{\varepsilon} n^{\prime \ell}$,
and we emphasize that these are derived from the fitted parameters.  
The contributions to $\lambda^{\mu}_i$\ 
and $\lambda^e_i$\ are $n^{\prime \mu} S^{\mu}_i$\ and $n^{\prime e} S^e_i$,
and their sum is shown in Fig.~\ref{fig:plot_sum}(a). 
The Monte Carlo simulation also determined the expected fraction  
of electron signal events, 
$R^{\varepsilon} = 0.583 \pm 0.043$, which contributes a constraining 
Gaussian probability factor to $\cal{L}$
\begin{equation}
	P(r^{\varepsilon}) 
	= \frac{1}{\sqrt{2 \pi} \Delta R^{\varepsilon} }
	\exp \left( -\frac{ ( r^{\varepsilon} - R^{\varepsilon})^2}
	{2(\Delta R^{\varepsilon})^2} \right).
\label{eq:Gaus}
\end{equation}

{\bf  False Muon Background.}
Fig.~\ref{fig:jpsi_track_mu}(a) shows the mass distribution for the 
subset of \Jpsi\ + track events that satisfied 
the purely geometric criteria for third-track muons.   
The bin contents of this distribution are represented by $J^{f\mu}_i$.  
This sample formed the parent distribution
for calculating the false muon contributions from punch-through and
decay-in-flight, and we combined these two sources of background
into a single distribution.  We calculated the bin-by-bin sums 
over the distributions in Figs.~\ref{fig:fake_mu}(a) and (b) 
and normalized each distribution to unit area.
To allow for a shape difference from the parent distribution,
we calculated the bin-by-bin fraction $F^{f\mu}_i \pm \Delta F^{f\mu}_i$
of the parent distribution.  We also scaled these fractions so that the 
resulting products $F^{f\mu}_i  J^{f\mu}_i$ summed to 1.0.
The quantities  $F^{f\mu}_i$\ account for any shape difference 
between the parent \Jpsi\ + track distribution ($J^{f\mu}_i$),
and the false-lepton distribution ($F^{f\mu}_i J^{f\mu}_i$), 
and they normalize the latter to unit area.
$N^{\prime f \mu} = 11.4 \pm 2.4$\ 
is the total number of background events that 
satisfied all the muon identification criteria.
In order to allow the fit to vary within the uncertainties in 
these measurements, we replaced them by parameters.
The fitted parameters $j^{f\mu}_i$\ 
were constrained by Poisson contributions to $\cal{L}$.
There was a similar constraint for $N^{\prime f\mu}$.
The fitted parameters $f^{f\mu}_i$\ were constrained 
through Gaussian factors in $\cal{L}$.
The parameter, $n^{\prime f\mu}$, is constrained to 
$N^{\prime f\mu}$\ 
by Gaussian factor in $\cal{L}$.
The contribution to $\lambda^{\mu}_i$\ from false muon backgrounds
is  $n^{\prime f\mu} f^{f\mu}_i j^{f\mu}_i$.

{\bf  False Electron Background.}
Our background estimate for false electrons used another subset of 
the \Jpsi\ + track distribution that satisfied the purely geometric 
criteria for third-track electrons.  This parent distribution is
$J^{fe}_i$.  (Below we discuss correlations between $J^{f\mu}_i$ and $J^{fe}_i$.)
The remaining input information and parameters 
for the false electron background are formally identical to 
those for the false muon background: $F^{fe}_i$, $f^{fe}_i$, $j^{fe}_i$, 
$N^{\prime fe} = 4.2 \pm 0.4$\ and $n^{\prime fe}$.
The contribution to $\lambda^e_i$\ from false electron backgrounds
is  $n^{\prime fe} f^{fe}_i j^{fe}_i$.

{\bf  Conversion Electron Background.}
We measured the number of identified conversion-electron background events 
to be $N^{\prime ce} = 2$. 
We represent this by a parameter $n^{\prime ce}$\ constrained 
to $N^{\prime ce}$\
by a Poisson factor in $\cal{L}$.
With the hybrid Monte Carlo calculation, we determined the ratio
of residual (not identified) conversions to identified conversions to be
$R^{ce} = 1.06 \pm 0.36$.
The corresponding fitted parameter is $r^{ce}$\ constrained by 
a Gaussian factor in the likelihood function.
We re-normalized the mass distribution for 
residual conversions in Fig.~\ref{fig:fake_e_conv}(b) to unit area,
represented by $J^{ce}_i \pm \Delta J^{ce}_i$.
The corresponding parameters are $j^{ce}_i$\ with 
constraining Gaussian probability factors in $\cal{L}$.
The contribution to $\lambda^e_i$\ from residual conversions 
is  $n^{\prime ce} r^{ce} j^{ce}_i$.

{\bf  \BBbar\ Backgrounds.}
We used a Monte Carlo procedure to calculate independently 
the \BBbar\ background contributions to \Jpsimu\ and \Jpsie.
The shapes of these were found to be identical and a single 
parent distribution distribution, $J^B_i \pm \Delta J^B_i$, normalized to 
unit area, was adopted for both.  It is represented by 
the parameters $j^B_i$ that are constrained by Gaussian 
terms in the likelihood function.
The Monte Carlo results for total numbers of events are:
$N^{\prime Be} = 2.3 \pm 0.9$\ for \Jpsie\ and
$N^{\prime B\mu} = 1.44 \pm 0.25$\ for \Jpsimu.
The corresponding parameters are $n^{\prime Be}$\ and $n^{\prime B\mu}$.
The contribution to $\lambda^{\mu}_i$\ and $\lambda^e_i$\ 
from  \BBbar\ background are, respectively, 
$n^{\prime B\mu} j^B_i$ and $n^{\prime Be} j^B_i$.

{\bf  Sums.}
We present here  
the two functions that, through their parameters, 
are adjusted for the best fit to the data distributions,
$D^{\mu}_i$ and $D^e_i$.
\begin{eqnarray}
  \lambda^{\mu}_i & = & (1 - r^{\varepsilon}) n^{\prime \ell} S^{\mu}_i 
			+ n^{\prime f\mu} f^{f\mu}_i j^{f\mu}_i
			+ n^{\prime B\mu} j^B_i\\
  \lambda^e_i &     = & r^{\varepsilon} n^{\prime \ell} S^e_i 
			+ n^{\prime fe} f^{fe}_i j^{fe}_i 
			+ n^{\prime ce} r^{ce} j^{ce}_i
			+ n^{\prime Be} j^B_i
 \end{eqnarray}

\subsection{ The Normalized Log-Likelihood Function }
\label{subapp:bin_m}
It is easy to show that a likelihood function ${\cal L}$,
which is the 
product of factors of the form given in Eq.~\protect\ref{eq:Gaus}, 
leads to $\xi^2 = \chi^2$\ through Eq.~\protect\ref{eq:xi2}.
For ${\cal L}$ composed of Poisson factors like those in 
Eq.~\protect\ref{eq:Pois} we find the corresponding factors in ${\cal L}_0$
to be
\begin{equation}
	P_{D^{\mu}_i}(D^{\mu}_i) = 
	\frac{(D^{\mu}_i)^{D^{\mu}_i}}{D^{\mu}_i !} 
	e^{-D^{\mu}_i}
\end{equation}
In the ratio
$P_{D^{\mu}_{i}}(\lambda^{\mu}_i)/P_{D^{\mu}_i}(D^{\mu}_i)$ 
the factorials cancel, 
and the contributions to the normalized log-likelihood function
are quite simple.
\begin{eqnarray}
  \xi^{\prime \, 2} & =	& -2 \ln \left( \frac{\cal L }
			{{\cal L}_0}\right) \\
	& =	& 2 \sum_i \left[ ( \lambda^{\mu}_i - D^{\mu}_i )
		- D^{\mu}_i \ln \left( \frac{\lambda^{\mu}_i }
		{D^{\mu}_i} \right)  \right]
\end{eqnarray}
where $\xi^{\prime \, 2}$\ is the first part $\xi^2$\ 
which we now write down in full.

\begin{mathletters}
\begin{eqnarray}
\xi^2 = &  2 \sum_i & \left\{ 
		\left[ ( \lambda^{\mu}_i - D^{\mu}_i )
		- D^{\mu}_i \ln \left( \frac{\lambda^{\mu}_i }
		{D^{\mu}_i} \right)  \right]
		+
		\left[ ( \lambda^e_i - D^e_i )
		- D^e_i \ln \left( \frac{\lambda^e_i }
		{D^e_i} \right)  \right]
		\label{eq:LL1}
		\right. 
\\ & \mbox{}+ & 
		\left. \left[ ( j^{f \mu}_i - J^{f \mu}_i )
		- J^{f \mu}_i \ln \left( \frac{j^{f \mu}_i }
		{J^{f \mu}_i} \right)  \right]
		+ \left[ ( j^{f e}_i - J^{f e}_i )
		- J^{f e}_i \ln \left( \frac{j^{f e}_i }
		{J^{f e}_i} \right)  \right] \right\}
		\label{eq:LL2}
\\ & + \sum_i &
		\left\{   
		\left(\frac{j^{ce}_i-J^{ce}_i}{\Delta J^{ce}_i}\right)^2
	+	\left(\frac{j^B_i-J^B_i}{\Delta J^B_i}\right)^2
	+	\left(\frac{f^{f\mu}_i-F^{f\mu}_i}{\Delta F^{f\mu}_i}\right)^2
	+	\left(\frac{f^{fe}_i-F^{fe}_i}{\Delta F^{fe}_i}\right)^2
		\right\}
		\label{eq:LL3}
\\ & + &
		\left(\frac{n^{\prime f\mu}-N^{\prime f\mu}}{\Delta N^{\prime f\mu}}\right)^2
	+	\left(\frac{n^{\prime fe}-N^{\prime fe}}{\Delta N^{\prime fe}}\right)^2
	+	\left(\frac{n^{\prime B\mu}-N^{\prime B\mu}}{\Delta N^{\prime B\mu}}\right)^2
	+	\left(\frac{n^{\prime Be}-N^{\prime Be}}{\Delta N^{\prime Be}}\right)^2
		\label{eq:LL4}
\\ & + &
		2 \left[ ( n^{\prime ce} - N^{\prime ce} )
		- N^{\prime ce} \ln \left( \frac{n^{\prime ce} }
		{N^{\prime ce}} \right)  \right]
	+	\left(\frac{r^{ce}-R^{ce}}{\Delta R^{ce}}\right)^2
	+	\left(\frac{r^{\varepsilon} - R^{\varepsilon}}
		{\Delta R^{\varepsilon}}\right)^2
		\label{eq:LL5}
\end{eqnarray}
\end{mathletters}
Line~\protect\ref{eq:LL1} is the fit to the \Bc\ candidate distributions.
Lines~\protect\ref{eq:LL2} and \protect\ref{eq:LL3} constrain the parent
distributions for the various backgrounds 
and the shape-dependent fractions for the false lepton distributions. 
Lines~\protect\ref{eq:LL4} and \protect\ref{eq:LL5} constrain the normalizations 
for the five background distributions, the Monte Carlo calculation
of the expected ratio of electron to muon \Bc\ events and the 
calculated ratio of residual to identified conversion-electron
background events.

\subsection{ Log-Likelihood for the Mass Analysis }
\label{subapp:unbin_m}
In the \Bc\ mass analysis, we performed an unbinned likelihood fit 
to the observed \Jpsil\ mass distribution. 
The unbinned likelihood function for this analysis was the product of 
the probability distributions for 
the \Jpsil\  mass for the \Bc\ signal and the backgrounds.
The individual contributions to the probability distribution played 
a role similar to that defined for the bin fractions in
App.~\ref{app:like_def} except that the bin index $i$\ was replaced
by $m_i$, the \Jpsil\ mass for the $i^{th}$\ event.  Further, the 
signal distribution differs for each assumed \Bc\ mass $M(B_c)$.
\begin{itemize}
 \item	$S_i^{\mu} \rightarrow S^{\mu}(m_{i},M_{B_c})$\ and
 	$S_i^{e} \rightarrow S^{e}(m_{i},M_{B_c})$ 
	represent the normalized signal distributions.
 \item	$F^{\mu}(m_i)$\ and $F^{e}(m_i)$\ represent the normalized 
	false $\mu$\ and false $e$\ background distributions.
 \item	$J^B(m_i) $\ represents the distribution of the \BBbar\ background 
	obtained from Monte Carlo calculations. 
 \item	$J^{ce}(m_i)$ represents the distribution for conversion and 
	Dalitz decay electrons. 
\end{itemize}
The preliminary version of 
each of the above functions was as a smooth spline fit to the 
appropriate binned distribution.  The fit was done 
prior to excluding events within 50 MeV of the $B^+$\ mass 
to eliminate \BJpsiK.  
The final version of the probability distribution was provided 
by a computer algorithm which,
given a specific value for $m_i$, 
returned the value of the spline function except when 
$m_i$\ was within the excluded region for \BJpsiK, in which
case it returned zero. 
The areas of the final probability distributions 
were each normalized to unity over the range 3.35 to 11.0 GeV/$c^2$.
$N'^{\mu}$ and  $N'^{e}$ are the total numbers 
of $\mu$ events and $e$ events.

The normalized probabilities for the muon and electron 
distributions are $\lambda^{\mu}/D^{\mu}$\ and
$\lambda^{e}/D^{e}$, where 
\begin{mathletters}
\begin{eqnarray}
        \lambda^{\mu}(m_{i},M_{B_c}) &=& 
	(1-r^{\varepsilon})n'^{\ell} S^{\mu}(m_{i},M_{B_c}) +  
	n'^{f\mu} F^{\mu}(m_i) + n'^{B\mu} J^B(m_i) 
  \\
	\lambda^{e}(m_{j},M_{B_c}) &=& r^{\varepsilon}n'^{\ell} 
		S^{e}(m_j,M_{B_c}) 
                + n'^{fe} F^{e}(m_j) 
                + n'^{Be} J^{Be}(m_j)+ n'^{ce} J^{ce}(m_j) 
  \\
	D^{\mu} &=&  
		(1 - r^{\varepsilon})n'^{\ell} + n'^{f\mu} + n'^{B\mu}
  \\
	D^{e} &=&  
		r^{\varepsilon}n'^{\ell} + n'^{fe}  + n'^{Be} + n'^{ce}
\end{eqnarray}
\end{mathletters}
The unbinned likelihood function contains the product of these 
probabilities for all the events.  The parameters in the probability
functions were adjusted for the best fit to the data.
The likelihood function also contained constraints on the parameters
determined independently of the candidate events.
We define the log-likelihood function by
  \begin{eqnarray}
        \xi^{2}_{m} = -2 \ln \left( \frac{ {\cal L}} { {\cal L}_{min} }\right)
  \end{eqnarray}
It is given by
\begin{mathletters}
\begin{eqnarray}
	\xi^{2}_{m} & = & -2 
	\left\{ 
	    \sum_{i}\left[ \ln\left( \frac{\lambda^{\mu}(m_{i},M_{B_c}) }
	    { D^{\mu} } \right)\right] 
	    + \sum_{j}\left[ \ln\left( \frac{
	    \lambda^{e}(m_{j},M_{B_c}) }{ D^{e} }\right)\right] 
	\right\}
   \label{eq:mline_1} \\
		& & 
		-2 \left\{
                - D^{\mu} + N'^{\mu} \ln D^{\mu} - D^{e} + N'^{e} \ln D^{e}
		\right\}
   \label{eq:mline_2} \\
                & & + \left( \frac{r^{\varepsilon} - R^{\varepsilon}}
                {\Delta R^{\varepsilon}} \right)^2
                + \left( \frac{n'^{f\mu} - N'^{f\mu}}{\Delta N'^{f\mu}}
	\right)^2 + \left( \frac{n'^{B\mu}-N'^{B\mu}}{\Delta
	  N'^{B\mu}} \right)^2
   \label{eq:mline_3} \\
                & & + \left(\frac{n'^{fe} - N'^{fe}}{\Delta N'^{fe}}\right)^2
                + \left(\frac{n'^{ce}-N'^{ce}}{\Delta N'^{ce}}\right)^2
                + \left(\frac{n'^{Be}-N'^{Be}}{\Delta N'^{Be}}
		  \right)^2 
   \label{eq:mline_4} \\
                & & + \; C 
\end{eqnarray}
\end{mathletters}
where $C$ was chosen so that 
$\xi^{2}_{m} = 0$  at ${\cal L} =  {\cal L}_{min}. $
Line~\protect\ref{eq:mline_1} is the fit to the \Bc\ candidate distributions.
Line~\protect\ref{eq:mline_2} 
is the constraint to the total numbers of \Jpsimu\ 
and \Jpsie\  events.
Lines~\protect\ref{eq:mline_3} and \protect\ref{eq:mline_4} 
constrain the ratio of $e$ to $\mu$  signals
and the number of background events for each background. 

\subsection{ Log-Likelihood for the Lifetime Analysis }
\label{subapp:unbin_ct}
The unbinned likelihood function used to fit the \Bc\ lifetime
was a product over the 42 \Jpsie\ and 
29 \Jpsimu\ candidates of the probability distribution
for \ctstar.  

The normalized probabilities which combine both signal and 
background distributions in $x_i = ct^{\ast}_i$\ 
for the \Jpsimu\ and \Jpsie\ 
are $\Lambda^{\mu}/D^{\prime \prime \mu}$\ and
$\Lambda^{e}/D^{\prime \prime e}$, where 
\begin{mathletters}
\begin{eqnarray}
        \Lambda^{\mu}(x_{i}, c \tau) &=& 
	(1-r^{\varepsilon})n^{\prime\prime \ell}
	{\cal F}_{sig}^{\mu}(x_{i},c \tau ) 
	+ n^{\prime \prime f\mu} {\cal F}^{f\mu}(x_i) 
	+ n^{\prime \prime B\mu} {\cal F}^{B\mu}(x_i) 
  \\
	\Lambda^{e}(x_{j}, c \tau ) &=& 
	r^{\varepsilon}n^{\prime \prime \ell}
	{\cal F}_{sig}^{e}(x_j, c \tau) 
        + n^{\prime \prime fe} {\cal F}^{fe}(x_j) 
        + n^{\prime \prime Be} {\cal F}^{Be}(x_j) 
	+ n^{\prime \prime ce} {\cal F}^{ce}(x_j) 
  \\
	D^{\prime \prime \mu} &=&  
		(1 - r^{\varepsilon})n^{\prime \prime \ell} 
		+ n^{\prime \prime f\mu} 
		+ n^{\prime \prime B\mu}
  \\
	D^{\prime \prime e} &=&  
		r^{\varepsilon}n^{\prime \prime \ell} 
		+ n^{\prime \prime fe} 
		+ n^{\prime \prime Be} 
		+ n^{\prime \prime ce}.
\end{eqnarray}
\end{mathletters}
The symbols are defined in Sec.~\ref{sec:life}.
The $\Lambda$-functions, of course, depend on all the fitted 
parameters, but we choose to emphasize the dependence on 
\ctau\, which is the only unconstrained parameter.

These probabilities are functions of the parameters given
in Table~\ref{tab:ct_bg} which describe the various 
backgrounds.  For each background, the shape parameters
were determined by a background fit that yielded the values in the 
table,  which we represent by $A^j_k$, 
and a variance matrix $V^j_{kl}$, where $j$\ is the
background label and $k$\ and $l$\ label the three or four
shape parameters.  
The lifetime fit varied a parameter, $a^j_k$\ corresponding 
to each of the  $A^j_k$, and these were constrained by 
a contribution to the log-likelihood function
\begin{equation}
	\chi^2_j = \sum_{k,l}(a^j_k - A^j_k)\cdot (V^j_{kl})^{-1} 
		\cdot (a^j_l - A^j_l)
\end{equation}

The number of events in each signal and backround contribution
was subjected to a Gaussian or Poisson constraint as in the previous 
parts of this Appendix.

We considered differences between the prompt background and 
that in the B-like region with $ct^* > 60$\ \um.
Our $dE/dx$\ measurements
indicated that for \Jpsi\ + track events,
the pion fraction for the third tracks in the prompt region 
was $f^p_{\pi} = 74 \pm 4$\% compared with 
$f^b_{\pi} = 56 \pm 3.4$\% noted in Sec.~\ref{subsec:punch}
for the B-like region.
These uncertainties are statistical only.
In order to account for systematic uncertainties, we
assumed $\rho_0 = f^b_{\pi}/f^p_{\pi} = 0.75 \pm 0.25$.
In the fit, we also allowed a variation in the relative 
probability $\omega_0 = 3.3 \pm 0.4$\ 
for pions and kaons to be mistakenly
identified as electrons.
The effect of this is to modify the values of 
some of $f^j_{\pm}$, which become cumbersome 
algebraic functions of the fitting parameters $\rho$\ and $\omega$.  
For clarity in the equations, we omit these details.

The log-likelihood function%
\footnote
{The log-likelihood function used here 
has a minimum of $-2 \ln {\cal L}^{comb} = -430$.
This value depends on the fact that \ctstar\ was expressed in 
cm in the computer program, 
although we have used \um\ in this report.
Had the programs used \um, the value would 
have been higher by $2 N^{\dpr \ell} \ln (10^4) = 1308$.}
that combines the 
unbinned fit to the \ctstar\ values for the candidate events
and constraints on the parameters describing the probabilities is 
\begin{mathletters}
\begin{eqnarray}
-2 \ln {\cal L}^{comb} 
&=& -2 \ln ( {\cal L}^{e} {\cal L}^{\mu})   \\
&=& -2\sum_{i}^{N^{\dpr e}} \ln \Lambda^e(x_i)
-2\sum_{i}^{N^{\dpr \mu}} \ln \Lambda^{\mu}(x_i)
\label{eq:like_ct1} \\
& & + 2 \left[ n^{\dpr B_c} + n^{\dpr fe} + n^{\dpr ce} r^{ce} 
	+ n^{\dpr Be}+ n^{\dpr f\mu} + n^{\dpr B\mu} 
	+ \ln (N^{\dpr e} !) + \ln (N^{\dpr \mu} !)  \right] 
\label{eq:like_ct2} \\
& & + \left ( \frac{r^{\varepsilon} -  R^{\varepsilon} }
	{\Delta R^{\varepsilon}} \right)^2
\label{eq:like_ct3} \\
& & + 2(n^{ce}- N^{\dpr ce} \ln n^{ce} + \ln (N^{\dpr ce}!)) 
+ \left ( \frac{r^{ce} - R^{ce} }{\Delta R^{ce}} \right)^2 
\label{eq:like_ct4} \\
& & + \left ( \frac{n^{f\mu} - N^{\dpr f\mu} }{\Delta N^{\dpr f\mu}} \right)^2
+ \left ( \frac{n^{B\mu} - N^{\dpr B \mu} }{\Delta N^{\dpr B \mu}} \right)^2 
+ \left ( \frac{n^{\dpr fe} - N^{\dpr fe} }{\Delta N^{\dpr fe}} \right)^2
+ \left ( \frac{n^{Be} - N^{\dpr Be} }{\Delta N^{\dpr Be}} \right)^2 
\label{eq:like_ct5} \\
& & + \left ( \frac{\rho -  \rho_0 }{\Delta \rho_0} \right)^2
+ \left ( \frac{\omega - \omega_0 }{\Delta \omega_0} \right)^2
+ \chi^2_  {fe} + \chi^2_{ce} + \chi^2_{f\mu}.
\label{eq:like_ct6} 
\end{eqnarray}
\end{mathletters}
Note that terms $N^{\dpr e}\ln D^{\dpr e}$\ 
and $N^{\dpr \mu}\ln D^{\dpr \mu}$\ 
do not appear because they cancel between  
the denominator of the log-probability sum 
(Line~\protect\ref{eq:like_ct1}) and the numerator of the Poisson constraint on 
the numbers of \Jpsie\ and \Jpsimu\ events (Line~\protect\ref{eq:like_ct2}). 
Line~\protect\ref{eq:like_ct3} is the constraint on the \Jpsie\ fraction in
the number of \Bc\ events.
Line~\protect\ref{eq:like_ct4} contains the Poisson constraint on the 
number of detected conversion electron background events and the 
Gaussian constraint on the ratio of undetected to detected background.
Line~\protect\ref{eq:like_ct5} contains Gaussian constraints on the 
numbers of other types of background events.
Finally, Line~\protect\ref{eq:like_ct6} provides constraints on 
$\rho$, $\omega$, and the shape parameters for the background
probability functions.

% ========================================================

%\input{ref}

\clearpage
%\bibliography{bc}

\begin{thebibliography}{10}

\bibitem{Lusignoli_decay}
M.~Lusignoli and M.~Masetti, Z. Phys. C {\bf 51}, 549 (1991).

\bibitem{isgw2}
N. Isgur, D. Scora, B. Grinstein, M. B. Wise, Phys. Rev. D {\bf 39}, 799
  (1989).

\bibitem{isgw}
D. Scora and N. Isgur, Phys. Rev. D {\bf 52}, 2783 (1995).

\bibitem{Chang_decay}
C.~H. Chang and Y.~Q. Chen, Phys. Rev. D {\bf 49}, 3399 (1994).

\bibitem{Kwong}
W.~Kwong and J.~Rosner, Phys. Rev. D {\bf 44}, 212 (1991).

\bibitem{Eichten}
E. Eichten and C. Quigg, Phys. Rev. D {\bf 49}, 5845 (1994).

\bibitem{Bigi}
I.~I.~Bigi, Phys. Lett. {\bf 371B}, 105 (1996).

\bibitem{Beneke}
M. Beneke and G. Buchalla, Phys. Rev. D {\bf 53}, 4991 (1996).

\bibitem{Gershtein}
S.~S. Gershtein {\it et al.}, Int. J. Mod. Phys. {\bf A6}, 2309 (1991).

\bibitem{Colangelo}
P. Colangelo {\it et al.}, Z. Phys. C {\bf 57}, 43 (1993).

\bibitem{Quigg}
C. Quigg, Proceedings of the Workshop on B Physics at Hadron Accelerators, ed.
  by P. McBride and C. Shekhar Mishra, Fermilab-CONF-93/267 (SSCL-SR-1225)
  (1994).

\bibitem{Choi}
Myoung-Taek Choi and Jae Kwan Kim, Phys. Rev. D {\bf 53}, 6670 (1996).

\bibitem{Lusignoli_prod}
M.~Lusignoli, M.~Masetti and S. Petrarca, Phys. Lett. B {\bf 266}, 142 (1991).

\bibitem{Braaten}
E. Braaten, K. Cheung and T.~C. Yuan, Phys. Rev. D {\bf 48}, R5049 (1993).

\bibitem{Chang_prod}
C.~H. Chang and Y.~Q. Chen, Phys. Rev. D {\bf 48}, 4086 (1993).

\bibitem{Oakes}
C.~H. Chang, Y.~Q. Chen and R.~J. Oakes, Phys. Rev. D {\bf 54}, 4344 (1996).

\bibitem{Masetti}
M.~Masetti and F.~Sartogo, Phys. Lett. {\bf 357B}, 659 (1995).

\bibitem{BcDELPHI}
P. Abreu {\it et al.}, The DELPHI Collaboration, Phys. Lett. {\bf 398B}, 207
  (1997).

\bibitem{BcOPAL}
K. Ackerstaff {\it et al.}, The OPAL Collaboration, CERN Preprint
  CERN-PPE/97-137, (submitted to Phys. Lett. B).

\bibitem{BcALEPH}
R.~Barate {\it et al.}, The ALEPH Collaboration, Phys.~Lett.~{\bf 402B}, 213
  (1997).

\bibitem{BcCDF}
F. Abe {\it et al.}, The CDF Collaboration, Phys. Rev. Lett. {\bf 77}, 5176
  (1996).

\bibitem{suzuki_thesis}
For details on the \Jpsie\ backgrounds and several aspects of the analysis of
  the \Jpsie\ channel see J. Suzuki, ``Observation of the \Bc\ Meson in 1.8-TeV
  Proton-Antiproton Collisions'', Ph.D. Dissertation, University of Tsukuba,
  1998.

\bibitem{singh_thesis}
For details on the \Jpsimu\ backgrounds and several aspects of the analysis of
  the \Jpsimu\ channel see Prem P. Singh, ``Search for the $B_c$ Meson at CDF",
  Ph.D. Dissertation, University of Pittsburgh, 1997.

\bibitem{CDF1}
F. Abe {\it et al.}, The CDF Collaboration, Nucl. Instrum. Methods Phys. Res.
  Sect. A {\bf 271}, 387 (1988).

\bibitem{CDF2}
F. Abe {\it et al.}, The CDF Collaboration, Phys. Rev. D {\bf 50}, 2966 (1994).
  Section 5.3 of this paper gives details of the electron and muon
  identification procedures similar to those used in the present analysis.

\bibitem{CDF3}
D.\ Amidei {\it et al.}, Nucl. Instrum. Methods Phys. Res. Sect. A {\bf 350},
  73 (1994).

\bibitem{CDF4}
P.~Azzi {\it et al.}, Nucl. Instrum. Methods Phys. Res. Sect. A {\bf 360}, 137
  (1995).

\bibitem{CFT}
Prompt signals from the CTC were sent to a fast track processor to provide
  measurements of track momenta with resolution $\delta p_T/p_T^2 = 0.03 ({\rm
  GeV/c})^{-1}$. For details, see G.~W.~Foster {\it et al.}, Nucl. Instrum.
  Methods Phys. Res. Sect. A {\bf 269}, 93 (1988).

\bibitem{B-lifetimes}
\Jpsi\ reconstruction in this analysis followed well established CDF procedures
  which have been reported elsewhere. See, {\it e.g.} F. Abe {\it et al.}, The
  CDF Collaboration, Submitted to Phys. Rev. D, Fermilab-Pub-97/352-E (1997).

\bibitem{Particledata}
``Review of Particle Physics," R.~M.~Barnett {\it et al.}, Phys. Rev. D {\bf
  54}, 1 (1996).

\bibitem{Mangano}
M. L. Mangano, P. Nason and G. Ridolfi, Nucl Phys. B {\bf 373}, 295 (1992).

\bibitem{Nason}
P. Nason, S. Dawson and R.~K. Ellis, Nucl. Phys. B {\bf 303}, 607 (1988); Nucl.
  Phys. B {\bf 327}, 49 (1989); Erratum {\bf 335}, 260 (1990).

\bibitem{MRSD}
A.~D. Martin, R.~G. Roberts and W.~J. Stirling, Phys. Lett. {\bf 306B}, 145
  (1993).

\bibitem{CTEQ}
James Bott {\it et al.}, Phys. Lett. {\bf B304}, 159 (1993).

\bibitem{CLEOMC}
P. Avery, K. Read and G. Trahern, Cornell Internal Note CSN-212, 1985
  (unpublished).

\bibitem{Scora}
``Semileptonic Hadron Decay in the Quark Potential Model'', D.~J. Scora, Ph.D.
  Dissertation, University of Toronto, (1992).

\bibitem{Peterson}
C. Peterson {\it et al.}, Phys. Rev. D {\bf 27}, 105 (1983).

\bibitem{Chrin}
J. Chrin, Z. Phys. C {\bf 36}, 163 (1987).

\bibitem{jetcone}
F. Abe {\it et al.}, The CDF Collaboration, Phys. Rev. D {\bf 45}, 1448 (1992).

\bibitem{bsemi}
J.D.~Richman and P.R.~Burchat, Rev. Mod. Phys. {\bf 67}, 893 (1995).

\end{thebibliography}
%\bibliographystyle{prsty}
% --------------  insert .bbl file here ------------------

% ========================================================

\newpage
% ---------------  \input{bctables}-----------------------
\mediumtext
\begin{table}
\caption{$B_c$ Signal and Background Summary:  The Counting Experiment}
\label{tab:count} 
\begin{tabular}{ lcc }
  & \multicolumn{2}{c}{ $4.0 < M(J \psi \, \ell) <  6.0$\ GeV/$c^2$ } \\
\tableline
  & \Jpsie\ results & \Jpsimu\ results  \\
\tableline
Misidentified leptons &  &  \\
\hspace{0.2in} False Electrons	& $2.6 \pm 0.05 \pm 0.3$ & \\
\hspace{0.2in} Conversions 	& $1.2 \pm 0.8 \pm 0.4$ & \\
\hspace{0.2in} Total False Muons	&	& $ 6.4 \pm 0.5 \pm 1.3$ \\
\hspace{0.4in} Punch-through & 		& $0.88 \pm 0.13 \pm 0.33$ \\
\hspace{0.4in} Decay-in-flight	&	& $ 5.5 \pm 0.5 \pm 1.3$ \\
$B \overline{B}$ bkg.	& $1.2 \pm 0.5$	& $0.7 \pm 0.3$ \\
\tableline
Total Background 
\tablenote{Upper limit on other backgrounds $<$\ 0.44.}
	 & $5.0 \pm 1.1$ & $7.1 \pm 1.5$ \\
\tableline
  Events observed in data & 19 & 12 \\
\tableline
Net Signal	& 14.0	& 4.9	\\
\multicolumn{1}{c}{Combined} & \multicolumn{2}{c}%
{$18.9$} \\
\tableline
$P_{Counting}$(Null) 
\tablenote{The probability that the background can account for the data
        in the absence of of a signal is based on a convolution of 
	Poisson uncertainties and Gaussian uncertainties of the backgrounds.}
	& $2.1 \times 10^{-5}$ & 0.084 \\
\end{tabular}
\end{table}

\newpage

\mediumtext
\begin{table}
\vspace*{-0.8in}
\caption{$B_c$ Signal and Background Summary: The Likelihood Analysis}
\label{tab:like} 
\begin{tabular}{ lcc }
 	& \multicolumn{2}{c}{ Input Constraint
\tablenote[1]{The numbers quoted here are for the mass range 
	$3.35 < M(J \psi \, \ell) <  11.0$\ GeV/$c^2$. }
	 } \\
 	& \multicolumn{2}{c}{ (Results of Fit)
\tablenotemark[1]
	} \\
\tableline
  & \Jpsie\ results & \Jpsimu\ results \\
\tableline
False Electrons	
	& $N^{\prime fe} = 4.2 \pm 0.4$		& \\
	& ($n^{\prime fe} = 4.2 \pm 0.4$) 	& \\
\tableline
Found Conversions
	& $N^{\prime ce} = 2$		& \\
	& ($n^{\prime ce} = 2.2 \pm 1.4$) 	& \\
\cline{2-3}
Conversion ratio
	& $R^{ce} = 1.06 \pm 0.36$ 		& \\
	& ($r^{ce} = 1.08 \pm 0.35$) 		& \\
\cline{2-3}
Unfound Conversions
\tablenote[2]{Derived from other parameters.}
	& $2.1  \pm 1.7$ 		& \\
	& ($2.4  \pm 1.7$) 		& \\
\tableline
False Muons
	& & $N^{\prime f \mu} = 11.4 \pm 2.4$ \\
	& & ($n^{\prime f \mu} = 9.2 \pm 2.3$) \\
\tableline
$B \overline{B}$ bkg.		
	& $N^{\prime Be} = 2.3 \pm 0.9$	
	& $N^{\prime B \mu} = 1.44 \pm 0.25$ \\
	& ($n^{\prime Be} = 2.6 \pm 0.9$) 
	& ($n^{\prime B \mu} = 1.42 \pm 0.25$) \\
\tableline
\hspace{0.2in} Total Background	
	& $8.6 \pm 2.0$		& $12.8 \pm 2.4$ \\
	& ($9.2 \pm 2.0$)	& ($10.6 \pm 2.3$) \\
\tableline
Total Signal
	& \multicolumn{2}{c}{($n^{\prime \ell} = 20.4^{+6.2}_{-5.6}$)} \\ 
\cline{2-3}
Electron Fraction	
	& \multicolumn{2}{c}{$R^{\varepsilon} = 0.58 \pm 0.04$} \\
	& \multicolumn{2}{c}{($r^{\varepsilon} = 0.59 \pm 0.04$)} \\
\cline{2-3}
$e$\ and $\mu$\ Signal
	& ($n^{\prime e} = 12.0^{+3.8}_{-3.2}$)	
	& ($n^{\prime \mu} = 8.4^{+2.7}_{-2.4}$) \\
\tableline
  Signal + Background	
	& 23			& 14 \\
	& ($21.2 \pm 4.3$)	& ($19.0 \pm 3.5$)\\
\tableline
$P$(Null)
\tablenote[3]{Probability that background alone can fluctuate
	   to produce an apparent signal of 20.4 events or
	   more, based on simulation of statistical fluctuations. }
	& \multicolumn{2}{c}{ $0.63 \times 10^{-6}$ } \\
\end{tabular}
\end{table}

\newpage
\mediumtext
\begin{table}
\caption{Parameters for Background Distributions in \ctstar}
\label{tab:ct_bg}
\begin{tabular}{ lccccc }
$j$	& $fe$	& $f\mu$ & $ce$	& $Be$	& $B\mu$	\\
\tableline
$N^{\prime \prime j}$	
	& $ 13.2 \pm 1.3$	& $12.6 \pm 2.8$	& 
\tablenote{The number of conversion background events 
	was calculated from identified conversions
	$N^{\prime \prime ce} = 3$\  and the ratio 
	$R^{ce} = 1.06 \pm 0.36$.  See App.~\ref{app:like_def}}
	& $1.5 \pm 1.1$	& $0.79 \pm 0.34$	\\
$f_+^j$	& $0.199 \pm 0.004$	& $0.36 \pm 0.01$	& $0.45 \pm 0.02$
	& $0.96 \pm 0.01$	& $0.98 \pm 0.06$	\\
$f_-^j$	& $0.032 \pm 0.004$	& $0.034 \pm 0.007$	& $0.12 \pm 0.02$
	& $1 - f_+^{Be}$	& $1 - f_+^{B\mu}$	\\
$\lambda_+^j$ (\um) & $371 \pm 15$	& $445 \pm 20$		& $382 \pm 27$
	& $371 \pm 15$		& $406 \pm 16$		\\
$\lambda_-^j$ (\um) & $103 \pm 9$	& $96 \pm 16$		& $138 \pm 27$
	& $65 \pm 15$		& $48 \pm 21$		\\
\end{tabular}
\end{table}

\newpage
\mediumtext
\begin{table}[hp]
\caption{ ${\cal R}({\rm f.s.}) =
	\frac{\sigma(B_c) \cdot BR(B_c \rightarrow {\rm f.s.})}
		   {\sigma(B) \cdot BR(B \rightarrow J \psi \, K)}$\ 
 Derived From Various Experimental Searches }
\label{tab:Br_limits} 
\begin{tabular}{ l | l | c }
Experiment
& \multicolumn{1}{c|}{final state (f.s.)} 
& ${\cal R}({\rm f.s.})$ \\
\hline \hline
DELPHI
\tablenote[1]{The ranges quoted for DELPHI and CDF Ref.~\cite{BcCDF} 
correspond to the assumed \Bc\ lifetime range 0.4 to 1.4 ps.}  
 Ref.~\cite{BcDELPHI} &
{$J/\psi \, \pi^{+}$}
& $<$ (0.9 to 0.7) (90\% C.L.) \\
  &
{$J/\psi \, \ell^{+}\nu$}
& $<$ (0.5 to 0.4) (90\% C.L.) \\
 &
{$J/\psi \,  \pi^{+} \pi^{-} \pi^{+}$}
& $<$\ 1.5 (90\% C.L.) \\
\hline
OPAL Ref.~\cite{BcOPAL} &
{$J/\psi \,  \pi^{+}$}
& $<$\ 0.6 (90\% C.L.) \\
 &
{$J/\psi \,  a^{+}_1$} 
& $<$\ 0.3 (90\% C.L.) \\
{} &
{$J/\psi \, \ell^{+}\nu$}
& $<$\ 0.4 (90\% C.L.) \\
\hline
ALEPH Ref.~\cite{BcALEPH} &
{$J/\psi \, \pi^{+}$}
& $<$\ 0.2 (90\% C.L.) \\
 &
{$J/\psi \, \ell^{+}\nu$}
& $<$\ 0.3 (90\% C.L.) \\
\hline
CDF Ref.~\cite{BcCDF}
\tablenotemark[1]
 & {$J/\psi \,  \pi^{+}$}
& $<$ (0.15 to 0.04) (95\% C.L.) \\
This Expt. &  {$J/\psi \, \ell^{+}\nu$} &
$ \sigmaBr $ 
\end{tabular}
\end{table}

\newpage
\begin{table}
\caption{\label{tab:secvtx} Calculated and observed false leptons
	in the background validation}
\begin{tabular}{lcc}
Tagged Sample
\tablenote{The numbers here are for events with 
dilepton mass $< 5$ GeV/$c^2$.}  
				& $e$		& $\mu$		\\
\hline
Observed $e$			& 33		& 37		\\
Expected background $+$ \BBbar 	& $43\pm10$	& $38\pm7$	\\
\hline
Observed $\mu$			& 43		& 63		\\
Expected background $+$ \BBbar 	& $41\pm4$	& $70\pm6$	\\
%\hline
\end{tabular}
\end{table}

% ========================================================

\newpage
%----------------\input{figuredef}----------------------

% detect.tex  fig 1  jpsi_mass.eps

% select.tex  fig 2   Jpsi_K.eps

% select.tex  fig 3  jpsi_track_e.eps

% select.tex  fig 4  jpsi_track_mu.eps

% select.tex  fig 5  pt_sum.eps

% background.tex  fig 6  fake_mu.eps

% background.tex  fig 7  DIF_pt.eps

% background.tex  fig 8  fake_ele.ps

% background.tex  fig 9  fake_e_conv.eps

% select.tex  fig 10 dedx_psi.eps

% background.tex  fig 11  psi_cv_pt.eps

% fit.tex  fig 12  plot_sum.eps

% fit.tex  fig 13 fit_emu.eps

% fit.tex  fig 14 fit_comb.eps

% fit.tex  fig 15 likelihood.eps

% fit.tex  fig 16 bc_like.ps

% fit.tex  fig 17 bc_sig.eps

% mass.tex  fig 18 bc_templates.eps

% mass.tex  fig 19 bcmass.eps

% life.tex  fig 20 kfig.eps

% life.tex  fig 21 bkg_shape.eps

% life.tex  fig 22 ct_fitk1.eps

% life.tex  fig 23 ct_fitk2.ps

% cross.tex  fig 24 lk.eps

% cross.tex  fig 25 ct_fit_test.eps

% cross.tex  fig 26 ct_crss1.ps

% valid.tex  fig 27 secvtx.eps

% valid.tex  fig 28 ip_bb.ps

% ========================================================

\end{document}